\RequirePackage{rotating}
\documentclass[acmsmall]{acmart}

\AtBeginDocument{%
  \providecommand\BibTeX{{%
    \normalfont B\kern-0.5em{\scshape i\kern-0.25em b}\kern-0.8em\TeX}}}

\setcopyright{acmcopyright}
\copyrightyear{2020}
\acmYear{2020}
\acmDOI{10.1145/1122445.1122456}

\acmJournal{JACM}
\acmVolume{37}
\acmNumber{4}
\acmArticle{111}
\acmMonth{8}

\usepackage{makecell}
\usepackage{pdflscape,array,booktabs}
\usepackage{enumitem}
\usepackage{outlines}
\usepackage{tabularx}
\usepackage{longtable}
\usepackage{mdframed}
\usepackage{soul}

\usepackage[textsize=scriptsize,backgroundcolor=yellow!40]{todonotes}

\begin{document}

\title{A Systematic Literature Review on Federated Machine Learning: From A Software Engineering Perspective}




\author{Sin Kit Lo}
\email{Kit.Lo@data61.csiro.au}
\affiliation{%
  \institution{Data61, CSIRO and University of New South Wales, Australia}
}

\author{Qinghua Lu}
\email{qinghua.lu@data61.csiro.au}
\affiliation{%
  \institution{Data61, CSIRO and University of New South Wales, Australia}
}

\author{Chen Wang}
\email{Chen.Wang@data61.csiro.au}
\affiliation{%
  \institution{Data61, CSIRO, Australia}
}

\author{Hye-young Paik}
\email{h.paik@unsw.edu.au}
\affiliation{%
  \institution{University of New South Wales, Australia}
}

\author{Liming Zhu}
\email{Liming.Zhu@data61.csiro.au}
\affiliation{%
  \institution{Data61, CSIRO and University of New South Wales, Australia}
}

\renewcommand{\shortauthors}{Lo, et al.}

\begin{abstract}
Federated learning is an emerging machine learning paradigm where clients 
train models locally and formulate a global model based on the local model updates. To identify the state-of-the-art in federated learning and explore how to develop federated learning systems, we perform a systematic literature review from a software engineering perspective, based on 231 primary studies. 
Our data synthesis covers the lifecycle of federated learning system development that includes background understanding, requirement analysis, architecture design, implementation, and evaluation. We highlight and summarise the findings from the results, and identify future trends to encourage researchers to advance their current work.
\end{abstract}



\maketitle

\section{Introduction}
Machine learning is adopted broadly in many areas, and data plays a critical role in machine learning systems due to its impact on the model performance. Although the widely deployed remote devices (e.g., mobile/IoT devices) generate massive amounts of data, data hungriness is still a challenge because of the increasing concern in data privacy (e.g., General Data Protection Regulation - GDPR~\cite{(gdpr)_2019}) 


To effectively address this challenge, federated learning was proposed by Google in 2016~\cite{mcmahan2016communicationefficient}. In federated learning, client devices perform model training locally and generate a global model collaboratively.
The data is stored locally and never transferred to the central server or other clients~\cite{kairouz2019advances, dinh2019federated}. Instead, only model updates are communicated for formulating the global model.

The growing interest in federated learning has increased the number of research projects 
and publications in the last four years. 
Although there are surveys conducted on this topic~\cite{kairouz2019advances, Li_2020, li2019survey}, there is still no systematic literature review on federated learning. It motivates us to perform a systematic literature review on federated learning to understand the state-of-the-art. 
Furthermore, client devices in federated learning systems form a large-scale distributed system. It calls for software engineering considerations apart from the core machine learning knowledge~\cite{8812912}. Thus, we explore how to develop federated learning systems by conducting a systematic literature review to provide a holistic and comprehensive view of the state-of-the-art federated learning research, especially from a software engineering perspective.




We perform a systematic literature review following Kitchenham's standard guideline~\cite{Kitchenham07guidelinesfor}. The objectives are to: (1) provide an overview of the research activities and diverse research topics in federated learning system development; (2) help practitioners understand challenges and approaches to develop a federated learning system.
The contributions of this paper are as follow:
\begin{itemize}

\item We present a comprehensive qualitative and quantitative synthesis reflecting the state-of-the-art in federated learning with data extracted from 231 primary studies. Our data synthesis investigates different stages of federated learning system development.

\item We provide all the empirical findings and identify future trends in federated learning research.
\end{itemize}

The remainder of the paper is organised as follows: Section~\ref{Section:ResearchMethodology} introduces the methodology. Section~\ref{Section:Results} presents the results and highlights the findings. 
Section~\ref{Section:Future} identifies future trends, followed by threats to validity in Section~\ref{Section:Threats}. Section~\ref{Section:RelatedWorks} discusses related work. Section~\ref{Section:Conclusion} concludes the paper.

\section{Methodology} \label{Section:ResearchMethodology}


Based on Kitchenham's  guideline~\cite{Kitchenham07guidelinesfor}, we developed the following protocol.

\begin{figure}[h]
  \includegraphics[width=0.9\linewidth]{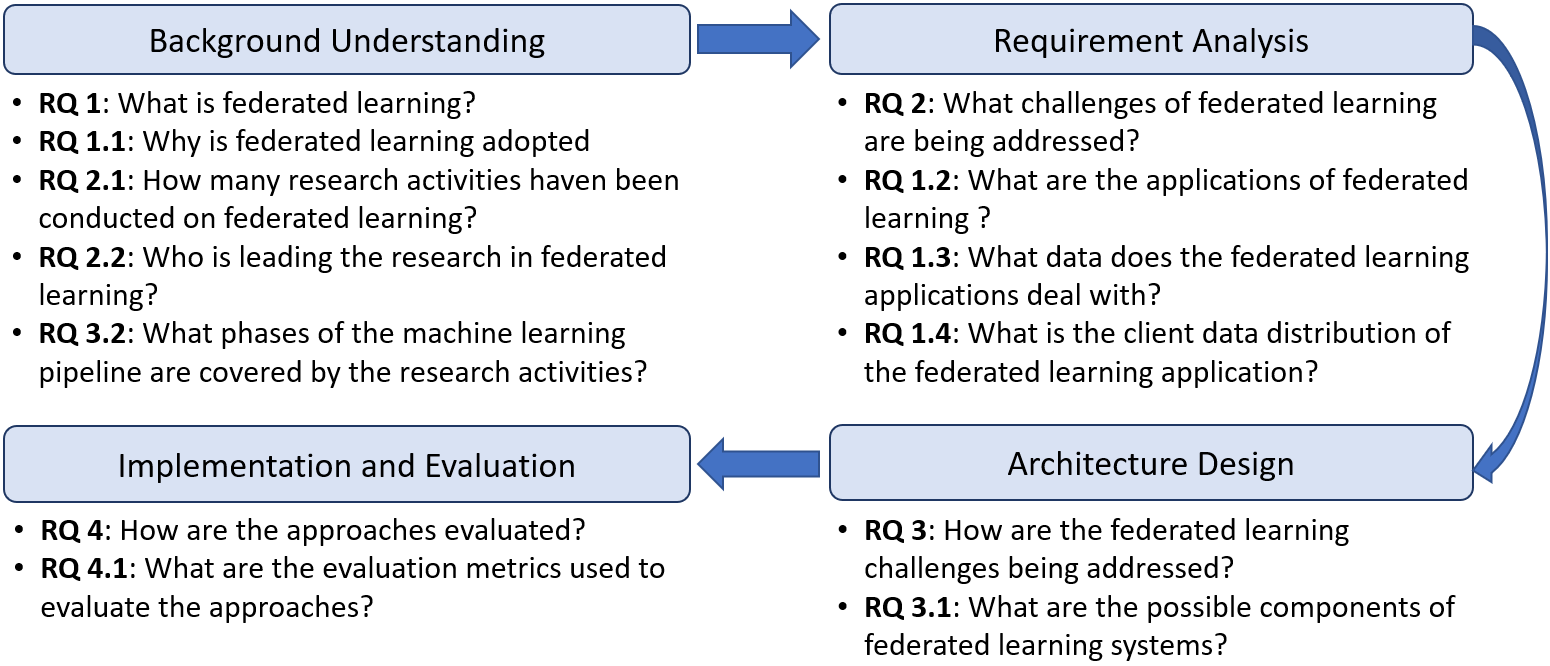}
  \caption{Research questions mapping with Software Development Life Cycle (SDLC)}
  \label{fig:RQ_SDLC}
\end{figure}

\subsection{Research Questions}
To provide a systematic literature review from the software engineering perspective, we view federated learning as a software system and study federated learning systems from the software development standpoint. As shown in Fig~\ref{fig:RQ_SDLC}, we adopt the software development practices of machine learning systems in~\cite{8812912} to describe the software development lifecycle (SDLC) for federated learning. 
In this section, we explain how each research question (RQ) is derived. 

\subsubsection{Background Understanding: }
To develop a federated learning system, we need to first know what federated learning is and when to adopt federated learning instead of centralised learning. Thus, we derive~\textbf{RQ 1 (What is federated learning?)} and~\textbf{RQ 1.1 (Why is federated learning adopted?)}. After learning the background and adoption objectives of federated learning, we intend to identify the research efforts and the experts in this field through~\textbf{RQ 2.1 (How many research activities have been conducted on federated learning?)} and~\textbf{RQ 2.2 (Who is leading the research in federated learning?)}. Lastly, we derive~\textbf{RQ 3.2 (What are the phases of the machine learning pipeline covered by the research activities?)} to examine the focused machine learning pipeline stages in the existing studies and understand the maturity of the area. 

\subsubsection{Requirement Analysis: }
After the background understanding stage, the requirements of  federated learning systems are analysed. We focus on the non-functional requirements since functional requirements are application-specific. We derive~\textbf{RQ 2 (What challenges of federated learning are being addressed?)} to identify the architectural drivers (i.e. non-functional requirements) of federated learning systems. ~\textbf{RQ 1.2 (What are the applications of federated learning?)},~\textbf{RQ 1.3 (What data does the federated learning applications deal with?)} and~\textbf{RQ 1.4 (What is the client data distribution of the federated learning applications?)} are designed to help  researchers and practitioners assess the suitability of federated learning for their systems, which is within the scope of the requirement analysis. 

\subsubsection{Architecture Design: }
After the requirement analysis, researchers and practitioners need to understand how to design the architecture. For this, we consider the approaches against each requirement. Hence, we derive~\textbf{RQ 3 (How are the federated learning challenges being addressed?)} and~\textbf{RQ 3.1 (What are the possible components of federated learning systems?)}. These 2 RQs aim (1) to  identify the possible approaches that address the challenges during the federated learning system development, and (2) to extract the software components for federated learning architecture design to fulfill the non-functional requirements (i.e. challenges).

\subsubsection{Implementation and Evaluation: }
After the architecture design stage, once the system is implemented, the federated learning systems including the built models need to be evaluated. Thus, we derive~\textbf{RQ 4 (How are the approaches evaluated?)} and~\textbf{RQ 4.1 (What are the evaluation metrics used to evaluate the approaches?)} to identify the methods and metrics for the evaluation of federated learning systems.

\subsection{Sources Selection and Strategy} \label{Section:Sourceselection}
We searched through the following search engines and databases: (1) \emph{ACM Digital Library}, (2) \emph{IEEE Xplorer}, (3) \emph{ScienceDirect} (4) \emph{Springer Link}, (5) \emph{ArXiv}, and (6) \emph{Google scholar}. 
The search time frame is set between 2016.01.01 and 2020.01.31. 
We screened and selected the papers from the initial search according to the preset inclusion and exclusion criteria elaborated in Section~\ref{Section:inclusionexclusion}.

\begin{figure}[h]
  \centering
  \resizebox{0.75\textwidth}{!}{
  \includegraphics[width=\linewidth]{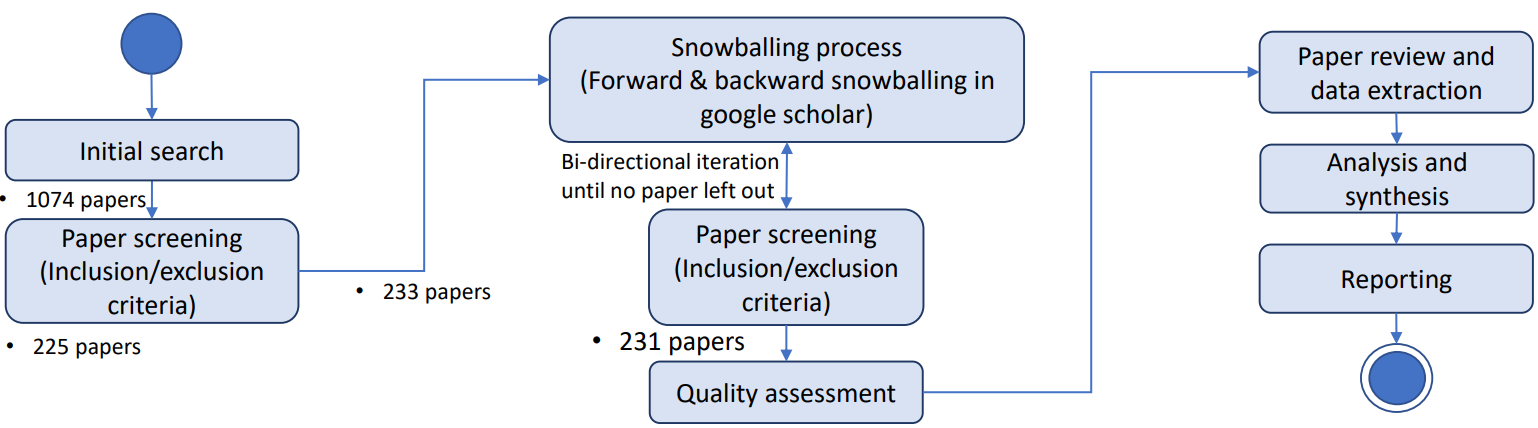}}
  \caption{Paper search and selection process map}
  \label{fig:searchprocess}
 \end{figure}

We then conducted forward and backward snowballing processes to search for any related papers that were left out from the initial search. 
The paper selection process consists of 2 phases: (1) The papers were first selected by two researchers through title and abstract screening independently, based on the inclusion and exclusion criteria. Then, the two researchers cross-checked the results and resolved any disagreement on the decisions. (2) The papers selected in the first phase were then assessed through full-text screening. The two researchers again cross-checked the selection results and resolved any disagreement on the selection decisions. Should any disagreement have occurred in either the first or the second phase, a third researcher (meta-reviewer) 
was consulted to finalise the decision. Fig.~\ref{fig:searchprocess} shows the paper search and selection process.


The initial search found 1074 papers, with 76 from~\emph{ACM Digital Library}, 320 from~\emph{IEEE Xplorer}, 5 from~\emph{ScienceDirect}, 85 from~\emph{Springer Link}, 256 from~\emph{ArXiv}, and 332 from~\emph{Google scholar}. After the paper screening, exclusion, and duplicates removal, we ended up with 225 papers. From there, we conducted the snowballing process and found 6 more papers. The final paper number for this review is 231. 
The number of papers per source are presented in Table 1. 

\begin{table*}[h]
  \caption{Number of selected publications per source}
  \label{paperpersource}
  \footnotesize
  \begin{tabular}{c|ccccccc}
    \toprule
    \textbf{Sources} & \textbf{ACM} & \textbf{IEEE} & \textbf{Springer} & \textbf{ScienceDirect} & \textbf{ArXiv} & \textbf{Google Scholar} & \textbf{Total}\\
    \midrule
    \textbf{Paper count} & \textbf{22} & \textup{74} & \textup{17} & \textup{4} & \textup{106} & \textup{8} & \textup{231}\\
    \bottomrule
  \end{tabular}
\end{table*}

\subsubsection{Search String Definition}
We used ``Federated Learning'' as the key term and included synonyms and abbreviations as supplementary terms to increase the search results. 
We designed the search strings for each primary source to check the title, abstract, and keywords. 
After completing the first draft of search strings, we examined the results of each search string against each database to check the effectiveness of the search strings. 
The finalised search terms are shown in Table~\ref{tab:key}. The search strings and the respective paper quantities of the initial search for each primary source are shown in Table~\ref{tab:ACM}, \ref{tab:IEEE}, \ref{tab:ScienceDirect}, \ref{tab:Springer}, \ref{tab:GoogleScholar}, and \ref{tab:ArXiv}.

\begin{table*}[h]
  \caption{Key and supplementary search terms}
  \label{tab:key}
  \footnotesize
  \begin{tabular}{l|l}
    \toprule
    \textbf{Key Term} & \textbf{Supplementary Terms}\\
    \midrule
    \textup{{\makecell[l]{Federated Learning}}} & \textup{{\makecell[l]{Federated Machine Learning, Federated ML, Federated Artificial Intelligence, Federated AI,\\Federated Intelligence, Federated Training}}}\\
    \bottomrule
  \end{tabular}
\end{table*}

\begin{table*}
  \caption{Search strings and quantity of \emph{ACM Digital Library}}
  \label{tab:ACM}
  \footnotesize
  \begin{tabular}{c|c}
    \toprule
    \textbf{Search string} & \textup{\makecell[l]{All: "federated learning"] OR [All: ,] OR [All: "federated machine learning"] OR [All: "federated ml\\"] OR [All: ,] OR [All: "federated intelligence"] OR [All: ,] OR [All: "federated training"] OR [All: ,]\\OR [All: "federated artificial intelligence"] OR [All: ,] OR [All: "federated ai"]\\AND [Publication Date: (01/01/2016 TO 01/31/2020)]}}\\
    \hline
    \textbf{\makecell{Result quantity}} & \textup{\makecell{76}}\\
    \hline
    \textbf{\makecell{Selected papers}} & \textbf{\makecell{22}}\\
    \bottomrule
  \end{tabular}
  \end{table*}
  
\begin{table*}
  \caption{Search strings and quantity of \emph{IEEE Xplorer}}
  \label{tab:IEEE}
  \footnotesize
  \begin{tabular}{c|c}
    \toprule
    \textbf{Search string} & \textup{\makecell[l]{("Document Title":"federated learning" OR "federated training" OR "federated intelligence" OR\\"federated machine learning" OR "federated ML" OR "federated artificial intelligence" OR\\"federated AI") OR ("Author Keywords":"federated learning" OR "federated training" OR\\"federated intelligence" OR "federated machine learning" "federated ML" OR "federated\\artificial intelligence" OR "federated AI")}}\\
    \hline
    \textbf{\makecell{Result quantity}} & \textup{\makecell{320}}\\
    \hline
    \textbf{\makecell{Selected papers}} & \textbf{\makecell{71}}\\
    \bottomrule
  \end{tabular}
  \end{table*}

\begin{table*}
  \caption{Search strings and quantity of \emph{ScienceDirect}}
  \label{tab:ScienceDirect}
  \footnotesize
  \begin{tabular}{c|c}
    \toprule
    \textbf{Search string} & \textup{\makecell[l]{"federated learning" OR "federated intelligence" OR "federated training" OR "federated machine\\learning" OR "federated ML" OR "federated artificial intelligence" OR "federated AI"}}\\
    \hline
    \textbf{\makecell{Result quantity}} & \textup{\makecell{5}}\\
    \hline
    \textbf{\makecell{Selected papers}} & \textbf{\makecell{4}}\\
    \bottomrule
  \end{tabular}
  \end{table*}

\begin{table*}
  \caption{Search strings and quantity of \emph{Springer Link}}
  \label{tab:Springer}
  \footnotesize
  \begin{tabular}{c|c}
    \toprule
    \textbf{Search string} & \textup{\makecell[l]{"federated learning" OR "federated intelligence" OR "federated training" OR "federated machine\\learning" OR "federated ML" OR "federated artificial intelligence" OR "federated AI"}}\\
    \hline
    \textbf{\makecell{Result quantity}} & \textup{\makecell{85}}\\
    \hline
    \textbf{\makecell{Selected papers}} & \textbf{\makecell{17}}\\
    \bottomrule
  \end{tabular}
  \end{table*}
  
\begin{table*}
  \caption{Search strings and quantity of \emph{ArXiv}}
  \label{tab:ArXiv}
  \footnotesize
  \begin{tabular}{c|c}
    \toprule
    \textbf{Search string} & \textup{\makecell[l]{order: -announced\_date\_first; size: 200; date\_range: from 2016-01-01 to 2020-01-31\; include\_cross\\\_list:True; terms: AND title=``federated learning'' OR ``federated intelligence'' OR ``federated\\training'' OR ``federated machine learning'' OR ``federated ML'' OR ``federated artificial\\intelligence'' OR ``federated AI''; OR abstract=``federated learning'' OR\\``federated intelligence'' OR ``federated training'' OR ``federated machine learning'' OR ``federated\\ML'' OR ``federated artificial intelligence'' OR ``federated AI''}}\\
    \hline
    \textbf{\makecell{Result quantity}} & \textup{\makecell{256}}\\
    \hline
    \textbf{\makecell{Remark}} & \textup{\makecell{Search title and abstract only (\emph{ArXiv} does not provide keyword search option)}}\\
    \hline
    \textbf{\makecell{Selected papers}} & \textbf{\makecell{103}}\\
    \bottomrule
  \end{tabular}
  \end{table*} 
  
\begin{table*}
  \caption{Search strings and quantity of \emph{Google scholar}}
  \label{tab:GoogleScholar}
  \footnotesize
  \begin{tabular}{c|c}
    \toprule
    \textbf{Search string} & \textup{\makecell[l]{"federated learning" OR "federated intelligence" OR "federated training" OR "federated machine\\learning" OR "federated ML" OR "federated artificial intelligence" OR "federated AI"}}\\
    \hline
    \textbf{\makecell{Result quantity}} & \textup{\makecell{332}}\\
    \hline
    \textbf{\makecell{Remark}} & \textup{\makecell{Search title only (\emph{Google scholar} does not provide abstract \& keyword search option)}}\\
    \hline
    \textbf{\makecell{Selected papers}} & \textbf{\makecell{8}}\\
    \bottomrule
  \end{tabular}
  \end{table*}

\subsubsection{Inclusion and Exclusion Criteria} \label{Section:inclusionexclusion}
The inclusion and exclusion criteria are formulated to effectively select relevant papers. After completing the first draft of the criteria, we conducted a pilot study on 20 randomly selected papers. Then, the two independent researchers cross-validated the papers selected by the other researcher and refined the criteria. 
The finalised inclusion criteria are as follow:

\begin{itemize}
 \item Both long and short papers that have elaborated on the component interactions of the federated learning system: We specifically focus on the research works that provide comprehensive explanations on the federated learning components functionalities and their mutual interactions. 
 
 \item Survey, review, and SLR papers: We included all the surveys and review papers to identify the open problems and future research trends in an objective manner. However, we excluded them from the stages for answering research questions.
 
 \item We included ArXiv and Google scholar's papers cited by the peer-reviewed papers published in the primary sources. 
\end{itemize}

The finalised exclusion criteria are as follow:

\begin{itemize}
    \item Papers that elaborate only low-level communication algorithms: The low-level communication algorithms or protocols between hardware devices are not the focus of this work.
    \item Papers that focus only on pure gradient optimisation. We excluded the papers that purely focus on the gradient and algorithm optimisation research. Our work focuses on the multi-tier processes and interactions of the federated learning software components.
    \item Papers that are not in English.
    \item Conference version of a study that has an extended journal version.
    \item PhD dissertations, tutorials, editorials and magazines.
\end{itemize}

\subsubsection{Quality Assessment}
A quality assessment scheme was developed to evaluate the quality of the papers. There are 4 quality criteria (QC) used to rate the papers. We used numerical scores ranging from 1.00 (lowest) to 5.00 (highest) to rate each paper. The average scores are calculated for each QC and the total score of all 4 QCs are obtained. We included papers that score greater than 1.00 to avoid missing out on studies that are insightful while maintaining the quality of the search pool. The QCs are as follow:

\begin{itemize}
    \item QC1: The citation rate. We identified this by checking the number of citations received by each paper according to ~\emph{Google scholar}.
    \item QC2: The methodology contribution. We identified the methodology contribution of the paper by asking 2 questions: (1) Is this paper highly relevant to the research? (2) Can we find a clear methodology that addresses its main research questions and goals?
    \item QC3: The sufficient presentation of the findings. Are there any solid findings/results and clear-cut outcomes? Each paper is evaluated based on the availability of results and the quality of findings.
    \item QC4: The future work discussions. We assessed each paper based on the availability of discussions on future work. 
\end{itemize}

\subsubsection{Data Extraction and Synthesis}
We downloaded all the selected papers and recorded all the essential information, including the title, source, year, paper type, venue, authors, affiliation, the number of citations of the paper, the score from all QCs, the answers for each RQ, and the research classification (refer to Appendix~\ref{appendix_1}\footnote{Data extraction sheet, \url{https://drive.google.com/file/d/10yYG8W1FW0qVQOru_kMyS86owuKnZPnz/view?usp=sharing}}). The following steps were followed to prevent any data extraction bias:

\begin{itemize}
    \item The two independent researchers conducted the data extraction of all the papers and cross checked the classification and discussed any dispute or inconsistencies in the extracted data.
    \item For any unresolved dispute on the extracted data, the first two authors tried to reach an agreement. When an agreement was not met, the meta reviewer reviewed the paper and finalised the decision together. 
    \item All the data was recorded in the Google sheet for analysis and synthesis processes.
\end{itemize}


\section{Results} \label{Section:Results}
In this section, the extracted results of each research question are summarised and analysed. 

  \begin{figure}[h]
  \centering
  \includegraphics[width=0.5\linewidth]{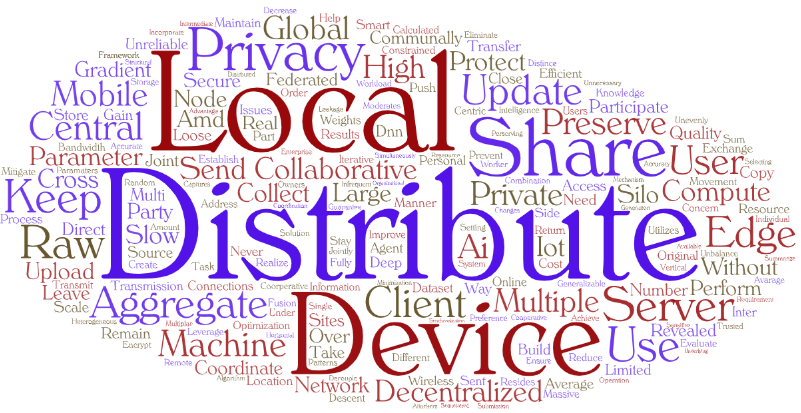}
  \caption{RQ 1: What is federated learning?}
  \label{fig:RQ1}
 \end{figure}

\begin{center}
\begin{table*}[h]
\caption{Characteristics of federated learning}
\scriptsize
  \label{tab:RQ1}
  \begin{tabular}{lll}
    \toprule
    \textbf{\thead{Category}} & \textbf{\thead{Characteristic}} & \textbf{\thead{Paper count}}\\
    \midrule
    \textup{\makecell[l]{Training settings (82\%)}} & \textup{\makecell[l]{Training a model on multiple clients\\Only sending model updates to the central server\\Producing a global model on the central server}} &  \textup{\makecell[r]{200\\133\\63}}\\
    \hline \\[-1.8ex]
        \textup{\makecell[l]{Data distribution (11\%)}} & \textup{\makecell[l]{Decentralised data storage\\Data generated locally}} &  \textup{\makecell[r]{40\\17}}\\
    \hline \\[-1.8ex]
\textup{\makecell[l]{Orchestration (3\%)}} & \textup{\makecell[l]{Training organised by a central server}} &  \textup{\makecell[r]{13}}\\
    \hline \\[-1.8ex]
    \textup{\makecell[l]{Client types (3\%)}} & \textup{\makecell[l]{Cross-device\\Cross-silo\\Both}} &  \textup{\makecell[r]{11\\1\\3}}\\
    \hline \\[-1.8ex]
    \textup{\makecell[l]{Data partitioning (<1\%)}} & \textup{\makecell[l]{Horizontal federated learning\\Vertical federated learning}} &  \textup{\makecell[r]{1\\1}}\\
    \bottomrule
\end{tabular}
\end{table*}
\end{center}

\subsection{RQ 1: What is federated learning?} \label{Section:RQ1}

The first research question (RQ 1) is "What is federated learning?". To answer RQ 1, the definition of federated learning reported by each study is recorded. 
This question helps the audience to understand: (1) what federated learning is, and (2) the perceptions of researchers on federated learning. Fig.~\ref{fig:RQ1} is a word cloud that shows the frequency of the words that appear in the original definition of federated learning in each study. The most frequently appeared words include: distribute, local, device, share, client, update, privacy, aggregate, edge, etc. To answer RQ1 more accurately,
we use these five categories to classify the definitions of federation learning: (1) training settings, (2) data distribution, (3) orchestration, (4) client types, and (5) data partitioning, as understood by the researchers, as shown in Table \ref{tab:RQ1}.

First, in the training settings, building a decentralised or distributed learning process over multiple clients is 
what most researchers conceived as federated learning. 
This can be observed as the most frequently mentioned keyword in RQ1 is ``training a model on multiple clients''. 
Other frequently mentioned keywords that describe the training settings are ``distributed'', ``collaborative'', and ``multiple parties/clients''. ``Only sending the model updates to the central server'' and ``producing a global model on the central server'' are the other two characteristics that describe how a federated learning system performs model training in a distributed manner. This also shows how researchers differentiate federated learning from conventional machine learning and distributed machine learning.



Secondly, federated learning can be explained in terms of the data distributions.
The keywords mentioned in the studies are ``data generated locally'' 
and ``data stored decentralised''. 
Data is collected and stored by client devices in different geographical locations. Hence, it exhibits non-IID (non-Identically \& Independently Distributed) and unbalanced data properties~\cite{kairouz2019advances, mcmahan2016communicationefficient}. Furthermore, the data is decentralised and is not shared with other clients to preserve data privacy. We will discuss more on the client data distribution in Section~\ref{Section:RQ1.4}.

Thirdly, researchers observe federated learning from the training process orchestration standpoint. In conventional federated learning, a central server orchestrates the training processes. The tasks consist of initialisation of a global model, distribution of the global models to participating client devices, collection of trained local models, and the aggregation of the collected local models to update the global model. Intuitively, researchers consider the usage of a single central server as a possible single-point-of-failure~\cite{8950073, 8892848}. Hence, decentralised approaches for the exchange of model updates are studied and the adoption of blockchains for decentralised data governance is introduced~\cite{8950073, 8892848}.


Fourthly, we observe two types of federated learning in terms of client types which are cross-device and cross-silo.
Cross-device federated learning deals with a massive number of smart devices, creating a large-scale distributed network to collaboratively train a model for the same applications~\cite{kairouz2019advances}. Some examples of the 
applications are mobile device keyboard word suggestions and human activity recognition. The setting are extended to cross-silo applications where data sharing between organisations is prohibited. 
For instance, data of a hospital is prohibited from exposure to other hospitals due to data security regulations. 
To enable machine learning under this environment, cross-silo federated learning conducts local model training using the data in each hospital (silo)~\cite{8818446, kairouz2019advances}.


Lastly, we found 3 data partitioning variations: horizontal, vertical, and federated transfer learning. Horizontal federated learning, or sample-based federated learning, is used when the datasets share the same feature space but different sample ID space~\cite{10.1145/3298981, liu2019communication}. Inversely, vertical federated learning, also known as feature-based federated learning, is applied to the cases where two or more datasets share the same sample ID space but different feature space. Federated transfer learning considers the data partitioning where two datasets only overlap partially in the sample space or the feature space. It aims to develop models that are applicable to both datasets~\cite{10.1145/3298981}. 

We summarised the findings as below. We connected each keyword and grouped them under the same definition. Finally, we arranged the definitions according to the frequency of the words that appeared.

\begin{figure*}[h!]
\begin{center}
\footnotesize
\begin{mdframed}[
    skipabove=1cm,    
    innerleftmargin =-1cm,
    innerrightmargin=-1cm,
    usetwoside=false,
]

\begin{center}
    \textbf{Findings of RQ 1: What is federated learning (FL)?}\\
\end{center} 

\begin{quotation}

\textit{\textbf{Federated learning}} is a type of distributed machine learning to preserve data privacy. Federated learning systems rely on a central server to coordinate the model training process on multiple, distributed client devices where the data are stored. The model training is performed locally on the client devices, without moving the data out of the client devices. Federated learning can also be performed in a decentralised manner.
\end{quotation}

\begin{quotation}
\textit{\textbf{Variations in federated learning:}} (1) centralised/decentralised federated learning, (2) cross-silo/device federated learning, (3) horizontal/vertical/transfer federated learning.
\end{quotation}

\begin{quotation}
\textbf{With regard to the software development lifecycle}, this question contributes to the \textit{background understanding} phase where we provide the definition of the fundamental settings and different variations of federated learning as reported by researchers and practitioners.
\end{quotation}

\end{mdframed}
\end{center} 
\end{figure*}

\subsection{RQ 1.1: Why is federated learning adopted?}\label{Section:RQ1.1}

\begin{figure}[h]
  \centering
  \includegraphics[width=0.65\linewidth]{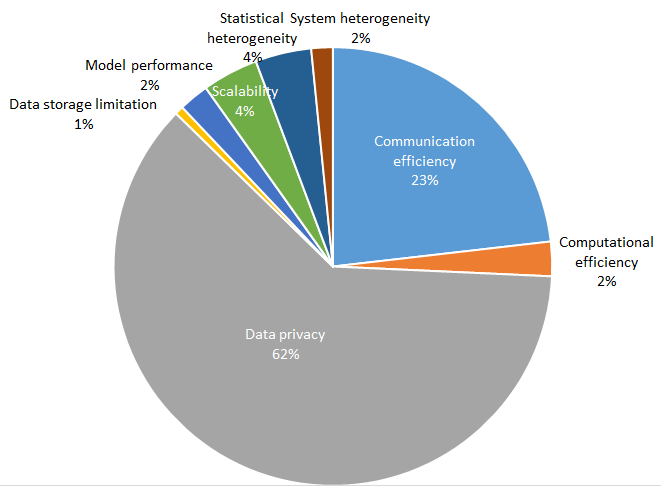}
  \caption{RQ 1.1: Why federated learning is adopted?}
  \label{fig:rq1.1}
 \end{figure}

The motivation of RQ 1.1 is to understand the advantages of federated learning. 
We classify the answers based on the non-functional requirements of federated learning adoption (illustrated in Fig.~\ref{fig:rq1.1}). Data privacy and communication efficiency are the two main motivations to adopt federated learning. Data privacy is preserved in federated learning as no raw local data moves out of the device~\cite{reisizadeh2019fedpaq, bonawitz2019federated}. Also, federated learning achieves higher communication efficiency by exchanging only model parameters or gradients~\cite{kairouz2019advances, 10.1145/3298981, mcmahan2016communicationefficient}. High data privacy and communication efficiency also promote scalability. Hence, more clients are motivated to join the training process~\cite{kairouz2019advances, 10.1145/3298981, mcmahan2016communicationefficient}.   

Statistical heterogeneity is defined as the data distribution with data volume and class distribution variance among devices (i.e. Non-IID). Essentially, the data are massively distributed across client devices, each only contains small amount of data~\cite{8884802, 8836609, roy2019braintorrent}, with unbalanced data classes~\cite{8884802} that are not representative of the overall data distribution~\cite{Hu2019}. 
When local models are trained independently on these devices, these models tend to be over-fitted to their local data~\cite{8836609, roy2019braintorrent, li2019fair}. Hence, federated learning is adopted to collaboratively trains the local models to form a generalised global model. System heterogeneity is defined as the property of devices having heterogeneous resources (e.g, computation, communication, storage, and energy). 
Federated learning can tackle this issue by enabling local model training and only communicates the model updates, which reduce the bandwidth footprint and energy consumption~\cite{corinzia2019variational, li2019fair, 8884802, 8728285, thomas2018federated, wang2019federated, amiri2020update}.

Another motivation is high computation efficiency. With a large number of participating clients and the increasing computation capability of clients, federated learning can have high model performance~\cite{mcmahan2016communicationefficient, 8928018, 8844592, 8560084, Song2019b} and computation efficiency~\cite{Awan2019, 8836609, 8647649}. Data storage efficiency is ensured by independent on-client training using locally generated data~\cite{Shen2019, Zhu2019, yurochkin2019bayesian, mcmahan2016communicationefficient, caldas2018expanding}.

\begin{figure*}[h!]
\begin{center}
\footnotesize
\begin{mdframed}[
    skipabove=1cm,    
    innerleftmargin =-1cm,
    innerrightmargin=-1cm,
    usetwoside=false,
]

\begin{center}
    \textbf{Findings of RQ 1.1: Why is federated learning adopted?}\\
\end{center} 

\begin{quotation}

\textit{\textbf{Motivation for adoption:}} Data privacy and communication efficiency are the two main motivations. Only a small number of studies adopt federated learning because of model performance. With a large number of participating clients, federated learning is expected to achieve high model performance. However, the approach is still immature when dealing with non-IID and unbalanced data distribution.
\end{quotation}

\begin{quotation}
\textbf{With regard to the software development lifecycle}, this question contributes to the \textit{background understanding} phase where we identify the objectives of federated learning adoption.
\end{quotation}

\end{mdframed}
\end{center} 
\end{figure*}

\newpage
\begin{center}

\scriptsize
\begin{longtable}{lll}
  \caption{Data Types and Applications Distribution}
  \label{tab:fl_app_data}\\

    \toprule
    \textbf{\thead{{Data Types (RQ 1.3)}}} & \textbf{\thead{{Applications (RQ 1.2)}}} & \textbf{\thead{{Count}}}\\
    \midrule
    \textup{\makecell[l]{Graph data\\(5\%)}} & \textup{\makecell[l]{Generalized Pareto Distribution parameter estimation\\Incumbent signal detection model\\Linear model fitting\\Network pattern recognition\\Computation resource management\\Waveform classification}} &  \textup{\makecell[r]{1\\1\\1\\8\\1\\1}}\\
    \hline \\[-1.8ex] 
    \textup{\makecell[l]{Image data\\(49\%)}} & \textup{\makecell[l]{Autonomous driving\\Healthcare (Bladder contouring, whole-brain segmentation)\\Clothes type recognition\\Facial recognition\\Handwritten character/digit recognition\\Human action prediction\\Image processing (classification/defect detection)\\Location prediction\\Phenotyping system\\Content recommendation}} &  \textup{\makecell[r]{5\\2\\14\\2\\109\\2\\4\\1\\1\\2}}\\
     \hline \\[-1.8ex]
     \textup{\makecell[l]{Sequential\\data\\(4\%)}} & \textup{\makecell[l]{Game AI model\\Network pattern recognition\\Content recommendation\\Robot system navigation\\Search rank system\\Stackelberg competition model}} &  \textup{\makecell[r]{2\\1\\1\\1\\6\\2}}\\
     \hline \\[-1.8ex]
     \textup{\makecell[l]{Structured\\data\\(21\%)}} & \textup{\makecell[l]{Air quality prediction\\Healthcare\\Credit card fraud detection\\Bankruptcy prediction\\Content recommendation (e-commerce)\\Energy consumption prediction\\Economic prediction (financial/house price/income/loan/market)\\Human action prediction\\Multi-site semiconductor data fusion\\Particle collision detection\\Industrial production recommendation\\Publication dataset binary classification\\Quality of Experience (QoE) prediction\\Search rank system\\Sentiment analysis\\System anomaly detection\\Song publishment year prediction
}} &  \textup{\makecell[r]{2\\25\\3\\5\\1\\1\\11\\4\\1\\1\\1\\1\\1\\1\\1\\1\\1}}\\
     \hline \\[-1.8ex]
     \textup{\makecell[l]{Text data\\(14\%)}} & \textup{\makecell[l]{Customer satisfaction prediction\\Keyboard suggestion (search/word/emoji)\\Movie rating prediction\\Out-of-Vocabulary word learning\\Suicidal ideation detection\\Product review prediction\\Resource management model\\Sentiment analysis\\Spam detection\\Speech recognition\\Text-to-Action decision model\\Content recommendation\\Wine quality prediction}} &  \textup{\makecell[r]{1\\21\\4\\1\\1\\1\\1\\5\\2\\1\\1\\1\\1}}\\
     \hline \\[-1.8ex]
     \textup{\makecell[l]{Time-series\\data\\(7\%)}} & \textup{\makecell[l]{Air quality prediction\\Automobile MPG prediction\\Healthcare (gestational weight gain / heart rate prediction)\\Energy prediction (consumption/demand/generation)\\Human action prediction\\Location prediction\\Network anomaly detection\\Resource management model\\Superconductors critical temperature prediction\\Vehicle type prediction}} & \textup{\makecell[r]{1\\1\\2\\3\\8\\1\\1\\2\\1\\1}}\\
    
    \bottomrule

\end{longtable}

\end{center}

\subsection{RQ 1.2 What are the applications of federated learning? \& RQ 1.3: What data does the federated learning applications deal with?}\label{Section:RQ1.2}

We study the federated learning applications and the types of data used in those applications through RQ 1.2 and RQ 1.3. 
The data types and applications are listed in Table~\ref{tab:fl_app_data}. 
The most widely used data types are image data, structured data, and text data, while the most popular application is image classification. 
In fact, MNIST\footnote{The MNIST database of handwritten digits, \url{http://yann.lecun.com/exdb/mnist/}} is the most frequently used dataset. More studies are needed to deal with IoT time-series data. 
Both graph data and sequential data are not popularly used in federated learning due to their data characteristics. We observed that the federated learning is widely adopted in applications that infer personal data, such as images, personal medical or financial data, and text recorded by personal mobile devices. 

\begin{figure*}[h!]
\begin{center}
\footnotesize
\begin{mdframed}[
    skipabove=1cm,    
    innerleftmargin =-1cm,
    innerrightmargin=-1cm,
    usetwoside=false,
]
\begin{center}
    \textbf{Findings of RQ 1.2: What are the applications of federated learning? \& }\\
    \textbf{RQ 1.3: What data does the federated learning applications deal with?}
\end{center} 

\begin{quotation}

\textit{\textbf{Applications and data:}} Federated learning is widely adopted in applications that deal with image data, structured data, and text data. Both graph data and sequential data are not popularly used due to their data characteristics (e.g., non-linear data structure). Also, there is only a few production-level applications. Most applications are still proof-of-concept prototypes or simulations.

\end{quotation}

\begin{quotation}
\textbf{With regard to the software development lifecycle}, this question contributes to the \textit{requirement analysis} phase where we identify the different applications and data types that have applied federated learning as a reference for researchers and practitioners.
\end{quotation}

\end{mdframed}
\end{center} 
\end{figure*}



\subsection{RQ 1.4: What is the client data distribution of the federated learning applications?}\label{Section:RQ1.4}

Table~\ref{tab:nonIID} shows the client data distribution types found in the studies. 
24\% of the studies have adopted non-IID data or have addressed the non-IID issue in their work. 23\% of the studies have adopted IID data. 13\% of the studies have compared the two types of client data distributions (Both), whereas 40\% of the studies have not specified which data distribution they have adopted (N/A). These studies ignored the effect of data distribution on the federated model performance. In the simulated non-IID data distribution settings, researchers mainly split the dataset by class and store each class into different client devices (e.g.,~\cite{zhao2018federated, HUANG2019103291, chen2019communicationefficient, sun2019energy, Jalalirad2019}), or by sorting the data accordingly before distributing them to each client device~\cite{8917724}. Furthermore, the data volume is uneven for each client device~\cite{8893114, Jalalirad2019}. For use case evaluations, the non-IID data are generated or collected by local devices, such as~\cite{hard2018federated, ramaswamy2019federated, koskela2019learning, Fan2019, duan2019astraea}. For IID data distribution settings, the data are randomised and distributed evenly to each client device (e.g.,~\cite{zhao2018federated, sun2019energy, 8917724, bakopoulou2019federated, jiang2019model}). 

\begin{figure*}[h!]
\begin{center}
\footnotesize
\begin{mdframed}[
    skipabove=1cm,    
    innerleftmargin =-1cm,
    innerrightmargin=-1cm,
    usetwoside=false,
]
\begin{center}
    \textbf{Findings of RQ 1.4:\\What is the client data distribution of the federated learning applications?} \\
\end{center} 

\begin{quotation}

\textit{\textbf{Client data distribution:}} The client data distribution influences the federated learning model performance. Model aggregation should consider that the distribution of the dataset on each client is different. Many studies are conducted on Non-IID issues, particularly on the FedAvg algorithm extensions for model aggregation.

\end{quotation}

\begin{quotation}
\textbf{With regard to the software development lifecycle}, this question contributes to the \textit{requirement analysis} phase where we identify the characteristics of the different types of data distribution that affect the federated learning model performance.
\end{quotation}

\end{mdframed}
\end{center} 
\end{figure*}

\begin{table*}[h]
  \caption{Client data distribution types of federated learning applications}
  \label{tab:nonIID}
  \footnotesize
  \begin{tabular}{l|cccc}
    \toprule
    
    \textbf{{Data distribution types}} & \textbf{{Non-IID}}& \textbf{{IID}}& \textbf{{Both}}& \textbf{{N/A}}\\
    \midrule
    \textbf{{\makecell[l]{Percentages}}} & \texttt{{\makecell{24\%}}}& \texttt{{\makecell{23\%}}}& \texttt{{\makecell{13\%}}}& \texttt{{\makecell{40\%}}}\\
    \bottomrule
  \end{tabular}
\end{table*}

    

\subsection{RQ 2: What challenges of federated learning are being addressed? \& RQ 2.1: How many research activities have been conducted on federated learning?}\label{Section:RQ2}

\begin{figure}
  \centering
  \includegraphics[width=0.7\linewidth]{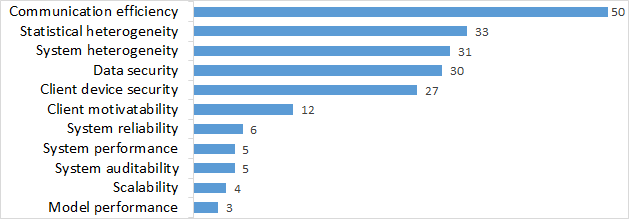}
  \caption{Challenges of federated learning}
  \label{fig:rq2}
\end{figure}

The motivation of RQ 2 is to identify the challenges of federated learning that are addressed by the studies. As shown in Fig.~\ref{fig:rq2}, we group the answers into categories based on ISO/IEC 25010 System and Software Quality model~\cite{iso25010.com} and ISO/IEC 25012 Data Quality model~\cite{iso25012.com}. 
We can observe that the communication efficiency of federated learning received the most attentions from researchers, followed by statistical heterogeneity and system heterogeneity. 

\begin{figure}
  \centering
  \includegraphics[width=\linewidth]{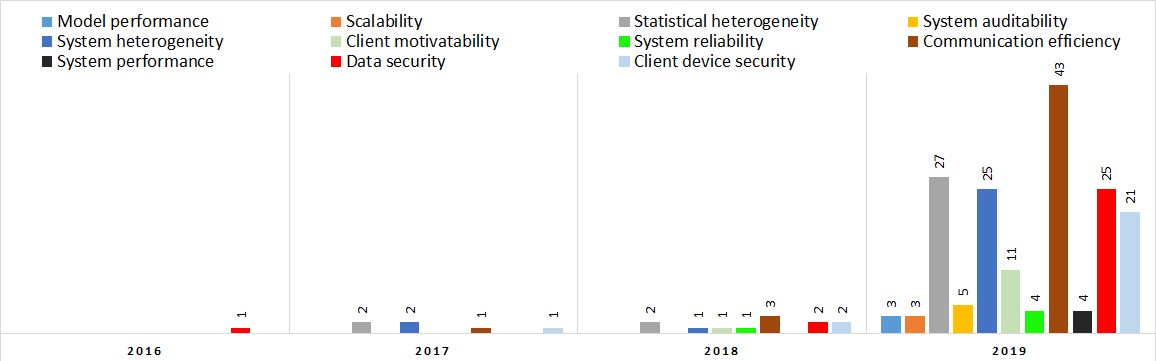}
  \caption{Research Area Trend}
  \label{RQ2:Research_trend}
\end{figure}

To explore the research interests evolved from 2016 to 2019, we cross-examined the results for RQ 2 and RQ 2.1 as illustrated in Fig.~{\ref{RQ2:Research_trend}}. Note that we included the results from 2016-2019 as we only searched studies up to 31.01.2020 and the trend of one month does not represent the trend of the entire year. We can see that the number of studies on communication efficiency, statistical heterogeneity, system heterogeneity, data security, and client device security surged drastically in 2019 compared to the years before.

Although transferring model updates instead of raw data can reduce communication costs, federated learning systems still perform multiple update iterations to reach convergence. Therefore, ways to reduce communication rounds are studied (e.g.,~\cite{8917724, 8759317, du2018efficient}). Furthermore, cross-device federated learning needs to accommodate a large number of client devices. Hence, some clients may drop out due to bandwidth limitation~\cite{8843451, 8765347}. These dropouts could negatively impact the outcome of federated learning systems in two ways: (i) reduce the amount of data available for training~\cite{8843451}, (ii) increase the overall training time~~\cite{shi2019device, Mandal2019}. The mechanism in ~\cite{liu2019boosting} addresses the dropout problem by abandoning the global model aggregated from  a low number of local models.


Federated learning is effective in dealing with statistical and system heterogeneity issues through the aggregation of local models trained locally on client devices~\cite{corinzia2019variational, li2019fair}.
However, the approach is still immature as open questions exist in handling the non-IID while maintaining the model performance~\cite{Ye2020, Truex2019, HUANG2019103291, verma2019approaches, corinzia2019variational}. 

The interests in data security (e.g.,~\cite{Bonawitz2017, geyer2017differentially}) and client device security (e.g.,~\cite{fung2018mitigating, 8894364, choudhury2019differential}) are also high.
The data security in federated learning is the degree to which a system ensures data are accessible only by an authorised party~\cite{iso25012.com}. While federated learning systems restrict raw data from leaving local devices, it is still possible to extract private information through back-tracing the gradients. The studies under this category mostly express their concerns about the possibility of information leakage from the local model gradient updates. Client device security 
can be expressed as the degree of security against dishonest and malicious client devices~\cite{kairouz2019advances}. The existence of malicious devices in the training process could poison the overall model performance by disrupting the training process or providing false updates to the central server. Furthermore, the intentional or unintentional misbehavior of client devices could reduce system reliability.

Client motivatability is discussed as an aspect to be explored (e.g.,~\cite{Zhan2020, 8893114, 8733825}) since model performance relies greatly on the number of participating clients. More participating clients means more data and computation resources are contributed to the model training process.

System reliability concerns are mostly on the adversarial or byzantine attacks that target the central server, hence exposes the single-point-of-failure (e.g.,~\cite{8892848, lalitha2019decentralized, hu2019decentralized}). 
The system performance of federated learning systems is mentioned in some studies (e.g.~\cite{8843942, 8935424, xie2019asynchronous}), which includes the considerations on computation efficiency, energy usage efficiency, and storage efficiency. 
In resource-restricted environments, the efficient resource management of federated learning systems is more crucial to system performance.

To improve system auditability, auditing mechanisms (e.g.,~\cite{jiang2019model, Anelli2019, wei2019multi}) are used to track the client devices' behavior, local model performance, and system runtime performance. 
Scalability is also mentioned as a limitation in federated learning (e.g.,~\cite{8950073, 8889996, xie2019asynchronous, reisizadeh2019fedpaq}) and lastly, the model performance limitation is investigated (e.g.,~\cite{Ji2019, Hu2019, nock2018entity}). The model performance of federated learning systems is highly dependent on the number of participants and the data volume of each participating client in the training process. Moreover, there is model performance limitation due to the non-IID data.

\begin{figure*}[h!]
\begin{center}
\footnotesize
\begin{mdframed}[
    skipabove=1cm,    
    innerleftmargin =-1cm,
    innerrightmargin=-1cm,
    usetwoside=false,
]
\begin{center}
    \textbf{Findings of RQ 2: What challenges of federated learning are being addressed?}\\
\end{center} 

\begin{quotation}

\textit{\textbf{Motivation vs. challenges:}} Most of the known motivations of federated learning also appear to be the most studied federated learning limitations, including communication efficiency, system and statistical heterogeneity, model performance, and scalability. This reflects that federated learning is still under-explored.
\end{quotation}

\begin{quotation}
\textbf{With regard to the software development lifecycle}, this question contributes to the \textit{requirement analysis} phase where we identify the various requirements of a federated learning system to be considered during the development.
\end{quotation}

\end{mdframed}
\end{center} 
\end{figure*}


To answer RQ 2.1, we classify the papers according to the research classification criteria proposed by Wieringa \cite{10.1007/s00766-005-0021-6}, which includes: (1) evaluation research, (2) philosophical papers, (3) proposal of solution, and (4) validation research. We use this classification to distinguish the focus of each research activity. 
\textit{Evaluation research} is the investigation of a problem in software engineering practice.
In general, the research results in new knowledge of causal relationships among phenomena, or in new knowledge of logical relationships among propositions. The causal properties are studied through case or field studies, field experiments and surveys.
\textit{Philosophical papers} propose a new way of looking at things, for instance, a new conceptual framework.
\textit{Proposal of solution} papers propose solution techniques and argue for its relevance, but without a full-blown validation. 
Lastly, \textit{validation research} investigates the properties of a proposed solution that has not yet been implemented in  practice. The investigation uses a thorough, methodologically sound research setup (e.g., experiments, simulations).

The research classification results are presented in Table~\ref{tab:researchclassification}. By far the most common type of research is validation research, while the other types of are far less frequent. In particular, there are few philosophical papers that propose a new conceptual framework for federated learning. 

\begin{table*}[h]
  \caption{The research classification of the selected paper}
  \label{tab:researchclassification}
  \footnotesize
  \begin{tabular}{l|ccccc}
    \toprule
    \textbf{\makecell[l]{Research\\Classification}} & \textbf{\makecell{Evaluation research}} & \textbf{\makecell{Philosophical paper}} & \textbf{\makecell{Proposal of solution}} & \textbf{\makecell{Validation\\research}}\\
    \midrule
    \textbf{{\makecell[l]{Paper count}}} & \textup{{\makecell{13}}}& \textup{{\makecell{12}}}& \textup{{\makecell{25}}}& \textup{{\makecell{183}}}\\
   
    \bottomrule
  \end{tabular}
\end{table*}

\begin{figure*}[h!]
\begin{center}
\footnotesize
\begin{mdframed}[
    skipabove=1cm,    
    innerleftmargin =-1cm,
    innerrightmargin=-1cm,
    usetwoside=false,
]
\begin{center}
    \textbf{Findings of RQ 2.1:\\How many research activities have been conducted on federated learning?} \\
\end{center} 

\begin{quotation}

\textit{\textbf{Research activities:}} The number of studies that explored communication efficiency, statistical and system heterogeneity, data security, and client device security surged drastically in 2019 compared to the years before. The most conducted research activities are validation research, followed by proposal of solution, evaluation research, and philosophical papers.
\end{quotation}

\begin{quotation}
\textbf{With regard to the software development lifecycle}, this question contributes to the \textit{background understanding} phase where we identify the types of research activities on federated learning.
\end{quotation}

\end{mdframed}
\end{center} 
\end{figure*}

\subsection{RQ 2.2: Who is leading the research in federated learning?} 

The motivation of RQ 2.2 is to understand the research impact in the federated learning community. 
We also intend to help researchers identify the state-of-the-art research in federated learning by selecting the top affiliations. 
As shown in table~\ref{tab:top_10_Number_of_papers}, we listed the top 10 affiliations by the number of papers published and the number of citations. Google, IBM, CMU, and WeBank appear in both the top-10 lists.
From this table, we can identify the research institutions that made the most effort on federated learning and those that made the most impact on the research domain in terms of citations.

\begin{table*}
\footnotesize
  \caption{Research Impact Analysis}
  \label{tab:top_10_Number_of_papers}
  \resizebox{\textwidth}{!}{
  \begin{tabular}{ccc|ccc}
    \toprule
    \multicolumn{3}{c|}{\textbf{{Top 10 affiliations by number of papers}}} & \multicolumn{3}{c}{\textbf{{Top 10 affiliations by number of citations}}}\\
    
    \midrule
    \textbf{{Rank}} & \textbf{{Affiliations}} & \textbf{{Paper count}} & \textbf{{Rank}} & \textbf{{Affiliations}} & \textbf{{No. of citations}}\\
    \hline
    \textup{{\makecell[l]{1}}} & \textup{{\makecell[l]{Google}}} & \textup{{\makecell{21}}} & \textup{{\makecell[l]{1}}} & \textup{{\makecell[l]{Google}}} & \textup{{\makecell{2269}}}\\
    \textup{{\makecell[l]{2}}} & \textup{{\makecell[l]{IBM}}} & \textup{{\makecell{11}}} & \textup{{\makecell[l]{2}}} & \textup{{\makecell[l]{Stanford University}}} & \textup{{\makecell{217}}}\\
    \textup{{\makecell[l]{3}}} & \textup{{\makecell[l]{WeBank}}} & \textup{{\makecell{8}}} & \textup{{\makecell[l]{3}}} & \textup{{\makecell[l]{ETH Zurich}}} & \textup{{\makecell{130}}}\\
    \textup{{\makecell[l]{3}}} & \textup{{\makecell[l]{Nanyang Technological University}}} & \textup{{\makecell{8}}} & \textup{{\makecell[l]{4}}} & \textup{{\makecell[l]{IBM}}} & \textup{{\makecell{122}}}\\
    \textup{{\makecell[l]{5}}} & \textup{{\makecell[l]{Tsinghua University}}} & \textup{{\makecell{6}}} & \textup{{\makecell[l]{5}}} & \textup{{\makecell[l]{Cornell University}}} & \textup{{\makecell{101}}}\\
    \textup{{\makecell[l]{5}}} & \textup{{\makecell[l]{Carnegie Mellon University (CMU)}}} & \textup{{\makecell{6}}} & \textup{{\makecell[l]{6}}} & \textup{{\makecell[l]{Carnegie Mellon University (CMU)}}} & \textup{{\makecell{98}}}\\
    \textup{{\makecell[l]{7}}} & \textup{{\makecell[l]{Beijing University of Posts and Telecommunications}}} & \textup{{\makecell{5}}} & \textup{{\makecell[l]{7}}} & \textup{{\makecell[l]{ARM}}} & \textup{{\makecell{82}}}\\
    \textup{{\makecell[l]{7}}} & \textup{{\makecell[l]{Kyung Hee University}}} & \textup{{\makecell{5}}} & \textup{{\makecell[l]{8}}} & \textup{{\makecell[l]{Tianjin University}}} & \textup{{\makecell{76}}}\\
    \textup{{\makecell[l]{9}}} & \textup{{\makecell[l]{Chinese Academy of Sciences
}}} & \textup{{\makecell{4}}} & \textup{{\makecell[l]{9}}} & \textup{{\makecell[l]{University of Oulu}}} & \textup{{\makecell{73}}}\\
    \textup{{\makecell[l]{9}}} & \textup{{\makecell[l]{Imperial College London}}} & \textup{{\makecell{4}}} & \textup{{\makecell[l]{10}}} & \textup{{\makecell[l]{WeBank}}} & \textup{{\makecell{66}}}\\
    \bottomrule
  \end{tabular}}
  \end{table*}

\begin{figure*}[h!]
\begin{center}
\footnotesize
\begin{mdframed}[
    skipabove=1cm,    
    innerleftmargin =-1cm,
    innerrightmargin=-1cm,
    usetwoside=false,
]
\begin{center}
    \textbf{Findings of RQ 2.2: Who is leading the research in federated learning?} \\
\end{center} 

\begin{quotation}

\textit{\textbf{Affiliations:}} Google, IBM, CMU, and WeBank appear in the top 10 affiliations list both by the number of papers and by the number of citations, which reflect that they made the most efforts on federated learning and also the most impact on the research domain.

\end{quotation}

\begin{quotation}
\textbf{With regard to the software development lifecycle}, this question contributes to the \textit{background understanding} phase where we provide the list of leading affiliations to help researchers and practitioners identify the state-of-the-art of federated learning.
\end{quotation}

\end{mdframed}
\end{center} 
\end{figure*}

\subsection{RQ 3: How are the federated learning challenges being addressed?}

\begin{figure}
  \includegraphics[width=0.75\linewidth]{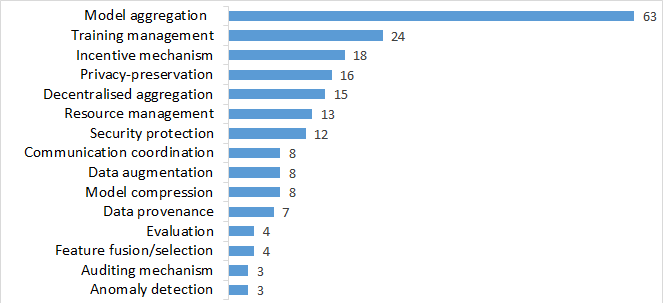}
  \caption{RQ 3: How are the federated learning challenges being addressed?}
  \label{fig:rq3}
\end{figure}

After looking at the challenges, we studied the approaches proposed by the researchers to address these challenges. Fig.~\ref{fig:rq3} shows the existing approaches: model aggregation (63), training management (24), incentive mechanisms (18), privacy-preserving mechanisms (16), decentralised aggregation (15), security management (12), resource management (13), communication coordination (8), data augmentation (7), data provenance (7), model compression (8), feature fusion/selection (4), auditing mechanisms (4), evaluation (4), and anomaly detection (3). We also mapped these approaches to the challenges of federated learning in Table~\ref{tab:rq2vs3}. Notice that we did not include every challenge mentioned in the collected studies but only those that have a proposed solution. 



As shown in Fig.~\ref{fig:rq3}, model aggregation mechanisms are the most proposed solution by the studies. From Table~\ref{tab:rq2vs3}, we can see that model aggregation mechanisms are applied to address communication efficiency, statistical heterogeneity, system heterogeneity, client device security, data security, system performance, scalability, model performance, system auditability, and system reliability issues. 

Researchers have proposed various kinds of aggregation methods, including selective aggregation~\cite{Ye2020, 8994206}, aggregation scheduling~\cite{yang2019age, sun2019energy}, asynchronous aggregation~\cite{xie2019asynchronous, chen2019asynchronous}, temporally weighted aggregation~\cite{8761315}, controlled averaging algorithms~\cite{karimireddy2019scaffold}, iterative round reduction~\cite{8917724, 8759317}, and shuffled model aggregation~\cite{ghazi2019scalable}. These approaches aim to: (1) reduce communication cost and latency for better communication efficiency and scalability; (2) manage the device computation and energy resources to solve system heterogeneity and system performance issues, and (3) select high quality models for aggregation based on the model performance. Some researchers have proposed secure aggregation to solve data security and client device security issues~\cite{bonawitz2019federated, bonawitz2016practical, bonawitz2019towards}. 

Decentralised aggregation is a type of model aggregations which removes the central server from the federated learning systems and is mainly adopted to solve system reliability limitations.
The decentralised aggregation can be realised through a peer-to-peer system~\cite{shayan2018biscotti, roy2019braintorrent}, one-hop neighbours collective learning~\cite{lalitha2019peer}, and Online Push-Sum (OPS) methods~\cite{he2019central}. 

Training management is the second most proposed approach in the studies. It is used to address statistical heterogeneity, system heterogeneity, communication efficiency, client motivatability, and client device security issues. To deal with the statistical heterogeneity issue, researchers have proposed various training methods, such as the clustering of training data~\cite{caldas2018federated, sattler2019clustered}, multi-stage training and fine-tuning of models~\cite{jiang2019improving}, brokered learning~\cite{fung2018dancing}, distributed multitask learning~\cite{corinzia2019variational, Smith2017}, and edge computing~\cite{DBLP:journals/corr/abs-1905-06641}. The objectives are to increase the training data and reduce data skewness effect while maintaining a distributed manner. Furthermore, to address system heterogeneity issues, some methods balance the tasks and resources of the client nodes for the training process~\cite{dinh2019federated, han2020adaptive, li2019fedmd}. 

Another frequently proposed approach is the incentive mechanism, which is a solution to increase client motivatability. 
There is no obligation for data or device owners to join the training process if there are no benefits for them to contribute their data and resources. Incentive mechanisms can attract data and device owners to join the model training. Incentives can be given based on the amount of computation, communication, and energy resources provided~\cite{wang2019measure, 8893114}, the local model performance~\cite{8932389}, the quality of the data provided~\cite{yurochkin2019bayesian}, the honest behavior of client devices~\cite{preuveneers2018chained}, and the dropout rate of a client node~\cite{Pandey2019}. The proposed incentive mechanisms can be hosted by either a central server~\cite{Zhan2020, 8851649}
or a blockchain~\cite{8843900, 8733825}.

Resource management is introduced to control and optimize computation resources, bandwidth, and energy consumption of participating devices in a training process to address system heterogeneity~\cite{8664630, 8761315}
and communication efficiency~\cite{sattler2019clustered, abad2019hierarchical}. The proposed approaches use control algorithms~\cite{8664630}, reinforcement learning~\cite{8791693, Zhan2020ExperienceDrivenCR}, and edge computing methods~\cite{8761315} to optimise the resource usage and improve the system efficiency. 
To address communication efficiency, model compression methods are utilised to reduce the data size and lower the communication cost that occurs during the model updates~\cite{8885054, 8928018, 8889996, caldas2018expanding, Konecny2016, li2019endtoend}. Furthermore, model compression can also promote scalability as it is applicable to bandwidth limited and latency-sensitive scenarios~\cite{8889996}. 

Privacy preservation methods are introduced to maintain (1) data security: prevents information (model parameters or gradients) leakage to unauthorised parties, and (2) client devices security: prevents the system from being compromised by dishonest nodes. One well-cited method used for data security maintenance is differential privacy~\cite{Xu2019a, ryffel2018generic},
such as gaussian noise addition to the features before model updates~\cite{Ma2019}. For client device security maintenance, secure multiparty computations method such as homomorphic encryption is used together with local differential privacy method~\cite{Xu2019a, ryffel2018generic, Truex2019}. While homomorphic encryption only allows the central server to compute the global model homomorphically based on the encrypted local updates, the local differential privacy method protects client data privacy by adding noise to model parameter data sent by each client \cite{Truex2019}.
Security protection method is also proposed to solve the issues for both client device security~\cite{8765347, Liu2019, liu2018secure} and data security~\cite{8836609, 8859260}, which includes encrypting model parameters or gradients prior to exchanging the models between the server and client nodes. 

Communication coordination methods are introduced to improve communication efficiency~\cite{8952884, 8891310}
, and system performance~\cite{8935424}. Specifically, communication techniques such as multi-channel random access communication~\cite{8935424} or over-the-air computation methods~\cite{8952884, amiri2019federated} enable wireless federated training process to achieve faster convergence.

To address the model performance issues~\cite{Hu2019}, statistical heterogeneity problems~\cite{verma2019approaches}, system auditability limitations~\cite{wang2019interpret} and communication efficiency limitations~\cite{8803001}, feature fusion or selection methods are used. Feature selection methods are used to improve the convergence rate~\cite{Hu2019} by only aggregating models with the selected features. Feature fusion is used to reduce the impact of non-IID data on the model performance~\cite{verma2019approaches}. Moreover, the feature fusion method reduces the dimension of data to be transferred to speed up the training process and increases the communication efficiency~\cite{8803001}. ~\cite{wang2019interpret} proposed a feature separation method that measures the importance level of the feature in the training process, used for system interpretation and auditing purposes.

Data provenance mechanisms are used to govern the data and information interactions between nodes. They are effective in preventing single-point-of-failures and adversarial attacks that intend to tamper the data. One way to implement a data provenance mechanism in a federated learning system is through blockchains. Blockchains record all the events instead of the raw data for audit and data provenance, and only permits authorised parties to access the information~\cite{8843900, Yin2020}. Furthermore, blockchains are used for incentive provisions in~\cite{8893114, 8733825, 8905038, 8894364}, also for storing global model updates in Merkle trees~\cite{8892848}.

Data augmentation methods are introduced to address data security~\cite{Triastcyn2019, peterson2019private} and client device security issues~\cite{orekondy2018gradientleaks}. The approaches use privacy-preserving generative adversarial network (GAN) models to locally reproduce the data samples of all devices~\cite{jeong2018communication} for data release~\cite{Triastcyn2019} and debugging processes~\cite{augenstein2019generative}. These methods provide extra protection to shield the actual data from exposure. Furthermore, data augmentation methods are also effective in solving statistical heterogeneity issues through  reproduction of an IID dataset for better training performance~\cite{jeong2018communication}.

Auditing mechanisms are proposed as a solution for the lack of system auditability and client device security limitations. They are responsible for assessing the honest or semi-honest behavior of a client node during training and detecting any anomalies~\cite{augenstein2019generative, 8893114, Zhao2020}. 
Some researchers have also introduced anomaly detection mechanisms~\cite{preuveneers2018chained, li2019abnormal, fung2018mitigating}, specifically to penalise adversarial or misbehaving nodes.

To properly assess the behavior of a federated learning system, some researchers have proposed evaluation mechanisms. The mechanisms are intended to solve the system auditability and statistical heterogeneity issues. For instance, in~\cite{wei2019multi}, a visualisation platform to illustrate the federated learning system is presented. In~\cite{caldas2018leaf}, a benchmarking platform is presented to realistically capture the characteristics of a federated scenario. Furthermore,~\cite{hsu2019measuring} introduces a method to measure the performance of the federated averaging algorithm.

\begin{figure*}[h!]
\begin{center}
\footnotesize
\begin{mdframed}[
    skipabove=1cm,    
    innerleftmargin =-1cm,
    innerrightmargin=-1cm,
    usetwoside=false,
]
\begin{center}
    \textbf{Findings of RQ 3: How are the federated learning challenges being addressed?} \\
\end{center} 

\begin{quotation}

\textit{\textbf{Top 5 proposed approaches:}} The top 5 proposed approaches are model aggregation, training management, incentive mechanisms, privacy-preserving methods, and resource management. These approaches mainly aim to solve issues such as communication efficiency, statistical and system heterogeneity, client motivatability, and data privacy. 
\end{quotation}

\begin{quotation}

\textit{\textbf{Remaining proposed approaches:}} A few papers worked on anomaly detection, auditing mechanisms, feature fusion/selection, evaluation mechanisms and data provenance.

\end{quotation}

\begin{quotation}
\textbf{With regard to the software development lifecycle}, this question contributes to the \textit{architecture design} phase where we summarise different approaches proposed to address the identified requirements of federated learning systems.
\end{quotation}

\end{mdframed}
\end{center} 
\end{figure*}

\begin{sidewaystable*}

  \caption{What challenges of federated learning are being addressed (RQ 2) vs. How are the federated learning challenges being addressed (RQ 3)}
  \label{tab:rq2vs3}
  \large
  \scalebox{0.57}{%
  \begin{tabular}{cccccccccccc}
    \toprule
    \textbf{\makecell{Challenges vs\\ Approaches}} & \textbf{\makecell{Model\\performance}} & \textbf{\makecell{Scalability}} & \textbf{\makecell{Statistical\\heterogeneity}} & \textbf{\makecell{System\\auditability}} & \textbf{\makecell{System\\heterogeneity}} & \textbf{\makecell{Client\\motivatability}} & \textbf{\makecell{System\\reliability}} & \textbf{\makecell{Communication\\efficiency}} & \textbf{\makecell{System\\performance}} & \textbf{\makecell{Data\\security}} & \textbf{\makecell{Client device\\security}}\\
    \midrule
    
    \textbf{\makecell{Model\\aggregation}} & \textup{\makecell{\cite{Ji2019}}} & \textup{\makecell{\cite{xie2019asynchronous},\cite{reisizadeh2019fedpaq}}} &\textup{\makecell{\cite{8818446},\cite{Ye2020},\cite{Daga2019},\\\cite{mohri2019agnostic},\cite{xie2019asynchronous},\cite{chen2019asynchronous},\\\cite{DBLP:journals/corr/abs-1905-06641},\cite{mcmahan2016communicationefficient},\cite{sun2019energy},\\\cite{peng2019federated},\cite{li2018federated},\cite{bonawitz2019towards},\\\cite{karimireddy2019scaffold}}} &\textup{\makecell{\cite{Anelli2019}}} &\textup{\makecell{\cite{Ye2020},\cite{du2018efficient},\cite{mohri2019agnostic},\\\cite{chen2019asynchronous},\cite{yurochkin2019bayesian},\cite{li2018federated},\\\cite{bonawitz2016practical},\cite{wu2019safa},\cite{niu2019secure},\\\cite{bonawitz2019towards}}} &\textup{\makecell{-}} &\texttt{\makecell{\cite{8892848}}} &\textup{\makecell{\cite{8870236},\cite{8945292},\cite{8917724},\cite{8854053},\cite{8759317},\\\cite{8744465},\cite{du2018efficient},\cite{liu2019communication},\cite{yang2019quasi},\cite{yang2019age},\\\cite{xie2019asynchronous},\cite{chen2019asynchronous},\cite{mcmahan2016communicationefficient},\cite{jeong2018communication},\cite{sun2019energy},\\\cite{bonawitz2019federated},\cite{reisizadeh2019fedpaq},\cite{jiang2019model},\cite{guha2019one},\cite{bonawitz2016practical},\\\cite{bonawitz2019towards},\cite{Smith2017},\cite{konen2016federated}}} &\textup{\makecell{\cite{xie2019asynchronous},\cite{chen2019asynchronous},\\\cite{jiang2019model}}} &\textup{\makecell{\cite{qian2019active},\cite{bonawitz2016practical},\\\cite{ghazi2019scalable}}} &\textup{\makecell{\cite{8994206},\cite{pillutla2019robust},\\\cite{He2019},\cite{Chen2018}}}\\
    
    \textbf{\makecell{Training management}} & \textup{\makecell{-}} & \textup{\makecell{-}} &\textup{\makecell{\cite{10.1007/978-3-030-29516-5_48},\cite{HUANG2019103291},\cite{caldas2018federated},\\\cite{sattler2019clustered},\cite{dinh2019federated},\cite{Konecny2016},\\\cite{li2019fedmd},\cite{jiang2019improving},\cite{ghosh2019robust},\\\cite{han2019robust},\cite{corinzia2019variational},\cite{Smith2017},\\\cite{han2020adaptive},\cite{chen2018federated}}} &\textup{\makecell{-}} &\textup{\makecell{\cite{chen2019distributed},\cite{dinh2019federated},\cite{li2019fedmd},\\\cite{han2020adaptive}}} &\textup{\makecell{\cite{fung2018dancing}}} &\textup{\makecell{-}} &\textup{\makecell{\cite{8698609},\cite{Xu2018},\cite{Zhang2019a},\\\cite{amiri2019federated},}} &\textup{\makecell{-}} &\textup{\makecell{-}} &\textup{\makecell{\cite{10.1007/978-3-030-29516-5_48}}}\\
    
    \textbf{\makecell{Incentive mechanisms}}& \textup{\makecell{-}} & \textup{\makecell{-}} &\textup{\makecell{\cite{Zhan2020}}} &\textup{\makecell{-}} &\textup{\makecell{\cite{8875430},\cite{8867906},\\\cite{khan2019federated}}} &\textup{\makecell{\cite{Zhan2020},\cite{8893114},\cite{8733825},\\\cite{8894364},\cite{8905038},\cite{8851649},\\\cite{8832210},\cite{8875430},\cite{8945913},\\\cite{khan2019federated},\cite{wang2019measure}}} &\textup{\makecell{-}} &\textup{\makecell{\cite{Fadaeddini2019},\cite{Pandey2020},\\\cite{Pandey2019}}} &\textup{\makecell{-}} &\textup{\makecell{-}} &\textup{\makecell{-}}\\
    
    \textbf{\makecell{Privacy\\preservation}} & \textup{\makecell{-}} & \textup{\makecell{-}} &\textup{\makecell{-}} &\textup{\makecell{-}} &\textup{\makecell{-}} &\textup{\makecell{-}} &\textup{\makecell{-}} &\textup{\makecell{-}} &\textup{\makecell{-}} &\textup{\makecell{\cite{8843942},\cite{Xu2019a},\cite{Awan2019},\\\cite{Ma2019},\cite{10.1007/978-3-030-29516-5_48},\cite{Mo},\\\cite{ryffel2018generic},\cite{geyer2017differentially},\cite{li2019differentially},\\\cite{triastcyn2019federated},\cite{bhowmick2018protection},\cite{wei2019federated}}} &\textup{\makecell{\cite{Truex2019},\cite{Jiang2019},\\\cite{Zhang2019a},\cite{choudhury2019differential}}}\\
    
    \textbf{\makecell{Decentralised\\aggregation}} & \textup{\makecell{-}} & \textup{\makecell{\cite{8950073}}} &\textup{\makecell{-}} &\textup{\makecell{-}} &\textup{\makecell{\cite{he2019central}}} &\textup{\makecell{-}} &\textup{\makecell{\cite{8733825},\cite{8950073},\cite{lalitha2018fully},\\\cite{lalitha2019decentralized},\cite{hu2019decentralized}}} &\texttt{\makecell{\cite{he2019central}}} &\textup{\makecell{\cite{8843942}}} &\textup{\makecell{\cite{yang2019parallel}}} &\textup{\makecell{\cite{shayan2018biscotti},\cite{roy2019braintorrent},\\\cite{yang2019parallel},\cite{lalitha2019peer},\\\cite{he2019central}}}\\
    
    \textbf{\makecell{Resource\\management}}& \textup{\makecell{-}} & \textup{\makecell{-}} &\textup{\makecell{-}} &\textup{\makecell{-}} &\textup{\makecell{\cite{8664630},\cite{8761315},\\\cite{8716527},\cite{Xu2019},\\\cite{Zhan2020ExperienceDrivenCR},\cite{yang2019energy},\\\cite{li2019fair},\cite{amiri2020update},\\\cite{ren2019accelerating},\cite{DBLP:journals/corr/abs-1907-06040},\\\cite{Li2019}}} &\texttt{\makecell{-}} &\textup{\makecell{-}} &\textup{\makecell{\cite{sattler2019clustered},\cite{abad2019hierarchical}}} &\textup{\makecell{-}} &\textup{\makecell{-}} &\textup{\makecell{-}}\\
    
    \textbf{\makecell{Security protection}} & \textup{\makecell{-}} & \textup{\makecell{-}} &\textup{\makecell{-}} &\textup{\makecell{-}} &\textup{\makecell{-}} &\textup{\makecell{-}} &\textup{\makecell{-}} &\textup{\makecell{-}} &\textup{\makecell{-}} &\textup{\makecell{\cite{8836609},\cite{8859260},\cite{8761267},\\\cite{8765347},\cite{Mandal2019},\cite{liu2019boosting},\\\cite{hardy2017private},\cite{chai2019secure},\cite{cheng2019secureboost}}} &\textup{\makecell{\cite{8765347},\cite{Liu2019},\\\cite{liu2018secure}}}\\
    
    \textbf{\makecell{Communication\\coordination}} & \textup{\makecell{-}} & \textup{\makecell{-}} &\textup{\makecell{-}} &\textup{\makecell{-}} &\textup{\makecell{-}} &\textup{\makecell{-}} &\textup{\makecell{-}} &\textup{\makecell{\cite{8952884},\cite{8891310},\cite{8904164},\\\cite{chen2019joint},\cite{vu2019cell},\cite{shi2019device},\\\cite{amiri2019federated}}} &\textup{\makecell{\cite{8935424}}} &\textup{\makecell{-}} &\textup{\makecell{-}}\\
    
    \textbf{\makecell{Data\\augmentation}}& \textup{\makecell{\cite{nock2018entity}}} & \textup{\makecell{-}} &\textup{\makecell{\cite{duan2019astraea},\cite{jeong2018communication}}} &\textup{\makecell{-}} &\textup{\makecell{-}} &\textup{\makecell{-}} &\textup{\makecell{-}} &\textup{\makecell{\cite{zhao2018federated}}} &\textup{\makecell{-}} &\textup{\makecell{\cite{Triastcyn2019}, \cite{peterson2019private}}} &\textup{\makecell{\cite{orekondy2018gradientleaks},\cite{Zhao2020}}}\\
    
    \textbf{\makecell{Model compression}}& \textup{\makecell{-}} & \textup{\makecell{\cite{8889996}}} &\textup{\makecell{-}} &\textup{\makecell{\cite{Bonawitz2017}}} &\textup{\makecell{-}} &\textup{\makecell{-}} &\textup{\makecell{-}} &\textup{\makecell{\cite{8885054},\cite{8928018},\\\cite{8889996},\cite{caldas2018expanding},\\\cite{Konecny2016},\cite{li2019endtoend}}} &\texttt{\makecell{-}} &\texttt{\makecell{-}} &\textup{\makecell{-}}\\
    
    \textbf{\makecell{Data provenance}} & \textup{\makecell{-}} & \textup{\makecell{-}} &\textup{\makecell{-}} &\textup{\makecell{-}} &\textup{\makecell{-}} &\textup{\makecell{-}} &\textup{\makecell{-}} &\textup{\makecell{-}} &\textup{\makecell{-}} &\textup{\makecell{\cite{8843900},\cite{8892848},\cite{Yin2020}}} &\textup{\makecell{\cite{8894364},\cite{8932389},\\\cite{Zhu2019},\cite{zhao2019mobile}}}\\
    
    \textbf{\makecell{Evaluation}} & \textup{\makecell{-}} & \textup{\makecell{-}} &\textup{\makecell{\cite{caldas2018leaf},\cite{hsu2019measuring}}} &\textup{\makecell{\cite{wei2019multi}}} &\textup{\makecell{\cite{caldas2018leaf}}} &\textup{\makecell{-}} &\textup{\makecell{-}} &\textup{\makecell{-}} &\textup{\makecell{-}} &\textup{\makecell{-}} &\textup{\makecell{-}}\\
    
    \textbf{\makecell{Feature\\fusion/selection}} & \textup{\makecell{\cite{Hu2019}}} & \textup{\makecell{-}} &\textup{\makecell{\cite{verma2019approaches}}} &\textup{\makecell{\cite{wang2019interpret}}} &\textup{\makecell{-}} &\textup{\makecell{-}} &\textup{\makecell{-}} &\texttt{\makecell{\cite{8803001}}} &\textup{\makecell{-}} &\textup{\makecell{-}} &\textup{\makecell{-}}\\
    
    \textbf{\makecell{Auditing\\mechanisms}}& \textup{\makecell{-}} & \textup{\makecell{-}} &\textup{\makecell{-}} &\textup{\makecell{\cite{8905038},\cite{augenstein2019generative}}} &\textup{\makecell{-}} &\textup{\makecell{-}} &\textup{\makecell{-}} &\textup{\makecell{-}} &\textup{\makecell{-}} &\textup{\makecell{-}} &\textup{\makecell{\cite{8905038}}}\\
    
    \textbf{\makecell{Anomaly detection}} & \textup{\makecell{-}} & \textup{\makecell{-}} &\textup{\makecell{-}} &\textup{\makecell{-}} &\textup{\makecell{-}} &\textup{\makecell{-}} &\textup{\makecell{-}} &\textup{\makecell{-}} &\textup{\makecell{-}} &\textup{\makecell{-}} &\textup{\makecell{\cite{preuveneers2018chained}, \cite{li2019abnormal},\\\cite{fung2018mitigating}}}\\
    \bottomrule
  \end{tabular}}


\end{sidewaystable*}

\subsection{RQ 3.1: What are the possible components of federated learning systems?}
The motivation of RQ 3.1 is to identify the components in a federated learning system and their responsibilities. 
We classify the components of federated learning systems into 2 main categories: central server and client devices. A central server initiates and orchestrate the training process, whereas the client devices perform the actual model training~\cite{mcmahan2016communicationefficient, Konecny2016}. 

Apart from the central server and client devices, some studies added edge device to the system as an intermediate hardware layer between the central server and client devices. The edge device is introduced to increase the communication efficiency by reducing the training data sample size~\cite{8854053, jiang2019model} and increases the update speed~\cite{Daga2019, DBLP:journals/corr/abs-1905-06641}. Table~\ref{tab:Component} summarises the number of mentions of each component. We classify them into client-based, server-based, or both to identify where these components are hosted. We can see that the model trainer is the most mentioned component for the client-based components, and model aggregator is mentioned the most in the server-based components. Furthermore, we notice that the software components that manage the system (e.g., anomaly detector, data provenance, communication coordinator, resource manager \& client selector) are mostly server-based while the software components that enhance the model performance (e.g., feature fusion, data augmentation) are mostly client-based. Lastly, two-way operating software components such as model encryption, privacy preservation, and model compressor exist in both clients and the server.

\begin{table*}

  \caption{Summary of component designs}
  \label{tab:Component}
  \scriptsize
  \begin{tabular}{cccc}
    \toprule
    \textbf{\makecell{Sub-component}} & \textbf{\makecell{Client-based}} & \textbf{\makecell{Server-based}} &  \textbf{\makecell{Both}}\\
    \midrule
    \textup{{\makecell[l]{Anomaly detector}}} & \textup{{\makecell{-}}} & \textup{{\makecell{3}}} & \textup{{\makecell{-}}}\\ 
    \textup{{\makecell[l]{Auditing mechanism}}} & \textup{{\makecell{-}}} & \textup{{\makecell{2}}} & \textup{{\makecell{-}}}\\  
    \textup{{\makecell[l]{Data provenance}}} & \textup{{\makecell{7}}} & \textup{{\makecell{14}}} & \textup{{\makecell{-}}}\\
    \textup{{\makecell[l]{Client selector}}} & \textup{{\makecell{-}}} & \textup{{\makecell{1}}} & \textup{{\makecell{-}}}\\ 
    \textup{{\makecell[l]{Communication coordinator}}} & \textup{{\makecell{-}}} & \textup{{\makecell{6}}} & \textup{{\makecell{-}}}\\ 
    \textup{{\makecell[l]{Data augmentation}}} & \textup{{\makecell{4}}} & \textup{{\makecell{3}}} & \textup{{\makecell{-}}}\\ 
    \textup{{\makecell[l]{Encryption mechanism}}} & \textup{{\makecell{3}}} & \textup{{\makecell{-}}} & \textup{{\makecell{12}}}\\
    \textup{{\makecell[l]{Feature fusion mechanism}}} & \textup{{\makecell{6}}} & \textup{{\makecell{-}}} & \textup{{\makecell{-}}}\\
    \textup{{\makecell[l]{Incentive mechanism}}} & \textup{{\makecell{2}}} & \textup{{\makecell{9}}} & \textup{{\makecell{-}}}\\
    \textup{{\makecell[l]{Model aggregator}}} & \textup{{\makecell{-}}} & \textup{{\makecell{188}}} & \textup{{\makecell{-}}}\\
    \textup{{\makecell[l]{Model compressor}}} & \textup{{\makecell{-}}} & \textup{{\makecell{-}}} & \textup{{\makecell{6}}}\\
    \textup{{\makecell[l]{Model trainer}}} & \textup{{\makecell{211}}} & \textup{{\makecell{-}}} & \textup{{\makecell{-}}}\\
    \textup{{\makecell[l]{Privacy preservation mechanism}}} & \textup{{\makecell{1}}} & \textup{{\makecell{-}}} & \textup{{\makecell{14}}}\\
    \textup{{\makecell[l]{Resource manager}}} & \textup{{\makecell{2}}} & \textup{{\makecell{9}}} & \textup{{\makecell{-}}}\\
    \textup{{\makecell[l]{Training manager}}} & \textup{{\makecell{1}}} & \textup{{\makecell{9}}} & \textup{{\makecell{-}}}\\
    \bottomrule
  \end{tabular}
\end{table*}

\subsubsection{Central server:} Physically, the central servers are usually hosted on a local machine~\cite{8647616, 8683546}, cloud server~\cite{thomas2018federated, Xu2018, 8761267}, mobile edge computing platform~\cite{8761315}, edge gateway~\cite{Shen2019, 8560084} or base station~\cite{8716527, 8843942, 8952884, 8891310}. The central server is a hardware component that creates the first global model by randomly initialises the model parameters or gradient\cite{mcmahan2016communicationefficient, chen2019joint, 8683546, 8859260, 8944302, 8917724}. Besides randomising the initial model, central servers can pre-train the global model, either using a self-generated sample dataset or a small amount of data collected from each client device~\cite{qian2019active, 8854053}. We can define this as the server-initialised model training setting. Note that not all federated learning systems initialise their global model on the central server. In decentralised federated learning, clients initialise the global model~\cite{8950073, 8905038, 8892848}.

After global model initialisation, the central servers broadcast the global model that includes the model parameters or gradients to the participating client devices. The global model can be broadcasted to all the participating client devices every round~\cite{8870236, 8891310, 8889996, 8904164}, or only to specific client devices, either randomly~\cite{8945292, 8917724, 8683546, 8952884, Ulm2019} or through selection based on the model training performance~\cite{8854053, Xu2019} and the resources availability~\cite{8664630, 8928018}. Similarly, the trained local models are also collected from either all the participating client devices~\cite{8664630, 8885054, 8836609} or only from selected client devices~\cite{Bonawitz2017, Mandal2019}. The collection of models can either be in an asynchronous~\cite{8772088, Hu2019, xie2019asynchronous, chen2019asynchronous} or synchronous manner~\cite{duan2019astraea, fung2018mitigating}. Finally, the central server performs model aggregations when it receives all or a specific amount of updates, followed by the redistribution of the updated global model to the client devices. This entire process continues until convergence is achieved.

Apart from orchestrating the model parameters and gradients exchange, the central server also hosts other software components, such as encryption/decryption mechanisms for model encryption and decryption~\cite{Awan2019, Bonawitz2017, Mandal2019}, and resource management mechanisms to optimise the resource consumption~\cite{8761315, 8716527}.
Evaluation framework is proposed to evaluate the model and system performance~\cite{caldas2018leaf, hsu2019measuring, wei2019multi}, while the client and model selector is introduced to select appropriate clients for model training and select high quality models for global aggregation~\cite{Xu2019, 8854053, 8664630, 8928018}. Feature fusion mechanisms are proposed to combine essential model features and reduces communication cost~\cite{Hu2019, verma2019approaches, wang2019interpret, 8803001}. Incentive mechanisms are utilised to motivate the clients' participation rate~\cite{Zhan2020, 8875430, 8867906, khan2019federated, wang2019measure, Pandey2020, Pandey2019}. An anomaly detector is introduced to detect system anomaly~\cite{fung2018mitigating}, while the model compressor compresses the model to reduce the data size~\cite{8885054, 8928018, 8889996, caldas2018expanding, Konecny2016}. Communication coordinator manages the multi-channel communication between the central server and client devices~\cite{8952884, 8891310, 8904164, chen2019joint, vu2019cell, shi2019device, amiri2019federated}. Lastly, auditing mechanisms audit the training processes~\cite{8905038, augenstein2019generative, 8905038, Zhao2020}.   

\subsubsection{Client devices:} The client devices are the hardware component that conducts model training using the locally available datasets. Firstly, each client device collects and pre-processes the local data (data cleaning, labeling, feature extraction, etc.). All client devices receive the initial global model and initiates the operations. The client devices decrypts and extracts the global model parameters. After that, they perform local model training. The received global model is also used for data inference and prediction by the client devices. 

The local model training minimises the loss function and optimises the local model performance. Typically, the model is trained for multiple rounds and epochs~\cite{mcmahan2016communicationefficient}, before being uploaded back to the central server for model aggregations. To reduce the number of communication rounds,~\cite{guha2019one} proposed to perform local training on multiple local mini-batch of data. The method only communicates with the central server after the model achieved convergence. 

After that, the client devices send the training results (model parameters or gradients) back to the central server. Before  uploading, client devices evaluate the local model performance and only upload when an agreed level of performance is achieved~\cite{hsu2019measuring}. The results are encrypted using the encryption mechanism before uploading to preserve the data security and prevent information leakage~\cite{Awan2019, Bonawitz2017, Mandal2019}. Furthermore, the model is compressed before uploaded to the central server to reduce communication cost~\cite{8885054, 8928018, 8889996, caldas2018expanding, Konecny2016}. In certain scenarios, not all devices are required to upload their results. Only the selected client devices are required to upload the results, depending on the selection criteria set by the central server. The criteria evaluates the available resources of the client devices and the model performance~\cite{8761315, 8716527, Zhan2020ExperienceDrivenCR}.
The client devices can host data augmentation mechanism~\cite{jeong2018communication, Triastcyn2019}, and feature fusion mechanisms that correlate to the central server in the same federated learning systems. After the completion of one model training round, the training result is uploaded to the central server for global aggregation.

For decentralised federated learning systems, the client devices communicate among themselves without the orchestration of the central server. The removed central server is mostly replaced by a blockchain as a software component for model and information provenance~\cite{8843900, 8893114, 8733825, 8894364, 8905038, 8892848, 8832210, Awan2019}. The blockchain is also responsible for incentive provision and differential private multiparty data model sharing. The initial model is created locally by each client device using local datasets. The models are updated using a consensus-based approach that enables devices to send model updates and receive gradients from neighbour nodes~\cite{8950073, lalitha2019decentralized, lalitha2019peer}. The client devices are connected through a peer-to-peer network~\cite{shayan2018biscotti, roy2019braintorrent, hu2019decentralized}. Each device has the model update copies of all the client devices. After reaching consensus, all the client devices will conduct model training using the new gradient. 

In cross-device settings, the system has a high client device population where each device is the data owner. In contrast, in the cross-silo setting, the client network is formed by several companies or organisations, regardless of the number of individual devices they own. The number of data silos is significantly smaller compared to the cross-device setting~\cite{kairouz2019advances}. 
Therefore, the cross-device system creates models for large-scale distributed data on the same application~\cite{8818446}, while the cross-silo system creates models to accommodate data that is heterogeneous in terms of its content and semantic in both features and sample space~\cite{8818446}. The data partitioning is also different. For the cross-device setting, the data is partitioned automatically by example (horizontal data partitioning). The data partitioning of cross-silo setting is fixed either by feature (vertical) or by example (horizontal)~\cite{kairouz2019advances}. 

\begin{figure*}[h!]
\begin{center}
\footnotesize
\begin{mdframed}[
    skipabove=1cm,    
    innerleftmargin =-1cm,
    innerrightmargin=-1cm,
    usetwoside=false,
]
\begin{center}
    \textbf{Findings of RQ 3.1: What are the possible components of federated learning?} \\
\end{center} 

\begin{quotation}

\textit{\textbf{Mandatory components (clients):}} data collection, data preprocessing, feature engineering, model training, and inference.
\end{quotation}

\begin{quotation}
\textit{\textbf{Mandatory components (server):}} model aggregation, evaluation.
\end{quotation}

\begin{quotation}
\textit{\textbf{Optional components (clients):}} anomaly detection, model compression, auditing mechanisms, data augmentation, feature fusion/selection, security protection, privacy preservation, data provenance.
\end{quotation}

\begin{quotation}
\textit{\textbf{Optional components (server):}} advanced model aggregation, training management, incentive mechanism, resource management, communication coordination.
\end{quotation}

\begin{quotation}
\textbf{With regard to the software development lifecycle}, this question contributes to the \textit{architecture design} phase where we discuss the roles, responsibilities, and the interactions of the different components from an architecture design perspective.
\end{quotation}

\begin{quotation}
\textit{*Mandatory components - Components that perform the main federated learning operations.}
\end{quotation}

\begin{quotation}
\textit{*Optional components - Components that assist/enhance the federated learning operations.}
\end{quotation}

\end{mdframed}
\end{center} 
\end{figure*}

\subsection{RQ 3.2: What are the phases of the machine learning pipeline covered by the research activities?}
Table~\ref{tab:MLpipeline} presents a summary of machine learning pipeline phases covered by the studies. Notice that only phases that are specifically elaborated in the papers are included. The top 3 most mentioned machine learning phases are "model training" (161 mentions), followed by "data collection" (22 mentions) and "data cleaning" (18 mentions). These 3 stages of federated learning are mostly similar to the approaches in conventional machine learning systems. The key differences are the distributed model training tasks, decentralised data storage, and non-IID data distribution. Notice that only Google mentioned model inference and deployment, specifically for Google keyboard applications. The on-device inference supported by TensorFlow Lite is mentioned in~\cite{hard2018federated, ramaswamy2019federated}, and~\cite{yang2018applied} mentioned that a model checkpoint from the server is used to build and deploy the on-device inference model. It uses the same featurisation flow which originally logged training examples on-device. However, the deployed model monitoring (e.g., dealing with performance degradation) and project management (e.g., model versioning) are not discussed in the existing studies. We infer that federated learning research is still in an early stage as most researchers focused on the data-processing and model training optimisation. 
\begin{figure*}[h!]
\begin{center}
\footnotesize
\begin{mdframed}[
    skipabove=1cm,    
    innerleftmargin =-1cm,
    innerrightmargin=-1cm,
    usetwoside=false,
]
\begin{center}
    \textbf{Findings of RQ 3.2:\\ What are the phases of the machine learning pipeline covered by the research activities?} \\
\end{center} 

\begin{quotation}

\textit{\textbf{Phases:}} The model training phase is most discussed. Only a few studies expressed data pre-processing, feature engineering, model evaluation, and only Google has discussions about  model deployment (e.g., deployment strategies) and model inference. Model monitoring (e.g., dealing with performance degradation), and project management (e.g., model versioning) are not discussed in the existing studies. More studies are needed for the development of production-level federated learning systems.
\end{quotation}

\begin{quotation}
\textbf{With regard to the software development lifecycle}, this question contributes to the \textit{background understanding} phase where we identify the machine learning pipeline phases focused by the federated learning studies.
\end{quotation}

\end{mdframed}
\end{center} 
\end{figure*}

\begin{table*}[h]

  \caption{Summary of machine learning pipeline phases}
  \label{tab:MLpipeline}
  \scriptsize
  \begin{tabular}{cccccccccc}
    \toprule
    \textbf{\makecell[l]{ML\\pipeline}} & \textup{\makecell{Data\\collection}} & \textup{\makecell{Data\\cleaning}} & \textup{\makecell{Data\\labelling}} & \textup{\makecell{Data\\augmentation}} & \textup{\makecell{Feature\\engineering}} & \textup{\makecell{Model\\training}} & \textup{\makecell{Model\\evaluation}} & \textup{\makecell{Model\\deployment}} & \textup{\makecell{Model\\inference}}\\
    \midrule
    \textbf{{\makecell[l]{Paper\\count}}} & \textup{{\makecell{22}}} & \textup{{\makecell{18}}} & \textup{{\makecell{13}}} & \textup{{\makecell{9}}} & \textup{{\makecell{8}}} & \textup{{\makecell{161}}} & \textup{{\makecell{10}}} & \textup{{\makecell{1}}} & \textup{{\makecell{2}}}\\
    \bottomrule
  \end{tabular}
\end{table*}




\subsection{RQ 4: How are the approaches evaluated?}

RQ 4 focuses on the evaluation approaches in the studies. We classify the evaluation approaches into two main groups: simulation and case study. 
For the simulation approach, image processing is the most common task, with 99 out of 197 cases, whereas the most implemented use cases are applications on mobile devices (11 out of 17 cases), such as word suggestion and human activity recognition. 

\begin{table*}[h]
  \caption{Evaluation approaches for federated learning}
  \label{tab:RQ4}
  \scriptsize
  \begin{tabular}{ccc}
    \toprule
    \textbf{\makecell{Evaluation methods}} & \textbf{\makecell{Application types}} & \textbf{\makecell{Paper count}}\\
    \midrule
    \textup{{\makecell[l]{Simulation (85\%)}}} & \textup{{\makecell[l]{Image processing\\Others}}} & \textup{{\makecell{99\\98}}}\\
    \hline \\[-1.8ex]
    \textup{{\makecell[l]{Case study (7\%)}}} & \textup{{\makecell[l]{Mobile device applications\\Healthcare\\Others}}} &\textup{{\makecell{11\\3\\3}}}\\
    \hline \\[-1.8ex]
    \textup{{\makecell[l]{Both (1\%)}}} & \textup{{\makecell[l]{Image processing\\Mobile device applications}}} &\textup{{\makecell{1\\1}}}\\
    \hline \\[-1.8ex]
    \textup{{\makecell[l]{No evaluation (7\%)}}} & \textup{{\makecell{-}}} &\textup{{\makecell{15}}}\\
    \bottomrule
  \end{tabular}
\end{table*}

\begin{figure*}[h!]
\begin{center}
\footnotesize
\begin{mdframed}[
    skipabove=1cm,    
    innerleftmargin =-1cm,
    innerrightmargin=-1cm,
    usetwoside=false,
]
\begin{center}
    \textbf{Findings of RQ 4: How are the approaches evaluated?} \\
\end{center} 
\begin{quotation}

\textit{\textbf{Evaluation:}} Researchers mostly evaluate their federated learning approaches by simulation using privacy-sensitive scenarios. There are only a few real-world case studies, e.g., Google's mobile keyboard prediction.

\end{quotation}

\begin{quotation}
\textbf{With regard to the software development lifecycle}, this question contributes to the \textit{implementation and evaluation} phase where we explore the different methods to evaluate the federated learning approaches.
\end{quotation}

\end{mdframed}
\end{center} 
\end{figure*}

\subsection{RQ 4.1: What are the evaluation metrics used to evaluate the approaches?}

Through RQ 4.1, we intend to identify the evaluation metrics for both qualitative and quantitative  methods adopted by federated learning systems.  We explain how each evaluation metric is used to assess the system and map these metrics to the quality attributes mentioned in RQ 2. The results are summarised in Table~{\ref{tab:rq4vsrq2}}.

First, the communication efficiency is evaluated by communication cost, dropout ratio, model performance, and system running time. The communication cost is quantified by the communication rounds against the learning accuracy~\cite{8885054, 8917724, 8945292, 8932389, 8672262, 8935424, liu2019communication}, the satisfying rate of communications~\cite{8944302},  communication overhead versus  number of clients~\cite{8761267, yang2019quasi}, the theoretical analysis of communication cost of data interchange between the server and clients~\cite{Song2019b, 8935736, Bonawitz2017},  data transmission rate~\cite{Xu2019a, 8765347}, bandwidth~\cite{Zhan2020ExperienceDrivenCR, QIAN2019562}, and latency of communication~\cite{8733825}. The dropout ratios are measured by the computation overhead against dropout ratios~\cite{niu2019secure, 8765347}. The results are showcased as the comparison between communication overhead for different dropout rates~\cite{8765347}, and the performance comparison against dropout rate~\cite{chen2019asynchronous, liu2019boosting, li2018federated}.

Secondly, the model performance is measured by the training loss~\cite{8672262, chen2018federated, liu2019communication}, AUC-ROC value~\cite{liu2019communication, 8843900, yang2019parallel}, F1-score~\cite{chen2019asynchronous, 8854245, bakopoulou2019federated}, root-mean-squared error (RMSE)~\cite{8851408, 8759317},
cross-entropy~\cite{Fan2019, xie2019asynchronous}, precision~\cite{8854245, Zhao2019, Yang2019}, recall~\cite{8854245, Zhao2019, Yang2019}, prediction error~\cite{Smith2017, konen2016federated}, mean absolute error~\cite{Hu2019}, dice coefficient~\cite{Sheller2019}, and perplexity value~\cite{sattler2019clustered}. 

Thirdly, the system scalability is evaluated by communication cost and system running time. For the system running time evaluation, the results are presented as the total execution time of the training protocol (computation time \& communication time)~\cite{8765347, 8843900, 8836609, 8945183, liu2019boosting}, the running time of different operation phases~\cite{8836609}, the running time of each round against the number of client devices~\cite{8761315, Bonawitz2017, preuveneers2018chained}, and the model training time~\cite{8894364, Xu2019a}.

The system performance is evaluated in multiple aspects, including system security (e.g., attack rate), scalability (e.g., communication and computation costs, dropout ratio), and system reliability (e.g., convergence rate, model performance, and system running time). The attack rate is measured as the proportion of attack targets that are incorrectly classified as the target label~\cite{fung2018mitigating}. Essentially, researchers use this metric to evaluate how effective the defence mechanisms are~\cite{fung2018dancing, Zhao2020, shayan2018biscotti, 8945183}. The types of attack are model poisoning attack, sybil attack, byzantine attack, and data reconstruction attack. Computation cost is the assessment of computation, storage, and energy resource usage of a system. The computation resources are quantified by the computation overhead against the number of client devices~\cite{niu2019secure, Zhan2020ExperienceDrivenCR, Mo, 8859260}, average computation time~\cite{jiang2019model, Li2019, yang2019energy}, computation throughput~\cite{8894364, 8716527}, computation latency~\cite{8733825}, computation utility, and the overhead of components~\cite{Zhan2020, chen2018federated}. The storage resources are evaluated by the memory and storage overhead~\cite{niu2019secure, Mandal2019, duan2019astraea}, and storage capacity~\cite{QIAN2019562}. The energy resources are calculated by the energy consumption for communication~\cite{8875430, DBLP:journals/corr/abs-1905-06641}, the energy consumption against computation time~\cite{Li2019, yang2019energy, DBLP:journals/corr/abs-1905-06641, Xu2019, Shen2019}, and the energy consumption against training dataset size~\cite{8875353}. Lastly, the convergence rate is quantified by the accuracy versus the communication rounds, system running time, and training data epochs~\cite{Benditkis2019, 8803001, 8867906}.

For statistical and system heterogeneity, qualitative analyses are conducted to verify if the proposed approaches have satisfied their purpose of addressing the limitations. The statistical heterogeneity is evaluated through formal verification such as model and equation proving~\cite{mohri2019agnostic, karimireddy2019scaffold, xie2019asynchronous}, whereas system heterogeneity is evaluated by the dropout ratio due to the limited resource and also formal verification through equation proving~\cite{mohri2019agnostic}.  

Client motivatability is measured by the incentive rate against different aspects of the system. The incentive rate is assessed by calculating the profit of the task publisher under different numbers of clients or accuracy levels~\cite{8832210}, the average reward based on the model performance~\cite{8807242}, and the relationship between the offered reward rate and the local accuracy over the communication cost~\cite{khan2019federated, Pandey2019, Pandey2020}. From the studies collected, there is no mention of any specific form of rewards provided as incentives. However, cryptocurrencies such as Bitcoin or tokens which can be converted to actual money are  common kinds of rewards.

Finally, both data security and client device security are measured by the attack rates and other respective qualitative evaluation metrics. Essentially, the data security analyses include analysis of encryption and verification process performance~\cite{8761267, ghazi2019scalable, bhowmick2018protection}, the differential privacy achievement~\cite{8843942, 8843900}, the effect of the removal of centralised trust~\cite{8843900}, and the guarantee of shared data quality~\cite{8843900}. The client device security analyses are the performance of encryption and verification process~\cite{8765347, liu2018secure, yang2019parallel}, the confidentiality guarantee for gradients, the auditability of gradient collection and update, and the fairness guarantee for model training~\cite{8894364}. Also, the data security is measured by privacy loss, which evaluates the privacy-preserving level of the proposed method~\cite{8807242}, derived from the differential average-case privacy~\cite{Triastcyn2019}.

\begin{table*}
  \caption{Quality Attributes vs Evaluation Metrics }
  \label{tab:rq4vsrq2}
  \resizebox*{\textwidth}{!}{%
  \begin{tabular}{lccccccccc}
    \toprule
    \textbf{\makecell{Quality\\attributes\\vs\\Evaluation\\metrics}} &  \textbf{\makecell{Communication\\efficiency}} & \textbf{\makecell{Model\\performance}} & \textbf{\makecell{Scalability}} & \textbf{\makecell{System\\performance}} & \textbf{\makecell{Statistical\\heterogeneity}} & \textbf{\makecell{System\\heterogeneity}} &
    \textbf{\makecell{Client\\motivatability}} &
    \textbf{\makecell{Data\\security}} &
    \textbf{\makecell{Client\\device\\security}}\\
    \midrule
    \textup{\makecell[l]{Attack rate}}  & \textup{\makecell{-}} &\textup{\makecell{-}} &\textup{\makecell{-}} &\textup{\makecell{1}}    &\textup{\makecell{-}}  &\textup{\makecell{-}} & \textup{\makecell{-}} &\textup{\makecell{1}} &\textup{\makecell{4}}\\

    \textup{\makecell[l]{Communication cost}}  & \textup{\makecell{58}} &\textup{\makecell{-}} &\textup{\makecell{3}} &\textup{\makecell{2}}    &\textup{\makecell{-}}  &\textup{\makecell{-}}  & \textup{\makecell{-}} &\textup{\makecell{-}} &\textup{\makecell{-}}\\
    
    \textup{\makecell[l]{Computation cost}}  & \textup{\makecell{-}} &\textup{\makecell{-}} &\textup{\makecell{-}} &\textup{\makecell{42}}   &\textup{\makecell{-}}  &\textup{\makecell{-}} & \textup{\makecell{-}}  &\textup{\makecell{-}} &\textup{\makecell{-}}\\
    
    \textup{\makecell[l]{Convergence rate}}  & \textup{\makecell{-}} &\textup{\makecell{-}} &\textup{\makecell{-}} & \textup{\makecell{15}}  &\textup{\makecell{-}} &\textup{\makecell{-}} &\textup{\makecell{-}} & \textup{\makecell{-}}\\

    \textup{\makecell[l]{Dropout ratio}} & \textup{\makecell{1}} & \textup{\makecell{-}} &\textup{\makecell{-}} &\textup{\makecell{2}} &\textup{\makecell{-}}&\textup{\makecell{2}}&\textup{\makecell{-}} &\textup{\makecell{1}} &\textup{\makecell{-}}\\
    
     \textup{\makecell[l]{Incentive rate}}  & \textup{\makecell{-}} &\textup{\makecell{-}} &\textup{\makecell{-}} &\textup{\makecell{-}}&\textup{\makecell{-}}&\textup{\makecell{-}} & \textup{\makecell{6}} &\textup{\makecell{-}} &\textup{\makecell{-}}\\

    \textup{\makecell[l]{Model performance}} & \textup{\makecell{1}} &\textup{\makecell{253}} &\textup{\makecell{-}} &\textup{\makecell{1}} &\textup{\makecell{-}}&\textup{\makecell{-}} & \textup{\makecell{2}} &\textup{\makecell{-}} &\textup{\makecell{-}}\\
    
    \textup{\makecell[l]{Privacy loss}} & \textup{\makecell{-}} &\textup{\makecell{-}} &\textup{\makecell{-}} &\textup{\makecell{-}} &\textup{\makecell{-}}&\textup{\makecell{-}}  & \textup{\makecell{-}} &\textup{\makecell{2}} &\textup{\makecell{-}}\\

    \textup{\makecell[l]{System running time}}  & \textup{\makecell{2}} &\textup{\makecell{-}} &\textup{\makecell{2}} &\textup{\makecell{20}}&\textup{\makecell{-}}&\textup{\makecell{-}} & \textup{\makecell{-}} &\textup{\makecell{-}} &\textup{\makecell{-}}\\
    
    \textup{\makecell[l]{Qualitative evaluation}} & \textup{\makecell{-}} &\textup{\makecell{-}} &\textup{\makecell{-}} &\textup{\makecell{-}} &\textup{\makecell{3}}&\textup{\makecell{1}} & \textup{\makecell{-}} &\textup{\makecell{7}} &\textup{\makecell{4}}\\
    
    \bottomrule
  \end{tabular}}

\end{table*}

\begin{figure*}[h!]
\begin{center}
\footnotesize
\begin{mdframed}[
    skipabove=1cm,    
    innerleftmargin =-1cm,
    innerrightmargin=-1cm,
    usetwoside=false,
]
\begin{center}
    \textbf{Findings of RQ 4.1: What are the evaluation metrics used to evaluate the approaches?} \\
\end{center} 
\begin{quotation}

\textit{\textbf{Evaluation:}} Both quantitative and qualitative analysis are used to evaluate the federated learning system.
\end{quotation}

\begin{quotation}
\textit{\textbf{Quantitative metrics examples:}} Model performance, communication \& computation cost, system running time, etc.
\end{quotation}

\begin{quotation}
\textit{\textbf{Qualitative metrics examples:}} Data security analysis on differential privacy achievement, Performance of encryption and verification process, confidentiality guarantee for gradients, the auditability of gradient collection and update, etc.

\end{quotation}

\begin{quotation}
\textbf{With regard to the software development lifecycle}, this question contributes to the \textit{implementation and evaluation} phase where we identify the different evaluation metrics used to assess the quality attributes of the federated learning systems.
\end{quotation}

\end{mdframed}
\end{center} 
\end{figure*}


\begin{figure}
  \includegraphics[width=\linewidth]{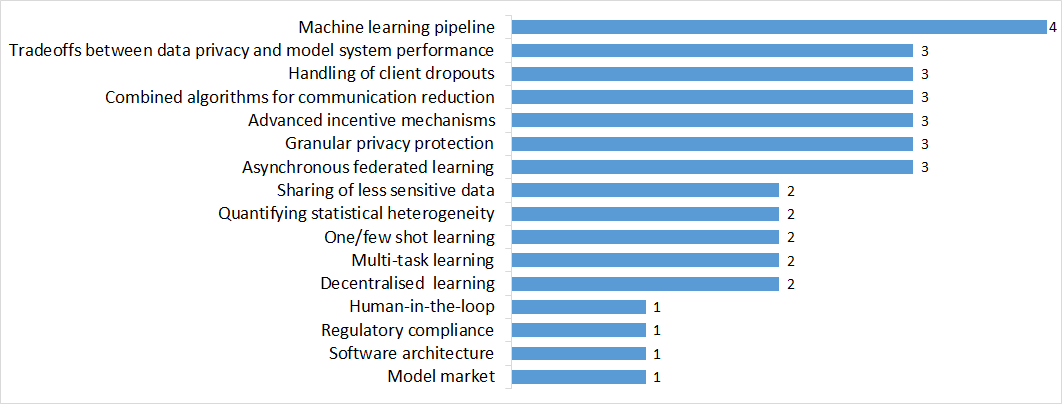}
  \caption{Open problems and future research trends highlighted by existing reviews/surveys}
  \label{fig:future}
\end{figure}

\subsection{Summary}
\begin{figure}
  \includegraphics[width=\linewidth]{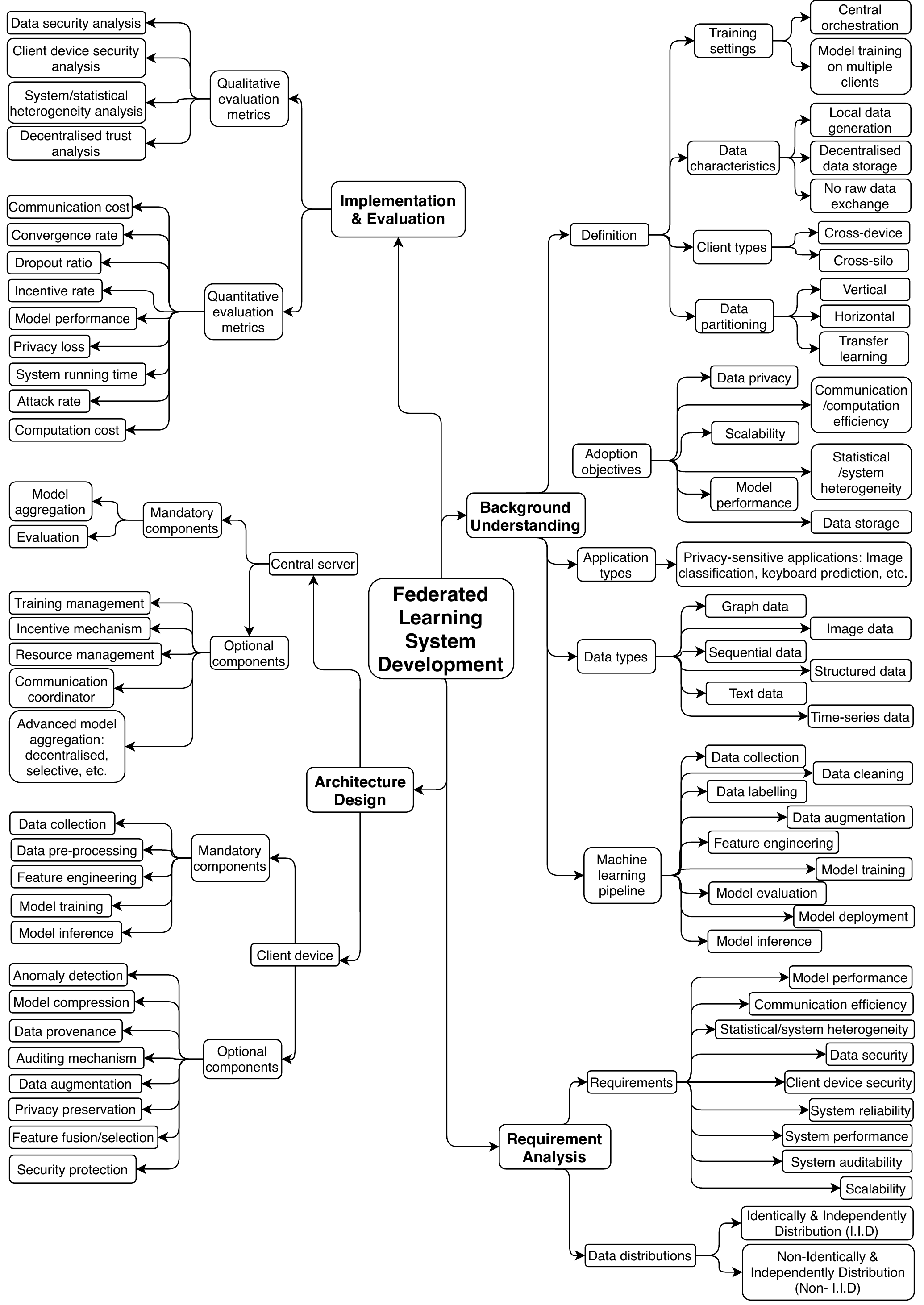}
  \caption{Mind-map summary of the findings}
  \label{fig:mind_map}
\end{figure}

We have presented all the findings and results extracted through each RQ. We summarise all the findings in a mind-map, as shown in Fig.~{\ref{fig:mind_map}}.

\section{Open Problems and Future Trends in Federated Learning} \label{Section:Future}
In this section, we discuss the open problems and future research trends from the survey and review papers we collected to provide unbiased analyses (refer Table~{\ref{tab:existingsurvey}}).
The findings are shown in Fig.~\ref{fig:future} and the detailed explanations are elaborated below:

\begin{itemize}
    \item \textbf{Enterprise and industrial-level implementation}. Being at the early stage of research~\cite{kairouz2019advances, Li_2020, li2019survey, lim2019federated, xu2019federated}, the only mentions of enterprise federated learning systems are the cross-silo settings and some possible applications on real-world use cases listed in Table~\ref{tab:fl_app_data}. The possible challenges are as follow: 
    \begin{itemize}
    \item \textbf{Machine learning pipeline}: Most existing studies in federated learning focus on the federated model training phase, without considering other machine learning pipeline stages (e.g., model deployment and monitoring) under the context of federated learning (e.g., new data lifecycles)~\cite{li2019survey, Li_2020, lim2019federated, xu2019federated}.
    
    \item \textbf{Benchmark}: Benchmarking schemes are needed to evaluate the system development under real-world settings, with rich datasets and representative workload~\cite{kairouz2019advances,li2019survey, Li_2020}.
    
    \item \textbf{Dealing with unlabeled client data}: 
    In practice, the data generated on clients may be mislabeled or unlabeled~\cite{kairouz2019advances, Li_2020, li2019survey, lim2019federated}. Some possible solutions are semi-supervised learning-based techniques, labeling the client data by learning the data label from other clients. However, these solutions may require  dealing with the data privacy, heterogeneity, and scalability issues.
    
    \item \textbf{Software architecture}: Federated learning still lacks systematic architecture design to guide methodical system development. A systematic architecture can provide design or algorithm alternatives for different system settings~\cite{li2019survey}.
    
    \item \textbf{Regulatory compliance}: The regulatory compliance issues for federated learning systems is under-explored~\cite{li2019survey} (e.g., whether data transfer limitations in GDPR is applied to model update transfer, ways to execute right-to-explainability to the global model, and whether the global model should be retrained if a client wants to quit). Machine learning and law enforcement communities are expected to cooperate to fill the gap between federated learning technology and the regulations in reality.
    
    \item \textbf{Human-in-the-loop}: Domain experts are expected to be involved in the federated learning process to provide professional guidance as end-users tend to trust the expert’s judgement more than the inference by machine learning algorithms.~\cite{xu2019federated}.

    \end{itemize}
    
    \item \textbf{Advanced privacy preservation methods}: More advanced privacy preservation methods are needed as the existing solutions still can reveal sensitive information~\cite{kairouz2019advances, li2019survey, lyu2020threats, lim2019federated,niknam2019federated, Li_2020}.
    \begin{itemize}
    
    \item \textbf{Tradeoffs between data privacy and model system performance}: The current approaches (e.g., differential privacy and collaborative training) sacrifice the model performance, and require significant computation cost~\cite{lim2019federated, lyu2020threats, niknam2019federated}. Hence, the design of federated learning systems needs to balance the tradeoffs between data privacy and model/system performance.
    
    \item \textbf{Granular privacy protection}: In reality, privacy requirements may differ across clients or across data samples on a single client. Therefore, it is necessary to protect data privacy in a more granular manner. Privacy heterogeneity should be considered in the design of federated learning systems for different privacy requirements (e.g., client device-specific or sample data specific). One direction of future work is extending differential privacy with granular privacy restriction (e.g., heterogeneous differential privacy)~\cite{Li_2020, lyu2020threats, li2019survey}.
    
    \item \textbf{Sharing of less sensitive model data}: Devices should only share less sensitive model data (e.g., inference results or signSGD), and this type of approaches may be considered in future work~\cite{lyu2020threats, kairouz2019advances}. 
    \end{itemize}
    
    \item \textbf{Improving system and model performance}. There are still some performance issues regarding federated learning systems, mainly on resource allocation (e.g., communication, computation, and energy efficiency). Moreover, model performance improvement through the non-algorithm or non-gradient optimisation approach (e.g., promoting more participants, the extension of the federated model training method) is also another future research trend.
    \begin{itemize}

    \item \textbf{Handling of client dropouts}: In practice, participating clients may drop out from the training process due to energy constraints or network connectivity issues \cite{kairouz2019advances, li2019survey, lim2019federated}. A large number of client dropouts can significantly degrade the model performance. Many of the existing studies do not consider the situation when the number of participating clients changes (i.e., departures or entries of clients). Hence, the design of a federated learning system should support the dynamic scheduling of model updates to tolerate client dropouts. Furthermore, new algorithms are needed to deal with the scenarios where only a small number of clients are left in the training rounds. The learning coordinator should provide stable network connections to the clients to avoid dropouts.
    
    \item \textbf{Advanced incentive mechanisms}: Without a well-designed incentive mechanism, potential clients may be reluctant to join the training process, which will discourage the adoption of federated learning~\cite{kairouz2019advances, li2019survey, xu2019federated}. In most of the current designs, the model owner pays for the participating clients based on some metrics (e.g., number of participating rounds or data size), which might not be effective in evaluating the incentive provision. 
    
    \item \textbf{Model markets}: A possible solution proposed to promote federated learning applications is the model market~\cite{li2019survey}. One can perform model prediction using the model purchased from the model market. In addition, the developed model can be listed on the model market with additional information (e.g., task, domain) for federated transfer learning. 
    
    \item \textbf{Combined algorithms for communication reduction}: The method to combine different communication reduction techniques (combining model compression technique with local updating) is an interesting direction to further improve the system's communication efficiency (e.g., optimise the size of model updates and communication instances)~\cite{kairouz2019advances, Li_2020, lim2019federated}. However, the feasibility of such a combination is still under-explored. In addition, the tradeoffs between model performance and communication efficiency are needed to be further examined (e.g., how to manage the tradeoffs under the change of training settings?).
    
    \item \textbf{Asynchronous federated learning}: Synchronous federated learning may have efficiency issues caused by stragglers.
    Asynchronous federated learning has been considered as a more practical solution, even allowing clients to participate in the training halfway \cite{kairouz2019advances, Li_2020, lim2019federated}. However, new asynchronous algorithms are still under explored to provide convergence guarantee.

    \item \textbf{Statistical heterogeneity quantification}: The quantification of the statistical heterogeneity into metrics (e.g., local dissimilarity) is needed to help improve the model performance for non-IID condition~\cite{Li_2020, lim2019federated}. However, the metrics can hardly be calculated before training. One important future direction is the design of efficient algorithms to calculate the degree of statistical heterogeneity that is useful for model optimisations.
    
    \item \textbf{Multi-task learning}: Most of the existing studies focus on training the same model on multiple clients. One interesting direction is to explore how to apply federated learning to train different models in the federated networks~\cite{kairouz2019advances, lyu2020threats}.
    
    \item \textbf{Decentralised learning}: As mentioned above, decentralised learning does not require a central server in the system~\cite{kairouz2019advances, lyu2020threats}, which prevents single-point-of-failure. Hence, it would be interesting to explore if there is any new attack or whether federated learning security issues still exist in this setting.
    
    \item \textbf{One/few shot learning}: Federated learning that executes less training iterations, such as one/few-shot learning, has been recently discussed for federated learning~\cite{kairouz2019advances, Li_2020}. However, more theoretical and empirical studies are needed in this direction.
    
    \item \textbf{Federated transfer learning}: For federated transfer learning, there are only 2 domains or data parties are assumed in most of the current literature. Therefore, expanding federated transfer learning to multiple domains, or data parties is an open problem~\cite{kairouz2019advances}.
     \end{itemize}
\end{itemize}









\section{Threats to Validity}  \label{Section:Threats}
We identified the threats to validity that might influence the outcomes of our research.
First, publication bias exists as most studies have positive results rather than negative results. 
While studies with positive results are much appealing to be published in comparison with studies with no or negative results, a tendency towards certain outcomes might leads to biased conclusions. The second threat is the exclusion of the studies which focus on pure algorithm improvements. 
The exclusion of respective studies may affect the completeness of this research as some discussion on the model performance and data heterogeneity issues might be relevant. 
The third threat is the incomplete search strings in the automatic search. We included all the possible supplementary terms related to federated learning while excluding keywords that return conventional machine learning publications. However, the search terms in the search strings may still be insufficient to search all the relevant work related to federated learning research topics.
The fourth threat is the exclusion of ArXiv and Google Scholar papers which are not cited by peer-reviewed papers. The papers from these sources are not peer-reviewed and we cannot guarantee the quality of these research works. However, we want to collect as many state-of-the-art studies in federated learning as possible. To maintain the quality of the search pool, we only include papers that are cited by peer-reviewed papers. 
The fifth threat is the time span of research works included in this systematic literature review. We only included papers published from 2016.01.01 to 2020.01.31. 
Since the data in 2020 does not represent the research trend of the entire year, we only include works from 2016 to 2019 for the research trend analysis (refer to Fig.~\ref{RQ2:Research_trend}). However, the studies collected in only January 2020 is equally significant to those from the previous years for identifying challenges and solutions. Hence, we keep the findings from the papers published in 2020 for the remaining discussions. 
The sixth threat is the study selection bias. 
To avoid the study selection bias between the 2 researchers, the cross-validation of the results from the pilot study is performed by the two independent researchers prior to the update of search terms and inclusion/exclusion criteria. The mutual agreement between the two independent researchers on the selections of paper is required. When a dispute on the decision occurs, the third researcher is consulted.
Lastly, there might be bias in data collection and analysis due to different background and experience of the researchers. 

\section{Related Work} \label{Section:RelatedWorks}
According to the protocol, we collected all the relevant surveys and reviews. There are 7 surveys and 1 review that studied the federated learning topic. 
To the best of our knowledge, there is still no systematic literature review conducted on federated learning.

Li et al.~\cite{li2019survey} propose a federated learning building blocks taxonomy that classifies the federated learning system into 6 aspects: data partitioning, machine learning model, privacy mechanism, communication architecture, the scale of the federation, and the motivation of federation. 
Kairouz et al.~\cite{kairouz2019advances} present a general survey on the advancement in research trends and the open problems suggested by researchers. Moreover, the paper covered detailed definitions of federated learning system components and different types of federated learning systems variations. Li et al.~\cite{Li_2020} present a review on federated learning's core challenges of federated learning in terms of communication efficiency, privacy, and future research directions.
Surveys on federated learning systems for specific research domains are also conducted. Niknam et al.~\cite{niknam2019federated} review federated learning in the context of wireless communications. The survey mainly investigates the data security and privacy challenges, algorithm challenges, and wireless setting challenges of federated learning systems. 
Lim et al.~\cite{lim2019federated} discuss federated learning papers in the mobile edge network.
Lyu et al.\cite{lyu2020threats} focus on the security threats and vulnerability challenges in federated learning systems whereas Xu and Wang~\cite{xu2019federated} explore the healthcare and medical informatics domain. A review of federated learning that focuses on data privacy and security aspects was conducted by~\cite{10.1145/3298981}. 

The comparisons of each survey and review papers with our systematic literature review are summarised in Table~\ref{tab:existingsurvey}.
We compare our work with the existing works in 4 aspects:
(1) Time frames: our review on the state-of-the-art is the most contemporary as it is the most up-to-date review. (2) Methodology: we followed Kitchenham's standard guideline~\cite{Kitchenham07guidelinesfor} to conduct this systematic literature review. Most of the existing works have no clear methodology, where information is collected and interpreted with subjective summaries of findings that may be subject to bias. (3) Comprehensiveness: the number of papers analysed in our work is higher than the existing reviews or surveys as we screened through relevant journals and conferences paper-by-paper. (4) Analysis: we provided 2 more detailed review on federated learning approaches, including (a) the context of studies (e.g., publication venues, year, type of evaluation metrics, method, and dataset used) for the state-of-the-art research identification and (b) the data synthesis of the findings through the lifecycle of federated learning systems. 

\begin{table*}[h]
  \caption{Comparison with existing reviews/surveys on federated learning}
  \label{tab:existingsurvey}
  \scriptsize
  \begin{tabular}{ccccc}
    \toprule
    \textbf{{Paper}}&
    \textbf{\makecell{Type}} & \textbf{\makecell{Time frames}} & \textbf{\makecell{Methodology}}    &\textbf{\makecell{Scoping}}\\
    \midrule
    \textup{{\makecell[l]{This study}}} &
    \textup{{\makecell{SLR}}} & \textup{{\makecell{2016-2020}}} & \textup{{\makecell{SLR guideline~\cite{Kitchenham07guidelinesfor}}}}  &
    \textup{{\makecell{Software engineering perspective}}}\\
    \textup{{\makecell[l]{Yang et al. (2019)~\cite{10.1145/3298981}}}} & 
    \textup{{\makecell{Survey}}} & \textup{{\makecell{2016-2018}}} & \textup{{\makecell{Undefined}}}  &
    \textup{{\makecell{General overview}}}\\
    \textup{{\makecell[l]{Kairouz et al. (2019)~\cite{kairouz2019advances}}}} &
    \textup{{\makecell{Survey}}} & \textup{{\makecell{2016-2019}}} & \textup{{\makecell{Undefined}}}  &\textup{{\makecell{General overview}}}\\
    \textup{{\makecell[l]{Li et al. (2020)~\cite{li2019survey}}}} &
    \textup{{\makecell{Survey}}} & \textup{{\makecell{2016-2019}}} & \textup{{\makecell{Customised}}}  &
    \textup{{\makecell{System view}}}\\
    \textup{{\makecell[l]{Li et al. (2019)~\cite{Li_2020}}}} &
    \textup{{\makecell{Survey}}} & \textup{{\makecell{2016-2020}}} & \textup{{\makecell{Undefined}}}  &
    \textup{{\makecell{General overview}}}\\
    \textup{{\makecell[l]{Niknam et al. (2020)~\cite{niknam2019federated}}}} &
    \textup{{\makecell{Review}}} & \textup{{\makecell{2016-2019}}} & \textup{{\makecell{Undefined}}}  &
    \textup{{\makecell{Wireless communications}}}\\
    \textup{{\makecell[l]{Lim et al. (2020)~\cite{lim2019federated}}}} &
    \textup{{\makecell{Survey}}} & \textup{{\makecell{2016-2020}}} & \textup{{\makecell{Undefined}}}  &
    \textup{{\makecell{Mobile edge networks}}}\\
    \textup{{\makecell[l]{Lyu et al. (2020)~\cite{lyu2020threats}}}} &
    \textup{{\makecell{Survey}}} & \textup{{\makecell{2016-2020}}} & \textup{{\makecell{Undefined}}}  &
    \textup{{\makecell{Vulnerabilities}}}\\
    \textup{{\makecell[l]{Xu and Wang (2019)~\cite{xu2019federated}}}} &
    \textup{{\makecell{Survey}}} & \textup{{\makecell{2016-2019}}} & \textup{{\makecell{Undefined}}} &
    \textup{{\makecell{Healthcare informatics}}}\\
    \bottomrule
  \end{tabular}
\end{table*}

\section{Conclusion} \label{Section:Conclusion}
Federated learning has attracted a broad range of interests from academia and industry. 
We performed a systematic literature review on federated learning from the software engineering perspective with 231 primary studies. The results show that most of the known motivations for using federated learning appear to be also the most studied research challenges in federated learning.
To tackle the challenges, the top five proposed approaches are model aggregation, training management, incentive mechanism, privacy preservation, and resource management. 
The research findings provide clear viewpoints on federated learning system development for production adoption. 
Finally, this paper sheds some light on the future research trends of federated learning and encourages researchers to extend and advance their current work.

\bibliographystyle{ACM-Reference-Format}

\bibliography{SLR_231}


\begin{thebibliography}{209}


\ifx \showCODEN    \undefined \def \showCODEN     #1{\unskip}     \fi
\ifx \showDOI      \undefined \def \showDOI       #1{#1}\fi
\ifx \showISBNx    \undefined \def \showISBNx     #1{\unskip}     \fi
\ifx \showISBNxiii \undefined \def \showISBNxiii  #1{\unskip}     \fi
\ifx \showISSN     \undefined \def \showISSN      #1{\unskip}     \fi
\ifx \showLCCN     \undefined \def \showLCCN      #1{\unskip}     \fi
\ifx \shownote     \undefined \def \shownote      #1{#1}          \fi
\ifx \showarticletitle \undefined \def \showarticletitle #1{#1}   \fi
\ifx \showURL      \undefined \def \showURL       {\relax}        \fi
\providecommand\bibfield[2]{#2}
\providecommand\bibinfo[2]{#2}
\providecommand\natexlab[1]{#1}
\providecommand\showeprint[2][]{arXiv:#2}

\bibitem[\protect\citeauthoryear{??}{iso}{2008}]%
        {iso25012.com}
 \bibinfo{year}{2008}\natexlab{}.
\newblock \bibinfo{title}{ISO/IEC 25012}.
\newblock
\newblock
\urldef\tempurl%
\url{https://iso25000.com/index.php/en/iso-25000-standards/iso-25012}
\showURL{%
\tempurl}


\bibitem[\protect\citeauthoryear{??}{iso}{2011}]%
        {iso25010.com}
 \bibinfo{year}{2011}\natexlab{}.
\newblock \bibinfo{title}{ISO/IEC 25010}.
\newblock
\newblock
\urldef\tempurl%
\url{https://iso25000.com/index.php/en/iso-25000-standards/iso-25010}
\showURL{%
\tempurl}


\bibitem[\protect\citeauthoryear{??}{(gd}{2019}]%
        {(gdpr)_2019}
 \bibinfo{year}{2019}\natexlab{}.
\newblock \bibinfo{title}{General Data Protection Regulation GDPR}.
\newblock
\newblock
\urldef\tempurl%
\url{https://gdpr-info.eu/}
\showURL{%
\tempurl}


\bibitem[\protect\citeauthoryear{Abad, Ozfatura, Gunduz, and Ercetin}{Abad
  et~al\mbox{.}}{2019}]%
        {abad2019hierarchical}
\bibfield{author}{\bibinfo{person}{Mehdi Salehi~Heydar Abad},
  \bibinfo{person}{Emre Ozfatura}, \bibinfo{person}{Deniz Gunduz}, {and}
  \bibinfo{person}{Ozgur Ercetin}.} \bibinfo{year}{2019}\natexlab{}.
\newblock \showarticletitle{Hierarchical federated learning across
  heterogeneous cellular networks}.
\newblock \bibinfo{journal}{\emph{arXiv preprint arXiv:1909.02362}}
  (\bibinfo{year}{2019}).
\newblock


\bibitem[\protect\citeauthoryear{{Ahn}, {Simeone}, and {Kang}}{{Ahn}
  et~al\mbox{.}}{2019}]%
        {8904164}
\bibfield{author}{\bibinfo{person}{J. {Ahn}}, \bibinfo{person}{O. {Simeone}},
  {and} \bibinfo{person}{J. {Kang}}.} \bibinfo{year}{2019}\natexlab{}.
\newblock \showarticletitle{Wireless Federated Distillation for Distributed
  Edge Learning with Heterogeneous Data}. In \bibinfo{booktitle}{\emph{2019
  IEEE 30th Annual International Symposium on Personal, Indoor and Mobile Radio
  Communications (PIMRC)}}. \bibinfo{pages}{1--6}.
\newblock
\showISSN{2166-9570}


\bibitem[\protect\citeauthoryear{Amiri and Gunduz}{Amiri and Gunduz}{2019}]%
        {amiri2019federated}
\bibfield{author}{\bibinfo{person}{Mohammad~Mohammadi Amiri} {and}
  \bibinfo{person}{Deniz Gunduz}.} \bibinfo{year}{2019}\natexlab{}.
\newblock \showarticletitle{Federated learning over wireless fading channels}.
\newblock \bibinfo{journal}{\emph{arXiv preprint arXiv:1907.09769}}
  (\bibinfo{year}{2019}).
\newblock


\bibitem[\protect\citeauthoryear{Amiri, Gunduz, Kulkarni, and Poor}{Amiri
  et~al\mbox{.}}{2020}]%
        {amiri2020update}
\bibfield{author}{\bibinfo{person}{Mohammad~Mohammadi Amiri},
  \bibinfo{person}{Deniz Gunduz}, \bibinfo{person}{Sanjeev~R. Kulkarni}, {and}
  \bibinfo{person}{H.~Vincent Poor}.} \bibinfo{year}{2020}\natexlab{}.
\newblock \bibinfo{title}{Update Aware Device Scheduling for Federated Learning
  at the Wireless Edge}.
\newblock
\newblock
\showeprint[arxiv]{2001.10402}~[cs.IT]


\bibitem[\protect\citeauthoryear{Anelli, Deldjoo, Di~Noia, and Ferrara}{Anelli
  et~al\mbox{.}}{2019}]%
        {Anelli2019}
\bibfield{author}{\bibinfo{person}{Vito~Walter Anelli}, \bibinfo{person}{Yashar
  Deldjoo}, \bibinfo{person}{Tommaso Di~Noia}, {and} \bibinfo{person}{Antonio
  Ferrara}.} \bibinfo{year}{2019}\natexlab{}.
\newblock \showarticletitle{Towards Effective Device-Aware Federated Learning}.
  \bibinfo{publisher}{Springer International Publishing},
  \bibinfo{address}{Cham}, \bibinfo{pages}{477--491}.
\newblock


\bibitem[\protect\citeauthoryear{{Anh}, {Luong}, {Niyato}, {Kim}, and
  {Wang}}{{Anh} et~al\mbox{.}}{2019}]%
        {8716527}
\bibfield{author}{\bibinfo{person}{T.~T. {Anh}}, \bibinfo{person}{N.~C.
  {Luong}}, \bibinfo{person}{D. {Niyato}}, \bibinfo{person}{D.~I. {Kim}}, {and}
  \bibinfo{person}{L. {Wang}}.} \bibinfo{year}{2019}\natexlab{}.
\newblock \showarticletitle{Efficient Training Management for Mobile
  Crowd-Machine Learning: A Deep Reinforcement Learning Approach}.
\newblock \bibinfo{journal}{\emph{IEEE Wireless Communications Letters}}
  \bibinfo{volume}{8}, \bibinfo{number}{5} (\bibinfo{date}{Oct}
  \bibinfo{year}{2019}), \bibinfo{pages}{1345--1348}.
\newblock
\showISSN{2162-2345}


\bibitem[\protect\citeauthoryear{Augenstein, McMahan, Ramage, Ramaswamy,
  Kairouz, Chen, Mathews, and y~Arcas}{Augenstein et~al\mbox{.}}{2019}]%
        {augenstein2019generative}
\bibfield{author}{\bibinfo{person}{Sean Augenstein},
  \bibinfo{person}{H.~Brendan McMahan}, \bibinfo{person}{Daniel Ramage},
  \bibinfo{person}{Swaroop Ramaswamy}, \bibinfo{person}{Peter Kairouz},
  \bibinfo{person}{Mingqing Chen}, \bibinfo{person}{Rajiv Mathews}, {and}
  \bibinfo{person}{Blaise~Aguera y Arcas}.} \bibinfo{year}{2019}\natexlab{}.
\newblock \bibinfo{title}{Generative Models for Effective ML on Private,
  Decentralized Datasets}.
\newblock
\newblock
\showeprint[arxiv]{1911.06679}~[cs.LG]


\bibitem[\protect\citeauthoryear{Awan, Li, Luo, and Liu}{Awan
  et~al\mbox{.}}{2019}]%
        {Awan2019}
\bibfield{author}{\bibinfo{person}{Sana Awan}, \bibinfo{person}{Fengjun Li},
  \bibinfo{person}{Bo Luo}, {and} \bibinfo{person}{Mei Liu}.}
  \bibinfo{year}{2019}\natexlab{}.
\newblock \showarticletitle{Poster: A Reliable and Accountable
  Privacy-Preserving Federated Learning Framework using the Blockchain}. In
  \bibinfo{booktitle}{\emph{Proceedings of the 2019 ACM SIGSAC Conference on
  Computer and Communications Security}}. \bibinfo{publisher}{Association for
  Computing Machinery}, \bibinfo{address}{London, United Kingdom},
  \bibinfo{pages}{2561–2563}.
\newblock


\bibitem[\protect\citeauthoryear{{Aïvodji}, {Gambs}, and {Martin}}{{Aïvodji}
  et~al\mbox{.}}{2019}]%
        {8844592}
\bibfield{author}{\bibinfo{person}{U.~M. {Aïvodji}}, \bibinfo{person}{S.
  {Gambs}}, {and} \bibinfo{person}{A. {Martin}}.}
  \bibinfo{year}{2019}\natexlab{}.
\newblock \showarticletitle{IOTFLA : A Secured and Privacy-Preserving Smart
  Home Architecture Implementing Federated Learning}. In
  \bibinfo{booktitle}{\emph{2019 IEEE Security and Privacy Workshops (SPW)}}.
  \bibinfo{pages}{175--180}.
\newblock


\bibitem[\protect\citeauthoryear{Bakopoulou, Tillman, and
  Markopoulou}{Bakopoulou et~al\mbox{.}}{2019}]%
        {bakopoulou2019federated}
\bibfield{author}{\bibinfo{person}{Evita Bakopoulou}, \bibinfo{person}{Balint
  Tillman}, {and} \bibinfo{person}{Athina Markopoulou}.}
  \bibinfo{year}{2019}\natexlab{}.
\newblock \showarticletitle{A federated learning approach for mobile packet
  classification}.
\newblock \bibinfo{journal}{\emph{arXiv preprint arXiv:1907.13113}}
  (\bibinfo{year}{2019}).
\newblock


\bibitem[\protect\citeauthoryear{{Bao}, {Su}, {Xiong}, {Huang}, and {Hu}}{{Bao}
  et~al\mbox{.}}{2019}]%
        {8905038}
\bibfield{author}{\bibinfo{person}{X. {Bao}}, \bibinfo{person}{C. {Su}},
  \bibinfo{person}{Y. {Xiong}}, \bibinfo{person}{W. {Huang}}, {and}
  \bibinfo{person}{Y. {Hu}}.} \bibinfo{year}{2019}\natexlab{}.
\newblock \showarticletitle{FLChain: A Blockchain for Auditable Federated
  Learning with Trust and Incentive}. In \bibinfo{booktitle}{\emph{2019 5th
  International Conference on Big Data Computing and Communications (BIGCOM)}}.
  \bibinfo{pages}{151--159}.
\newblock
\showISSN{null}


\bibitem[\protect\citeauthoryear{Benditkis, Keren, Mor-Yosef, Avidor, Shoham,
  and Tal-Israel}{Benditkis et~al\mbox{.}}{2019}]%
        {Benditkis2019}
\bibfield{author}{\bibinfo{person}{Daniel Benditkis}, \bibinfo{person}{Aviv
  Keren}, \bibinfo{person}{Liron Mor-Yosef}, \bibinfo{person}{Tomer Avidor},
  \bibinfo{person}{Neta Shoham}, {and} \bibinfo{person}{Nadav Tal-Israel}.}
  \bibinfo{year}{2019}\natexlab{}.
\newblock \showarticletitle{Distributed deep neural network training on edge
  devices}. In \bibinfo{booktitle}{\emph{Proceedings of the 4th ACM/IEEE
  Symposium on Edge Computing}}. \bibinfo{publisher}{Association for Computing
  Machinery}, \bibinfo{address}{Arlington, Virginia},
  \bibinfo{pages}{304–306}.
\newblock


\bibitem[\protect\citeauthoryear{Bhowmick, Duchi, Freudiger, Kapoor, and
  Rogers}{Bhowmick et~al\mbox{.}}{2018}]%
        {bhowmick2018protection}
\bibfield{author}{\bibinfo{person}{Abhishek Bhowmick}, \bibinfo{person}{John
  Duchi}, \bibinfo{person}{Julien Freudiger}, \bibinfo{person}{Gaurav Kapoor},
  {and} \bibinfo{person}{Ryan Rogers}.} \bibinfo{year}{2018}\natexlab{}.
\newblock \showarticletitle{Protection against reconstruction and its
  applications in private federated learning}.
\newblock \bibinfo{journal}{\emph{arXiv preprint arXiv:1812.00984}}
  (\bibinfo{year}{2018}).
\newblock


\bibitem[\protect\citeauthoryear{Bonawitz, Eichner, Grieskamp, Huba, Ingerman,
  Ivanov, Kiddon, Konecny, Mazzocchi, McMahan, et~al\mbox{.}}{Bonawitz
  et~al\mbox{.}}{2019a}]%
        {bonawitz2019towards}
\bibfield{author}{\bibinfo{person}{Keith Bonawitz}, \bibinfo{person}{Hubert
  Eichner}, \bibinfo{person}{Wolfgang Grieskamp}, \bibinfo{person}{Dzmitry
  Huba}, \bibinfo{person}{Alex Ingerman}, \bibinfo{person}{Vladimir Ivanov},
  \bibinfo{person}{Chloe Kiddon}, \bibinfo{person}{Jakub Konecny},
  \bibinfo{person}{Stefano Mazzocchi}, \bibinfo{person}{H~Brendan McMahan},
  {et~al\mbox{.}}} \bibinfo{year}{2019}\natexlab{a}.
\newblock \showarticletitle{Towards federated learning at scale: System
  design}.
\newblock \bibinfo{journal}{\emph{arXiv preprint arXiv:1902.01046}}
  (\bibinfo{year}{2019}).
\newblock


\bibitem[\protect\citeauthoryear{Bonawitz, Ivanov, Kreuter, Marcedone, McMahan,
  Patel, Ramage, Segal, and Seth}{Bonawitz et~al\mbox{.}}{2016}]%
        {bonawitz2016practical}
\bibfield{author}{\bibinfo{person}{Keith Bonawitz}, \bibinfo{person}{Vladimir
  Ivanov}, \bibinfo{person}{Ben Kreuter}, \bibinfo{person}{Antonio Marcedone},
  \bibinfo{person}{H~Brendan McMahan}, \bibinfo{person}{Sarvar Patel},
  \bibinfo{person}{Daniel Ramage}, \bibinfo{person}{Aaron Segal}, {and}
  \bibinfo{person}{Karn Seth}.} \bibinfo{year}{2016}\natexlab{}.
\newblock \showarticletitle{Practical secure aggregation for federated learning
  on user-held data}.
\newblock \bibinfo{journal}{\emph{arXiv preprint arXiv:1611.04482}}
  (\bibinfo{year}{2016}).
\newblock


\bibitem[\protect\citeauthoryear{Bonawitz, Ivanov, Kreuter, Marcedone, McMahan,
  Patel, Ramage, Segal, and Seth}{Bonawitz et~al\mbox{.}}{2017}]%
        {Bonawitz2017}
\bibfield{author}{\bibinfo{person}{Keith Bonawitz}, \bibinfo{person}{Vladimir
  Ivanov}, \bibinfo{person}{Ben Kreuter}, \bibinfo{person}{Antonio Marcedone},
  \bibinfo{person}{H.~Brendan McMahan}, \bibinfo{person}{Sarvar Patel},
  \bibinfo{person}{Daniel Ramage}, \bibinfo{person}{Aaron Segal}, {and}
  \bibinfo{person}{Karn Seth}.} \bibinfo{year}{2017}\natexlab{}.
\newblock \showarticletitle{Practical Secure Aggregation for Privacy-Preserving
  Machine Learning}. In \bibinfo{booktitle}{\emph{Proceedings of the 2017 ACM
  SIGSAC Conference on Computer and Communications Security}}.
  \bibinfo{publisher}{Association for Computing Machinery},
  \bibinfo{address}{Dallas, Texas, USA}, \bibinfo{pages}{1175–1191}.
\newblock


\bibitem[\protect\citeauthoryear{Bonawitz, Salehi, Kone{\v{c}}n{\`y}, McMahan,
  and Gruteser}{Bonawitz et~al\mbox{.}}{2019b}]%
        {bonawitz2019federated}
\bibfield{author}{\bibinfo{person}{Keith Bonawitz}, \bibinfo{person}{Fariborz
  Salehi}, \bibinfo{person}{Jakub Kone{\v{c}}n{\`y}}, \bibinfo{person}{Brendan
  McMahan}, {and} \bibinfo{person}{Marco Gruteser}.}
  \bibinfo{year}{2019}\natexlab{b}.
\newblock \showarticletitle{Federated learning with autotuned
  communication-efficient secure aggregation}.
\newblock \bibinfo{journal}{\emph{arXiv preprint arXiv:1912.00131}}
  (\bibinfo{year}{2019}).
\newblock


\bibitem[\protect\citeauthoryear{Caldas, Duddu, Wu, Li, Konečný, McMahan,
  Smith, and Talwalkar}{Caldas et~al\mbox{.}}{2018a}]%
        {caldas2018leaf}
\bibfield{author}{\bibinfo{person}{Sebastian Caldas}, \bibinfo{person}{Sai
  Meher~Karthik Duddu}, \bibinfo{person}{Peter Wu}, \bibinfo{person}{Tian Li},
  \bibinfo{person}{Jakub Konečný}, \bibinfo{person}{H.~Brendan McMahan},
  \bibinfo{person}{Virginia Smith}, {and} \bibinfo{person}{Ameet Talwalkar}.}
  \bibinfo{year}{2018}\natexlab{a}.
\newblock \bibinfo{title}{LEAF: A Benchmark for Federated Settings}.
\newblock
\newblock
\showeprint[arxiv]{1812.01097}~[cs.LG]


\bibitem[\protect\citeauthoryear{Caldas, Kone{\v{c}}ny, McMahan, and
  Talwalkar}{Caldas et~al\mbox{.}}{2018b}]%
        {caldas2018expanding}
\bibfield{author}{\bibinfo{person}{Sebastian Caldas}, \bibinfo{person}{Jakub
  Kone{\v{c}}ny}, \bibinfo{person}{H~Brendan McMahan}, {and}
  \bibinfo{person}{Ameet Talwalkar}.} \bibinfo{year}{2018}\natexlab{b}.
\newblock \showarticletitle{Expanding the Reach of Federated Learning by
  Reducing Client Resource Requirements}.
\newblock \bibinfo{journal}{\emph{arXiv preprint arXiv:1812.07210}}
  (\bibinfo{year}{2018}).
\newblock


\bibitem[\protect\citeauthoryear{Caldas, Smith, and Talwalkar}{Caldas
  et~al\mbox{.}}{2018c}]%
        {caldas2018federated}
\bibfield{author}{\bibinfo{person}{Sebastian Caldas}, \bibinfo{person}{Virginia
  Smith}, {and} \bibinfo{person}{Ameet Talwalkar}.}
  \bibinfo{year}{2018}\natexlab{c}.
\newblock \showarticletitle{Federated Kernelized Multi-Task Learning}. In
  \bibinfo{booktitle}{\emph{SysML Conference 2018}}.
\newblock


\bibitem[\protect\citeauthoryear{Chai, Wang, Chen, and Yang}{Chai
  et~al\mbox{.}}{2019}]%
        {chai2019secure}
\bibfield{author}{\bibinfo{person}{Di Chai}, \bibinfo{person}{Leye Wang},
  \bibinfo{person}{Kai Chen}, {and} \bibinfo{person}{Qiang Yang}.}
  \bibinfo{year}{2019}\natexlab{}.
\newblock \bibinfo{title}{Secure Federated Matrix Factorization}.
\newblock
\newblock
\showeprint[arxiv]{1906.05108}~[cs.CR]


\bibitem[\protect\citeauthoryear{Chen, Luo, Dong, Li, and He}{Chen
  et~al\mbox{.}}{2018a}]%
        {chen2018federated}
\bibfield{author}{\bibinfo{person}{Fei Chen}, \bibinfo{person}{Mi Luo},
  \bibinfo{person}{Zhenhua Dong}, \bibinfo{person}{Zhenguo Li}, {and}
  \bibinfo{person}{Xiuqiang He}.} \bibinfo{year}{2018}\natexlab{a}.
\newblock \bibinfo{title}{Federated Meta-Learning with Fast Convergence and
  Efficient Communication}.
\newblock
\newblock
\showeprint[arxiv]{1802.07876}~[cs.LG]


\bibitem[\protect\citeauthoryear{{Chen}, {Semiari}, {Saad}, {Liu}, and
  {Yin}}{{Chen} et~al\mbox{.}}{2020}]%
        {8851408}
\bibfield{author}{\bibinfo{person}{M. {Chen}}, \bibinfo{person}{O. {Semiari}},
  \bibinfo{person}{W. {Saad}}, \bibinfo{person}{X. {Liu}}, {and}
  \bibinfo{person}{C. {Yin}}.} \bibinfo{year}{2020}\natexlab{}.
\newblock \showarticletitle{Federated Echo State Learning for Minimizing Breaks
  in Presence in Wireless Virtual Reality Networks}.
\newblock \bibinfo{journal}{\emph{IEEE Transactions on Wireless
  Communications}} \bibinfo{volume}{19}, \bibinfo{number}{1}
  (\bibinfo{date}{Jan} \bibinfo{year}{2020}), \bibinfo{pages}{177--191}.
\newblock
\showISSN{1558-2248}


\bibitem[\protect\citeauthoryear{Chen, Yang, Saad, Yin, Poor, and Cui}{Chen
  et~al\mbox{.}}{2019}]%
        {chen2019joint}
\bibfield{author}{\bibinfo{person}{Mingzhe Chen}, \bibinfo{person}{Zhaohui
  Yang}, \bibinfo{person}{Walid Saad}, \bibinfo{person}{Changchuan Yin},
  \bibinfo{person}{H~Vincent Poor}, {and} \bibinfo{person}{Shuguang Cui}.}
  \bibinfo{year}{2019}\natexlab{}.
\newblock \showarticletitle{A joint learning and communications framework for
  federated learning over wireless networks}.
\newblock \bibinfo{journal}{\emph{arXiv preprint arXiv:1909.07972}}
  (\bibinfo{year}{2019}).
\newblock


\bibitem[\protect\citeauthoryear{Chen, Chen, Sun, Wu, and Hong}{Chen
  et~al\mbox{.}}{2019a}]%
        {chen2019distributed}
\bibfield{author}{\bibinfo{person}{Xiangyi Chen}, \bibinfo{person}{Tiancong
  Chen}, \bibinfo{person}{Haoran Sun}, \bibinfo{person}{Zhiwei~Steven Wu},
  {and} \bibinfo{person}{Mingyi Hong}.} \bibinfo{year}{2019}\natexlab{a}.
\newblock \bibinfo{title}{Distributed Training with Heterogeneous Data:
  Bridging Median- and Mean-Based Algorithms}.
\newblock
\newblock
\showeprint[arxiv]{1906.01736}~[cs.LG]


\bibitem[\protect\citeauthoryear{Chen, Ning, and Rangwala}{Chen
  et~al\mbox{.}}{2019b}]%
        {chen2019asynchronous}
\bibfield{author}{\bibinfo{person}{Yujing Chen}, \bibinfo{person}{Yue Ning},
  {and} \bibinfo{person}{Huzefa Rangwala}.} \bibinfo{year}{2019}\natexlab{b}.
\newblock \showarticletitle{Asynchronous Online Federated Learning for Edge
  Devices}.
\newblock \bibinfo{journal}{\emph{arXiv preprint arXiv:1911.02134}}
  (\bibinfo{year}{2019}).
\newblock


\bibitem[\protect\citeauthoryear{Chen, Su, and Xu}{Chen et~al\mbox{.}}{2018b}]%
        {Chen2018}
\bibfield{author}{\bibinfo{person}{Yudong Chen}, \bibinfo{person}{Lili Su},
  {and} \bibinfo{person}{Jiaming Xu}.} \bibinfo{year}{2018}\natexlab{b}.
\newblock \showarticletitle{Distributed Statistical Machine Learning in
  Adversarial Settings: Byzantine Gradient Descent}.
\newblock \bibinfo{journal}{\emph{SIGMETRICS Perform. Eval. Rev.}}
  \bibinfo{volume}{46}, \bibinfo{number}{1} (\bibinfo{year}{2018}),
  \bibinfo{pages}{96}.
\newblock
\urldef\tempurl%
\url{https://doi.org/10.1145/3292040.3219655}
\showURL{%
\tempurl}


\bibitem[\protect\citeauthoryear{Chen, Sun, and Jin}{Chen
  et~al\mbox{.}}{2019c}]%
        {chen2019communicationefficient}
\bibfield{author}{\bibinfo{person}{Yang Chen}, \bibinfo{person}{Xiaoyan Sun},
  {and} \bibinfo{person}{Yaochu Jin}.} \bibinfo{year}{2019}\natexlab{c}.
\newblock \bibinfo{title}{Communication-Efficient Federated Deep Learning with
  Asynchronous Model Update and Temporally Weighted Aggregation}.
\newblock
\newblock
\showeprint[arxiv]{1903.07424}~[cs.LG]


\bibitem[\protect\citeauthoryear{{Chen}, {Sun}, and {Jin}}{{Chen}
  et~al\mbox{.}}{2019}]%
        {8945292}
\bibfield{author}{\bibinfo{person}{Y. {Chen}}, \bibinfo{person}{X. {Sun}},
  {and} \bibinfo{person}{Y. {Jin}}.} \bibinfo{year}{2019}\natexlab{}.
\newblock \showarticletitle{Communication-Efficient Federated Deep Learning
  With Layerwise Asynchronous Model Update and Temporally Weighted
  Aggregation}.
\newblock \bibinfo{journal}{\emph{IEEE Transactions on Neural Networks and
  Learning Systems}} (\bibinfo{year}{2019}), \bibinfo{pages}{1--10}.
\newblock
\showISSN{2162-2388}


\bibitem[\protect\citeauthoryear{Cheng, Fan, Jin, Liu, Chen, and Yang}{Cheng
  et~al\mbox{.}}{2019}]%
        {cheng2019secureboost}
\bibfield{author}{\bibinfo{person}{Kewei Cheng}, \bibinfo{person}{Tao Fan},
  \bibinfo{person}{Yilun Jin}, \bibinfo{person}{Yang Liu},
  \bibinfo{person}{Tianjian Chen}, {and} \bibinfo{person}{Qiang Yang}.}
  \bibinfo{year}{2019}\natexlab{}.
\newblock \showarticletitle{SecureBoost: A Lossless Federated Learning
  Framework}.
\newblock \bibinfo{journal}{\emph{arXiv preprint arXiv:1901.08755}}
  (\bibinfo{year}{2019}).
\newblock


\bibitem[\protect\citeauthoryear{{Choi} and {Pokhrel}}{{Choi} and
  {Pokhrel}}{2019}]%
        {8935424}
\bibfield{author}{\bibinfo{person}{J. {Choi}} {and} \bibinfo{person}{S.~R.
  {Pokhrel}}.} \bibinfo{year}{2019}\natexlab{}.
\newblock \showarticletitle{Federated Learning with Multichannel ALOHA}.
\newblock \bibinfo{journal}{\emph{IEEE Wireless Communications Letters}}
  (\bibinfo{year}{2019}), \bibinfo{pages}{1--1}.
\newblock
\showISSN{2162-2345}


\bibitem[\protect\citeauthoryear{Choudhury, Gkoulalas-Divanis, Salonidis,
  Sylla, Park, Hsu, and Das}{Choudhury et~al\mbox{.}}{2019}]%
        {choudhury2019differential}
\bibfield{author}{\bibinfo{person}{Olivia Choudhury}, \bibinfo{person}{Aris
  Gkoulalas-Divanis}, \bibinfo{person}{Theodoros Salonidis},
  \bibinfo{person}{Issa Sylla}, \bibinfo{person}{Yoonyoung Park},
  \bibinfo{person}{Grace Hsu}, {and} \bibinfo{person}{Amar Das}.}
  \bibinfo{year}{2019}\natexlab{}.
\newblock \bibinfo{title}{Differential Privacy-enabled Federated Learning for
  Sensitive Health Data}.
\newblock
\newblock
\showeprint[arxiv]{1910.02578}~[cs.LG]


\bibitem[\protect\citeauthoryear{Corinzia and Buhmann}{Corinzia and
  Buhmann}{2019}]%
        {corinzia2019variational}
\bibfield{author}{\bibinfo{person}{Luca Corinzia} {and}
  \bibinfo{person}{Joachim~M Buhmann}.} \bibinfo{year}{2019}\natexlab{}.
\newblock \showarticletitle{Variational Federated Multi-Task Learning}.
\newblock \bibinfo{journal}{\emph{arXiv preprint arXiv:1906.06268}}
  (\bibinfo{year}{2019}).
\newblock


\bibitem[\protect\citeauthoryear{Daga, Nicholson, Gavrilovska, and
  Lugones}{Daga et~al\mbox{.}}{2019}]%
        {Daga2019}
\bibfield{author}{\bibinfo{person}{Harshit Daga}, \bibinfo{person}{Patrick~K.
  Nicholson}, \bibinfo{person}{Ada Gavrilovska}, {and} \bibinfo{person}{Diego
  Lugones}.} \bibinfo{year}{2019}\natexlab{}.
\newblock \showarticletitle{Cartel: A System for Collaborative Transfer
  Learning at the Edge}. In \bibinfo{booktitle}{\emph{Proceedings of the ACM
  Symposium on Cloud Computing}}. \bibinfo{publisher}{Association for Computing
  Machinery}, \bibinfo{address}{Santa Cruz, CA, USA}, \bibinfo{pages}{25–37}.
\newblock


\bibitem[\protect\citeauthoryear{{Deng}, {Chen}, {Zhang}, {Gong}, and
  {Zhu}}{{Deng} et~al\mbox{.}}{2019}]%
        {8928018}
\bibfield{author}{\bibinfo{person}{K. {Deng}}, \bibinfo{person}{Z. {Chen}},
  \bibinfo{person}{S. {Zhang}}, \bibinfo{person}{C. {Gong}}, {and}
  \bibinfo{person}{J. {Zhu}}.} \bibinfo{year}{2019}\natexlab{}.
\newblock \showarticletitle{Content Compression Coding for Federated Learning}.
  In \bibinfo{booktitle}{\emph{2019 11th International Conference on Wireless
  Communications and Signal Processing (WCSP)}}. \bibinfo{pages}{1--6}.
\newblock
\showISSN{2325-3746}


\bibitem[\protect\citeauthoryear{Dinh, Tran, Nguyen, Hong, Bao, Zomaya, and
  Gramoli}{Dinh et~al\mbox{.}}{2019}]%
        {dinh2019federated}
\bibfield{author}{\bibinfo{person}{Canh Dinh}, \bibinfo{person}{Nguyen~H Tran},
  \bibinfo{person}{Minh~NH Nguyen}, \bibinfo{person}{Choong~Seon Hong},
  \bibinfo{person}{Wei Bao}, \bibinfo{person}{Albert Zomaya}, {and}
  \bibinfo{person}{Vincent Gramoli}.} \bibinfo{year}{2019}\natexlab{}.
\newblock \showarticletitle{Federated Learning over Wireless Networks:
  Convergence Analysis and Resource Allocation}.
\newblock \bibinfo{journal}{\emph{arXiv preprint arXiv:1910.13067}}
  (\bibinfo{year}{2019}).
\newblock


\bibitem[\protect\citeauthoryear{{Doku}, {Rawat}, and {Liu}}{{Doku}
  et~al\mbox{.}}{2019}]%
        {8843451}
\bibfield{author}{\bibinfo{person}{R. {Doku}}, \bibinfo{person}{D.~B. {Rawat}},
  {and} \bibinfo{person}{C. {Liu}}.} \bibinfo{year}{2019}\natexlab{}.
\newblock \showarticletitle{Towards Federated Learning Approach to Determine
  Data Relevance in Big Data}. In \bibinfo{booktitle}{\emph{2019 IEEE 20th
  International Conference on Information Reuse and Integration for Data
  Science (IRI)}}. \bibinfo{pages}{184--192}.
\newblock


\bibitem[\protect\citeauthoryear{Du, Zeng, Yan, and Zhang}{Du
  et~al\mbox{.}}{2018}]%
        {du2018efficient}
\bibfield{author}{\bibinfo{person}{Wei Du}, \bibinfo{person}{Xiao Zeng},
  \bibinfo{person}{Ming Yan}, {and} \bibinfo{person}{Mi Zhang}.}
  \bibinfo{year}{2018}\natexlab{}.
\newblock \showarticletitle{Efficient Federated Learning via Variational
  Dropout}.
\newblock  (\bibinfo{year}{2018}).
\newblock


\bibitem[\protect\citeauthoryear{Duan}{Duan}{2019}]%
        {duan2019astraea}
\bibfield{author}{\bibinfo{person}{Moming Duan}.}
  \bibinfo{year}{2019}\natexlab{}.
\newblock \showarticletitle{Astraea: Self-balancing federated learning for
  improving classification accuracy of mobile deep learning applications}.
\newblock \bibinfo{journal}{\emph{arXiv preprint arXiv:1907.01132}}
  (\bibinfo{year}{2019}).
\newblock


\bibitem[\protect\citeauthoryear{{Duan}, {Zhang}, {Wang}, {Li}, and
  {Zhang}}{{Duan} et~al\mbox{.}}{2019}]%
        {8854245}
\bibfield{author}{\bibinfo{person}{S. {Duan}}, \bibinfo{person}{D. {Zhang}},
  \bibinfo{person}{Y. {Wang}}, \bibinfo{person}{L. {Li}}, {and}
  \bibinfo{person}{Y. {Zhang}}.} \bibinfo{year}{2019}\natexlab{}.
\newblock \showarticletitle{JointRec: A Deep Learning-based Joint Cloud Video
  Recommendation Framework for Mobile IoTs}.
\newblock \bibinfo{journal}{\emph{IEEE Internet of Things Journal}}
  (\bibinfo{year}{2019}), \bibinfo{pages}{1--1}.
\newblock
\showISSN{2372-2541}


\bibitem[\protect\citeauthoryear{Fadaeddini, Majidi, and Eshghi}{Fadaeddini
  et~al\mbox{.}}{2019}]%
        {Fadaeddini2019}
\bibfield{author}{\bibinfo{person}{Amin Fadaeddini}, \bibinfo{person}{Babak
  Majidi}, {and} \bibinfo{person}{Mohammad Eshghi}.}
  \bibinfo{year}{2019}\natexlab{}.
\newblock \showarticletitle{Privacy Preserved Decentralized Deep Learning: A
  Blockchain Based Solution for Secure AI-Driven Enterprise}.
  \bibinfo{publisher}{Springer International Publishing},
  \bibinfo{address}{Cham}, \bibinfo{pages}{32--40}.
\newblock


\bibitem[\protect\citeauthoryear{Fan, Song, Jiang, Chen, and Shibasaki}{Fan
  et~al\mbox{.}}{2019}]%
        {Fan2019}
\bibfield{author}{\bibinfo{person}{Zipei Fan}, \bibinfo{person}{Xuan Song},
  \bibinfo{person}{Renhe Jiang}, \bibinfo{person}{Quanjun Chen}, {and}
  \bibinfo{person}{Ryosuke Shibasaki}.} \bibinfo{year}{2019}\natexlab{}.
\newblock \showarticletitle{Decentralized Attention-based Personalized Human
  Mobility Prediction}.
\newblock \bibinfo{journal}{\emph{Proc. ACM Interact. Mob. Wearable Ubiquitous
  Technol.}} \bibinfo{volume}{3}, \bibinfo{number}{4} (\bibinfo{year}{2019}),
  \bibinfo{pages}{Article 133}.
\newblock


\bibitem[\protect\citeauthoryear{{Feng}, {Niyato}, {Wang}, {Kim}, and
  {Liang}}{{Feng} et~al\mbox{.}}{2019}]%
        {8875430}
\bibfield{author}{\bibinfo{person}{S. {Feng}}, \bibinfo{person}{D. {Niyato}},
  \bibinfo{person}{P. {Wang}}, \bibinfo{person}{D.~I. {Kim}}, {and}
  \bibinfo{person}{Y. {Liang}}.} \bibinfo{year}{2019}\natexlab{}.
\newblock \showarticletitle{Joint Service Pricing and Cooperative Relay
  Communication for Federated Learning}. In \bibinfo{booktitle}{\emph{2019
  International Conference on Internet of Things (iThings) and IEEE Green
  Computing and Communications (GreenCom) and IEEE Cyber, Physical and Social
  Computing (CPSCom) and IEEE Smart Data (SmartData)}}.
  \bibinfo{pages}{815--820}.
\newblock


\bibitem[\protect\citeauthoryear{Fung, Koerner, Grant, and Beschastnikh}{Fung
  et~al\mbox{.}}{2018a}]%
        {fung2018dancing}
\bibfield{author}{\bibinfo{person}{Clement Fung}, \bibinfo{person}{Jamie
  Koerner}, \bibinfo{person}{Stewart Grant}, {and} \bibinfo{person}{Ivan
  Beschastnikh}.} \bibinfo{year}{2018}\natexlab{a}.
\newblock \bibinfo{title}{Dancing in the Dark: Private Multi-Party Machine
  Learning in an Untrusted Setting}.
\newblock
\newblock
\showeprint[arxiv]{1811.09712}~[cs.CR]


\bibitem[\protect\citeauthoryear{Fung, Yoon, and Beschastnikh}{Fung
  et~al\mbox{.}}{2018b}]%
        {fung2018mitigating}
\bibfield{author}{\bibinfo{person}{Clement Fung}, \bibinfo{person}{Chris~JM
  Yoon}, {and} \bibinfo{person}{Ivan Beschastnikh}.}
  \bibinfo{year}{2018}\natexlab{b}.
\newblock \showarticletitle{Mitigating sybils in federated learning poisoning}.
\newblock \bibinfo{journal}{\emph{arXiv preprint arXiv:1808.04866}}
  (\bibinfo{year}{2018}).
\newblock


\bibitem[\protect\citeauthoryear{Geyer, Klein, and Nabi}{Geyer
  et~al\mbox{.}}{2017}]%
        {geyer2017differentially}
\bibfield{author}{\bibinfo{person}{Robin~C Geyer}, \bibinfo{person}{Tassilo
  Klein}, {and} \bibinfo{person}{Moin Nabi}.} \bibinfo{year}{2017}\natexlab{}.
\newblock \showarticletitle{Differentially private federated learning: A client
  level perspective}.
\newblock \bibinfo{journal}{\emph{arXiv preprint arXiv:1712.07557}}
  (\bibinfo{year}{2017}).
\newblock


\bibitem[\protect\citeauthoryear{Ghazi, Pagh, and Velingker}{Ghazi
  et~al\mbox{.}}{2019}]%
        {ghazi2019scalable}
\bibfield{author}{\bibinfo{person}{Badih Ghazi}, \bibinfo{person}{Rasmus Pagh},
  {and} \bibinfo{person}{Ameya Velingker}.} \bibinfo{year}{2019}\natexlab{}.
\newblock \bibinfo{title}{Scalable and Differentially Private Distributed
  Aggregation in the Shuffled Model}.
\newblock
\newblock
\showeprint[arxiv]{1906.08320}~[cs.LG]


\bibitem[\protect\citeauthoryear{Ghosh, Hong, Yin, and Ramchandran}{Ghosh
  et~al\mbox{.}}{2019}]%
        {ghosh2019robust}
\bibfield{author}{\bibinfo{person}{Avishek Ghosh}, \bibinfo{person}{Justin
  Hong}, \bibinfo{person}{Dong Yin}, {and} \bibinfo{person}{Kannan
  Ramchandran}.} \bibinfo{year}{2019}\natexlab{}.
\newblock \showarticletitle{Robust Federated Learning in a Heterogeneous
  Environment}.
\newblock \bibinfo{journal}{\emph{arXiv preprint arXiv:1906.06629}}
  (\bibinfo{year}{2019}).
\newblock


\bibitem[\protect\citeauthoryear{Guha, Talwlkar, and Smith}{Guha
  et~al\mbox{.}}{2019}]%
        {guha2019one}
\bibfield{author}{\bibinfo{person}{Neel Guha}, \bibinfo{person}{Ameet
  Talwlkar}, {and} \bibinfo{person}{Virginia Smith}.}
  \bibinfo{year}{2019}\natexlab{}.
\newblock \showarticletitle{One-Shot Federated Learning}.
\newblock \bibinfo{journal}{\emph{arXiv preprint arXiv:1902.11175}}
  (\bibinfo{year}{2019}).
\newblock


\bibitem[\protect\citeauthoryear{Han, Wang, and Leung}{Han
  et~al\mbox{.}}{2020}]%
        {han2020adaptive}
\bibfield{author}{\bibinfo{person}{Pengchao Han}, \bibinfo{person}{Shiqiang
  Wang}, {and} \bibinfo{person}{Kin~K. Leung}.}
  \bibinfo{year}{2020}\natexlab{}.
\newblock \bibinfo{title}{Adaptive Gradient Sparsification for Efficient
  Federated Learning: An Online Learning Approach}.
\newblock
\newblock
\showeprint[arxiv]{2001.04756}~[cs.LG]


\bibitem[\protect\citeauthoryear{Han and Zhang}{Han and Zhang}{2019}]%
        {han2019robust}
\bibfield{author}{\bibinfo{person}{Yufei Han} {and} \bibinfo{person}{Xiangliang
  Zhang}.} \bibinfo{year}{2019}\natexlab{}.
\newblock \showarticletitle{Robust Federated Training via Collaborative Machine
  Teaching using Trusted Instances}.
\newblock \bibinfo{journal}{\emph{arXiv preprint arXiv:1905.02941}}
  (\bibinfo{year}{2019}).
\newblock


\bibitem[\protect\citeauthoryear{{Hao}, {Li}, {Luo}, {Xu}, {Yang}, and
  {Liu}}{{Hao} et~al\mbox{.}}{2019a}]%
        {8859260}
\bibfield{author}{\bibinfo{person}{M. {Hao}}, \bibinfo{person}{H. {Li}},
  \bibinfo{person}{X. {Luo}}, \bibinfo{person}{G. {Xu}}, \bibinfo{person}{H.
  {Yang}}, {and} \bibinfo{person}{S. {Liu}}.} \bibinfo{year}{2019}\natexlab{a}.
\newblock \showarticletitle{Efficient and Privacy-enhanced Federated Learning
  for Industrial Artificial Intelligence}.
\newblock \bibinfo{journal}{\emph{IEEE Transactions on Industrial Informatics}}
  (\bibinfo{year}{2019}), \bibinfo{pages}{1--1}.
\newblock
\showISSN{1941-0050}


\bibitem[\protect\citeauthoryear{{Hao}, {Li}, {Xu}, {Liu}, and {Yang}}{{Hao}
  et~al\mbox{.}}{2019b}]%
        {8761267}
\bibfield{author}{\bibinfo{person}{M. {Hao}}, \bibinfo{person}{H. {Li}},
  \bibinfo{person}{G. {Xu}}, \bibinfo{person}{S. {Liu}}, {and}
  \bibinfo{person}{H. {Yang}}.} \bibinfo{year}{2019}\natexlab{b}.
\newblock \showarticletitle{Towards Efficient and Privacy-Preserving Federated
  Deep Learning}. In \bibinfo{booktitle}{\emph{ICC 2019 - 2019 IEEE
  International Conference on Communications (ICC)}}. \bibinfo{pages}{1--6}.
\newblock
\showISSN{1550-3607}


\bibitem[\protect\citeauthoryear{Hard, Rao, Mathews, Ramaswamy, Beaufays,
  Augenstein, Eichner, Kiddon, and Ramage}{Hard et~al\mbox{.}}{2018}]%
        {hard2018federated}
\bibfield{author}{\bibinfo{person}{Andrew Hard}, \bibinfo{person}{Kanishka
  Rao}, \bibinfo{person}{Rajiv Mathews}, \bibinfo{person}{Swaroop Ramaswamy},
  \bibinfo{person}{Fran{\c{c}}oise Beaufays}, \bibinfo{person}{Sean
  Augenstein}, \bibinfo{person}{Hubert Eichner}, \bibinfo{person}{Chlo{\'e}
  Kiddon}, {and} \bibinfo{person}{Daniel Ramage}.}
  \bibinfo{year}{2018}\natexlab{}.
\newblock \showarticletitle{Federated learning for mobile keyboard prediction}.
\newblock \bibinfo{journal}{\emph{arXiv preprint arXiv:1811.03604}}
  (\bibinfo{year}{2018}).
\newblock


\bibitem[\protect\citeauthoryear{Hardy, Henecka, Ivey-Law, Nock, Patrini,
  Smith, and Thorne}{Hardy et~al\mbox{.}}{2017}]%
        {hardy2017private}
\bibfield{author}{\bibinfo{person}{Stephen Hardy}, \bibinfo{person}{Wilko
  Henecka}, \bibinfo{person}{Hamish Ivey-Law}, \bibinfo{person}{Richard Nock},
  \bibinfo{person}{Giorgio Patrini}, \bibinfo{person}{Guillaume Smith}, {and}
  \bibinfo{person}{Brian Thorne}.} \bibinfo{year}{2017}\natexlab{}.
\newblock \showarticletitle{Private federated learning on vertically
  partitioned data via entity resolution and additively homomorphic
  encryption}.
\newblock \bibinfo{journal}{\emph{arXiv preprint arXiv:1711.10677}}
  (\bibinfo{year}{2017}).
\newblock


\bibitem[\protect\citeauthoryear{He, Tan, Tang, Qiu, and Liu}{He
  et~al\mbox{.}}{2019b}]%
        {he2019central}
\bibfield{author}{\bibinfo{person}{Chaoyang He}, \bibinfo{person}{Conghui Tan},
  \bibinfo{person}{Hanlin Tang}, \bibinfo{person}{Shuang Qiu}, {and}
  \bibinfo{person}{Ji Liu}.} \bibinfo{year}{2019}\natexlab{b}.
\newblock \showarticletitle{Central server free federated learning over
  single-sided trust social networks}.
\newblock \bibinfo{journal}{\emph{arXiv preprint arXiv:1910.04956}}
  (\bibinfo{year}{2019}).
\newblock


\bibitem[\protect\citeauthoryear{He, Ling, and Chen}{He et~al\mbox{.}}{2019a}]%
        {He2019}
\bibfield{author}{\bibinfo{person}{X. He}, \bibinfo{person}{Q. Ling}, {and}
  \bibinfo{person}{T. Chen}.} \bibinfo{year}{2019}\natexlab{a}.
\newblock \showarticletitle{Byzantine-Robust Stochastic Gradient Descent for
  Distributed Low-Rank Matrix Completion}, In \bibinfo{booktitle}{2019 IEEE
  Data Science Workshop (DSW)}.
\newblock \bibinfo{journal}{\emph{2019 IEEE Data Science Workshop (DSW)}},
  \bibinfo{pages}{322--326}.
\newblock


\bibitem[\protect\citeauthoryear{Hsu, Qi, and Brown}{Hsu et~al\mbox{.}}{2019}]%
        {hsu2019measuring}
\bibfield{author}{\bibinfo{person}{Tzu-Ming~Harry Hsu}, \bibinfo{person}{Hang
  Qi}, {and} \bibinfo{person}{Matthew Brown}.} \bibinfo{year}{2019}\natexlab{}.
\newblock \bibinfo{title}{Measuring the Effects of Non-Identical Data
  Distribution for Federated Visual Classification}.
\newblock
\newblock
\showeprint[arxiv]{1909.06335}~[cs.LG]


\bibitem[\protect\citeauthoryear{{Hu}, {Gao}, {Liu}, and {Ma}}{{Hu}
  et~al\mbox{.}}{2018}]%
        {8647649}
\bibfield{author}{\bibinfo{person}{B. {Hu}}, \bibinfo{person}{Y. {Gao}},
  \bibinfo{person}{L. {Liu}}, {and} \bibinfo{person}{H. {Ma}}.}
  \bibinfo{year}{2018}\natexlab{}.
\newblock \showarticletitle{Federated Region-Learning: An Edge Computing Based
  Framework for Urban Environment Sensing}. In \bibinfo{booktitle}{\emph{2018
  IEEE Global Communications Conference (GLOBECOM)}}. \bibinfo{pages}{1--7}.
\newblock


\bibitem[\protect\citeauthoryear{Hu, Jiang, and Wang}{Hu
  et~al\mbox{.}}{2019a}]%
        {hu2019decentralized}
\bibfield{author}{\bibinfo{person}{Chenghao Hu}, \bibinfo{person}{Jingyan
  Jiang}, {and} \bibinfo{person}{Zhi Wang}.} \bibinfo{year}{2019}\natexlab{a}.
\newblock \showarticletitle{Decentralized Federated Learning: A Segmented
  Gossip Approach}.
\newblock \bibinfo{journal}{\emph{arXiv preprint arXiv:1908.07782}}
  (\bibinfo{year}{2019}).
\newblock


\bibitem[\protect\citeauthoryear{Hu, Sun, Chen, and Lu}{Hu
  et~al\mbox{.}}{2019b}]%
        {Hu2019}
\bibfield{author}{\bibinfo{person}{Yao Hu}, \bibinfo{person}{Xiaoyan Sun},
  \bibinfo{person}{Yang Chen}, {and} \bibinfo{person}{Zishuai Lu}.}
  \bibinfo{year}{2019}\natexlab{b}.
\newblock \showarticletitle{Model and Feature Aggregation Based Federated
  Learning for Multi-sensor Time Series Trend Following}.
  \bibinfo{publisher}{Springer International Publishing},
  \bibinfo{address}{Cham}, \bibinfo{pages}{233--246}.
\newblock


\bibitem[\protect\citeauthoryear{{Hua}, {Yang}, and {Shi}}{{Hua}
  et~al\mbox{.}}{2019}]%
        {8891310}
\bibfield{author}{\bibinfo{person}{S. {Hua}}, \bibinfo{person}{K. {Yang}},
  {and} \bibinfo{person}{Y. {Shi}}.} \bibinfo{year}{2019}\natexlab{}.
\newblock \showarticletitle{On-Device Federated Learning via Second-Order
  Optimization with Over-the-Air Computation}. In
  \bibinfo{booktitle}{\emph{2019 IEEE 90th Vehicular Technology Conference
  (VTC2019-Fall)}}. \bibinfo{pages}{1--5}.
\newblock
\showISSN{1090-3038}


\bibitem[\protect\citeauthoryear{Huang, Shea, Qian, Masurkar, Deng, and
  Liu}{Huang et~al\mbox{.}}{2019}]%
        {HUANG2019103291}
\bibfield{author}{\bibinfo{person}{Li Huang}, \bibinfo{person}{Andrew~L. Shea},
  \bibinfo{person}{Huining Qian}, \bibinfo{person}{Aditya Masurkar},
  \bibinfo{person}{Hao Deng}, {and} \bibinfo{person}{Dianbo Liu}.}
  \bibinfo{year}{2019}\natexlab{}.
\newblock \showarticletitle{Patient clustering improves efficiency of federated
  machine learning to predict mortality and hospital stay time using
  distributed electronic medical records}.
\newblock \bibinfo{journal}{\emph{Journal of Biomedical Informatics}}
  \bibinfo{volume}{99} (\bibinfo{year}{2019}), \bibinfo{pages}{103291}.
\newblock
\showISSN{1532-0464}
\urldef\tempurl%
\url{http://www.sciencedirect.com/science/article/pii/S1532046419302102}
\showURL{%
\tempurl}


\bibitem[\protect\citeauthoryear{Jalalirad, Scavuzzo, Capota, and
  Sprague}{Jalalirad et~al\mbox{.}}{2019}]%
        {Jalalirad2019}
\bibfield{author}{\bibinfo{person}{Amir Jalalirad}, \bibinfo{person}{Marco
  Scavuzzo}, \bibinfo{person}{Catalin Capota}, {and} \bibinfo{person}{Michael
  Sprague}.} \bibinfo{year}{2019}\natexlab{}.
\newblock \showarticletitle{A Simple and Efficient Federated Recommender
  System}. In \bibinfo{booktitle}{\emph{Proceedings of the 6th IEEE/ACM
  International Conference on Big Data Computing, Applications and
  Technologies}}. \bibinfo{publisher}{Association for Computing Machinery},
  \bibinfo{address}{Auckland, New Zealand}, \bibinfo{pages}{53–58}.
\newblock


\bibitem[\protect\citeauthoryear{Jeong, Oh, Kim, Park, Bennis, and Kim}{Jeong
  et~al\mbox{.}}{2018}]%
        {jeong2018communication}
\bibfield{author}{\bibinfo{person}{Eunjeong Jeong}, \bibinfo{person}{Seungeun
  Oh}, \bibinfo{person}{Hyesung Kim}, \bibinfo{person}{Jihong Park},
  \bibinfo{person}{Mehdi Bennis}, {and} \bibinfo{person}{Seong-Lyun Kim}.}
  \bibinfo{year}{2018}\natexlab{}.
\newblock \showarticletitle{Communication-efficient on-device machine learning:
  Federated distillation and augmentation under non-iid private data}.
\newblock \bibinfo{journal}{\emph{arXiv preprint arXiv:1811.11479}}
  (\bibinfo{year}{2018}).
\newblock


\bibitem[\protect\citeauthoryear{Ji, Long, Pan, Zhu, Jiang, and Wang}{Ji
  et~al\mbox{.}}{2019}]%
        {Ji2019}
\bibfield{author}{\bibinfo{person}{Shaoxiong Ji}, \bibinfo{person}{Guodong
  Long}, \bibinfo{person}{Shirui Pan}, \bibinfo{person}{Tianqing Zhu},
  \bibinfo{person}{Jing Jiang}, {and} \bibinfo{person}{Sen Wang}.}
  \bibinfo{year}{2019}\natexlab{}.
\newblock \showarticletitle{Detecting Suicidal Ideation with Data Protection in
  Online Communities}. \bibinfo{publisher}{Springer International Publishing},
  \bibinfo{address}{Cham}, \bibinfo{pages}{225--229}.
\newblock


\bibitem[\protect\citeauthoryear{Jiang, Song, Tong, Wu, Zhao, Xu, and
  Yang}{Jiang et~al\mbox{.}}{2019b}]%
        {Jiang2019}
\bibfield{author}{\bibinfo{person}{Di Jiang}, \bibinfo{person}{Yuanfeng Song},
  \bibinfo{person}{Yongxin Tong}, \bibinfo{person}{Xueyang Wu},
  \bibinfo{person}{Weiwei Zhao}, \bibinfo{person}{Qian Xu}, {and}
  \bibinfo{person}{Qiang Yang}.} \bibinfo{year}{2019}\natexlab{b}.
\newblock \showarticletitle{Federated Topic Modeling}. In
  \bibinfo{booktitle}{\emph{Proceedings of the 28th ACM International
  Conference on Information and Knowledge Management}}.
  \bibinfo{publisher}{Association for Computing Machinery},
  \bibinfo{address}{Beijing, China}, \bibinfo{pages}{1071–1080}.
\newblock


\bibitem[\protect\citeauthoryear{Jiang, Kone{\v{c}}n{\`y}, Rush, and
  Kannan}{Jiang et~al\mbox{.}}{2019a}]%
        {jiang2019improving}
\bibfield{author}{\bibinfo{person}{Yihan Jiang}, \bibinfo{person}{Jakub
  Kone{\v{c}}n{\`y}}, \bibinfo{person}{Keith Rush}, {and}
  \bibinfo{person}{Sreeram Kannan}.} \bibinfo{year}{2019}\natexlab{a}.
\newblock \showarticletitle{Improving federated learning personalization via
  model agnostic meta learning}.
\newblock \bibinfo{journal}{\emph{arXiv preprint arXiv:1909.12488}}
  (\bibinfo{year}{2019}).
\newblock


\bibitem[\protect\citeauthoryear{Jiang, Wang, Ko, Lee, and Tassiulas}{Jiang
  et~al\mbox{.}}{2019c}]%
        {jiang2019model}
\bibfield{author}{\bibinfo{person}{Yuang Jiang}, \bibinfo{person}{Shiqiang
  Wang}, \bibinfo{person}{Bong~Jun Ko}, \bibinfo{person}{Wei-Han Lee}, {and}
  \bibinfo{person}{Leandros Tassiulas}.} \bibinfo{year}{2019}\natexlab{c}.
\newblock \showarticletitle{Model Pruning Enables Efficient Federated Learning
  on Edge Devices}.
\newblock \bibinfo{journal}{\emph{arXiv preprint arXiv:1909.12326}}
  (\bibinfo{year}{2019}).
\newblock


\bibitem[\protect\citeauthoryear{Kairouz, McMahan, Avent, Bellet, Bennis,
  Bhagoji, Bonawitz, Charles, Cormode, Cummings, D'Oliveira, Rouayheb, Evans,
  Gardner, Garrett, Gascón, Ghazi, Gibbons, Gruteser, Harchaoui, He, He, Huo,
  Hutchinson, Hsu, Jaggi, Javidi, Joshi, Khodak, Konečný, Korolova,
  Koushanfar, Koyejo, Lepoint, Liu, Mittal, Mohri, Nock, Özgür, Pagh,
  Raykova, Qi, Ramage, Raskar, Song, Song, Stich, Sun, Suresh, Tramèr,
  Vepakomma, Wang, Xiong, Xu, Yang, Yu, Yu, and Zhao}{Kairouz
  et~al\mbox{.}}{2019}]%
        {kairouz2019advances}
\bibfield{author}{\bibinfo{person}{Peter Kairouz}, \bibinfo{person}{H.~Brendan
  McMahan}, \bibinfo{person}{Brendan Avent}, \bibinfo{person}{Aurélien
  Bellet}, \bibinfo{person}{Mehdi Bennis}, \bibinfo{person}{Arjun~Nitin
  Bhagoji}, \bibinfo{person}{Keith Bonawitz}, \bibinfo{person}{Zachary
  Charles}, \bibinfo{person}{Graham Cormode}, \bibinfo{person}{Rachel
  Cummings}, \bibinfo{person}{Rafael G.~L. D'Oliveira},
  \bibinfo{person}{Salim~El Rouayheb}, \bibinfo{person}{David Evans},
  \bibinfo{person}{Josh Gardner}, \bibinfo{person}{Zachary Garrett},
  \bibinfo{person}{Adrià Gascón}, \bibinfo{person}{Badih Ghazi},
  \bibinfo{person}{Phillip~B. Gibbons}, \bibinfo{person}{Marco Gruteser},
  \bibinfo{person}{Zaid Harchaoui}, \bibinfo{person}{Chaoyang He},
  \bibinfo{person}{Lie He}, \bibinfo{person}{Zhouyuan Huo},
  \bibinfo{person}{Ben Hutchinson}, \bibinfo{person}{Justin Hsu},
  \bibinfo{person}{Martin Jaggi}, \bibinfo{person}{Tara Javidi},
  \bibinfo{person}{Gauri Joshi}, \bibinfo{person}{Mikhail Khodak},
  \bibinfo{person}{Jakub Konečný}, \bibinfo{person}{Aleksandra Korolova},
  \bibinfo{person}{Farinaz Koushanfar}, \bibinfo{person}{Sanmi Koyejo},
  \bibinfo{person}{Tancrède Lepoint}, \bibinfo{person}{Yang Liu},
  \bibinfo{person}{Prateek Mittal}, \bibinfo{person}{Mehryar Mohri},
  \bibinfo{person}{Richard Nock}, \bibinfo{person}{Ayfer Özgür},
  \bibinfo{person}{Rasmus Pagh}, \bibinfo{person}{Mariana Raykova},
  \bibinfo{person}{Hang Qi}, \bibinfo{person}{Daniel Ramage},
  \bibinfo{person}{Ramesh Raskar}, \bibinfo{person}{Dawn Song},
  \bibinfo{person}{Weikang Song}, \bibinfo{person}{Sebastian~U. Stich},
  \bibinfo{person}{Ziteng Sun}, \bibinfo{person}{Ananda~Theertha Suresh},
  \bibinfo{person}{Florian Tramèr}, \bibinfo{person}{Praneeth Vepakomma},
  \bibinfo{person}{Jianyu Wang}, \bibinfo{person}{Li Xiong},
  \bibinfo{person}{Zheng Xu}, \bibinfo{person}{Qiang Yang},
  \bibinfo{person}{Felix~X. Yu}, \bibinfo{person}{Han Yu}, {and}
  \bibinfo{person}{Sen Zhao}.} \bibinfo{year}{2019}\natexlab{}.
\newblock \bibinfo{title}{Advances and Open Problems in Federated Learning}.
\newblock
\newblock
\showeprint[arxiv]{1912.04977}~[cs.LG]


\bibitem[\protect\citeauthoryear{{Kang}, {Xiong}, {Niyato}, {Xie}, and
  {Zhang}}{{Kang} et~al\mbox{.}}{2019a}]%
        {8832210}
\bibfield{author}{\bibinfo{person}{J. {Kang}}, \bibinfo{person}{Z. {Xiong}},
  \bibinfo{person}{D. {Niyato}}, \bibinfo{person}{S. {Xie}}, {and}
  \bibinfo{person}{J. {Zhang}}.} \bibinfo{year}{2019}\natexlab{a}.
\newblock \showarticletitle{Incentive Mechanism for Reliable Federated
  Learning: A Joint Optimization Approach to Combining Reputation and Contract
  Theory}.
\newblock \bibinfo{journal}{\emph{IEEE Internet of Things Journal}}
  \bibinfo{volume}{6}, \bibinfo{number}{6} (\bibinfo{date}{Dec}
  \bibinfo{year}{2019}), \bibinfo{pages}{10700--10714}.
\newblock
\showISSN{2372-2541}


\bibitem[\protect\citeauthoryear{{Kang}, {Xiong}, {Niyato}, {Yu}, {Liang}, and
  {Kim}}{{Kang} et~al\mbox{.}}{2019b}]%
        {8851649}
\bibfield{author}{\bibinfo{person}{J. {Kang}}, \bibinfo{person}{Z. {Xiong}},
  \bibinfo{person}{D. {Niyato}}, \bibinfo{person}{H. {Yu}}, \bibinfo{person}{Y.
  {Liang}}, {and} \bibinfo{person}{D.~I. {Kim}}.}
  \bibinfo{year}{2019}\natexlab{b}.
\newblock \showarticletitle{Incentive Design for Efficient Federated Learning
  in Mobile Networks: A Contract Theory Approach}. In
  \bibinfo{booktitle}{\emph{2019 IEEE VTS Asia Pacific Wireless Communications
  Symposium (APWCS)}}. \bibinfo{pages}{1--5}.
\newblock


\bibitem[\protect\citeauthoryear{{Kang}, {Xiong}, {Niyato}, {Zou}, {Zhang}, and
  {Guizani}}{{Kang} et~al\mbox{.}}{2020}]%
        {8994206}
\bibfield{author}{\bibinfo{person}{J. {Kang}}, \bibinfo{person}{Z. {Xiong}},
  \bibinfo{person}{D. {Niyato}}, \bibinfo{person}{Y. {Zou}},
  \bibinfo{person}{Y. {Zhang}}, {and} \bibinfo{person}{M. {Guizani}}.}
  \bibinfo{year}{2020}\natexlab{}.
\newblock \showarticletitle{Reliable Federated Learning for Mobile Networks}.
\newblock \bibinfo{journal}{\emph{IEEE Wireless Communications}}
  \bibinfo{volume}{27}, \bibinfo{number}{2} (\bibinfo{year}{2020}),
  \bibinfo{pages}{72--80}.
\newblock


\bibitem[\protect\citeauthoryear{Karimireddy, Kale, Mohri, Reddi, Stich, and
  Suresh}{Karimireddy et~al\mbox{.}}{2019}]%
        {karimireddy2019scaffold}
\bibfield{author}{\bibinfo{person}{Sai~Praneeth Karimireddy},
  \bibinfo{person}{Satyen Kale}, \bibinfo{person}{Mehryar Mohri},
  \bibinfo{person}{Sashank~J Reddi}, \bibinfo{person}{Sebastian~U Stich}, {and}
  \bibinfo{person}{Ananda~Theertha Suresh}.} \bibinfo{year}{2019}\natexlab{}.
\newblock \showarticletitle{SCAFFOLD: Stochastic controlled averaging for
  on-device federated learning}.
\newblock \bibinfo{journal}{\emph{arXiv preprint arXiv:1910.06378}}
  (\bibinfo{year}{2019}).
\newblock


\bibitem[\protect\citeauthoryear{Khan, Tran, Pandey, Saad, Han, Nguyen, and
  Hong}{Khan et~al\mbox{.}}{2019}]%
        {khan2019federated}
\bibfield{author}{\bibinfo{person}{Latif~U Khan}, \bibinfo{person}{Nguyen~H
  Tran}, \bibinfo{person}{Shashi~Raj Pandey}, \bibinfo{person}{Walid Saad},
  \bibinfo{person}{Zhu Han}, \bibinfo{person}{Minh~NH Nguyen}, {and}
  \bibinfo{person}{Choong~Seon Hong}.} \bibinfo{year}{2019}\natexlab{}.
\newblock \showarticletitle{Federated Learning for Edge Networks: Resource
  Optimization and Incentive Mechanism}.
\newblock \bibinfo{journal}{\emph{arXiv preprint arXiv:1911.05642}}
  (\bibinfo{year}{2019}).
\newblock


\bibitem[\protect\citeauthoryear{{Kim}, {Park}, {Bennis}, and {Kim}}{{Kim}
  et~al\mbox{.}}{2019}]%
        {8733825}
\bibfield{author}{\bibinfo{person}{H. {Kim}}, \bibinfo{person}{J. {Park}},
  \bibinfo{person}{M. {Bennis}}, {and} \bibinfo{person}{S. {Kim}}.}
  \bibinfo{year}{2019}\natexlab{}.
\newblock \showarticletitle{Blockchained On-Device Federated Learning}.
\newblock \bibinfo{journal}{\emph{IEEE Communications Letters}}
  (\bibinfo{year}{2019}), \bibinfo{pages}{1--1}.
\newblock
\showISSN{2373-7891}


\bibitem[\protect\citeauthoryear{{Kim} and {Hong}}{{Kim} and {Hong}}{2019}]%
        {8893114}
\bibfield{author}{\bibinfo{person}{Y.~J. {Kim}} {and} \bibinfo{person}{C.~S.
  {Hong}}.} \bibinfo{year}{2019}\natexlab{}.
\newblock \showarticletitle{Blockchain-based Node-aware Dynamic Weighting
  Methods for Improving Federated Learning Performance}. In
  \bibinfo{booktitle}{\emph{2019 20th Asia-Pacific Network Operations and
  Management Symposium (APNOMS)}}. \bibinfo{pages}{1--4}.
\newblock
\showISSN{2576-8565}


\bibitem[\protect\citeauthoryear{Kitchenham and Charters}{Kitchenham and
  Charters}{2007}]%
        {Kitchenham07guidelinesfor}
\bibfield{author}{\bibinfo{person}{B. Kitchenham} {and} \bibinfo{person}{S
  Charters}.} \bibinfo{year}{2007}\natexlab{}.
\newblock \bibinfo{title}{Guidelines for performing Systematic Literature
  Reviews in Software Engineering}.
\newblock
\newblock


\bibitem[\protect\citeauthoryear{Kone{\v{c}}n{\`y}, McMahan, Yu, Richt{\'a}rik,
  Suresh, and Bacon}{Kone{\v{c}}n{\`y} et~al\mbox{.}}{2016}]%
        {Konecny2016}
\bibfield{author}{\bibinfo{person}{Jakub Kone{\v{c}}n{\`y}},
  \bibinfo{person}{H~Brendan McMahan}, \bibinfo{person}{Felix~X Yu},
  \bibinfo{person}{Peter Richt{\'a}rik}, \bibinfo{person}{Ananda~Theertha
  Suresh}, {and} \bibinfo{person}{Dave Bacon}.}
  \bibinfo{year}{2016}\natexlab{}.
\newblock \showarticletitle{Federated learning: Strategies for improving
  communication efficiency}.
\newblock \bibinfo{journal}{\emph{arXiv preprint arXiv:1610.05492}}
  (\bibinfo{year}{2016}).
\newblock


\bibitem[\protect\citeauthoryear{Konečný, McMahan, Ramage, and
  Richtárik}{Konečný et~al\mbox{.}}{2016}]%
        {konen2016federated}
\bibfield{author}{\bibinfo{person}{Jakub Konečný},
  \bibinfo{person}{H.~Brendan McMahan}, \bibinfo{person}{Daniel Ramage}, {and}
  \bibinfo{person}{Peter Richtárik}.} \bibinfo{year}{2016}\natexlab{}.
\newblock \bibinfo{title}{Federated Optimization: Distributed Machine Learning
  for On-Device Intelligence}.
\newblock
\newblock
\showeprint[arxiv]{1610.02527}~[cs.LG]


\bibitem[\protect\citeauthoryear{Koskela and Honkela}{Koskela and
  Honkela}{2019}]%
        {koskela2019learning}
\bibfield{author}{\bibinfo{person}{Antti Koskela} {and} \bibinfo{person}{Antti
  Honkela}.} \bibinfo{year}{2019}\natexlab{}.
\newblock \showarticletitle{Learning Rate Adaptation for Federated and
  Differentially Private Learning}.
\newblock \bibinfo{journal}{\emph{stat}}  \bibinfo{volume}{1050}
  (\bibinfo{year}{2019}), \bibinfo{pages}{31}.
\newblock


\bibitem[\protect\citeauthoryear{Lalitha, Kilinc, Javidi, and
  Koushanfar}{Lalitha et~al\mbox{.}}{2019a}]%
        {lalitha2019peer}
\bibfield{author}{\bibinfo{person}{Anusha Lalitha},
  \bibinfo{person}{Osman~Cihan Kilinc}, \bibinfo{person}{Tara Javidi}, {and}
  \bibinfo{person}{Farinaz Koushanfar}.} \bibinfo{year}{2019}\natexlab{a}.
\newblock \showarticletitle{Peer-to-peer Federated Learning on Graphs}.
\newblock \bibinfo{journal}{\emph{arXiv preprint arXiv:1901.11173}}
  (\bibinfo{year}{2019}).
\newblock


\bibitem[\protect\citeauthoryear{Lalitha, Shekhar, Javidi, and
  Koushanfar}{Lalitha et~al\mbox{.}}{2018}]%
        {lalitha2018fully}
\bibfield{author}{\bibinfo{person}{Anusha Lalitha}, \bibinfo{person}{Shubhanshu
  Shekhar}, \bibinfo{person}{Tara Javidi}, {and} \bibinfo{person}{Farinaz
  Koushanfar}.} \bibinfo{year}{2018}\natexlab{}.
\newblock \showarticletitle{Fully decentralized federated learning}. In
  \bibinfo{booktitle}{\emph{Third workshop on Bayesian Deep Learning
  (NeurIPS)}}.
\newblock


\bibitem[\protect\citeauthoryear{Lalitha, Wang, Kilinc, Lu, Javidi, and
  Koushanfar}{Lalitha et~al\mbox{.}}{2019b}]%
        {lalitha2019decentralized}
\bibfield{author}{\bibinfo{person}{Anusha Lalitha}, \bibinfo{person}{Xinghan
  Wang}, \bibinfo{person}{Osman Kilinc}, \bibinfo{person}{Yongxi Lu},
  \bibinfo{person}{Tara Javidi}, {and} \bibinfo{person}{Farinaz Koushanfar}.}
  \bibinfo{year}{2019}\natexlab{b}.
\newblock \bibinfo{title}{Decentralized Bayesian Learning over Graphs}.
\newblock
\newblock
\showeprint[arxiv]{1905.10466}~[stat.ML]


\bibitem[\protect\citeauthoryear{{Leroy}, {Coucke}, {Lavril}, {Gisselbrecht},
  and {Dureau}}{{Leroy} et~al\mbox{.}}{2019}]%
        {8683546}
\bibfield{author}{\bibinfo{person}{D. {Leroy}}, \bibinfo{person}{A. {Coucke}},
  \bibinfo{person}{T. {Lavril}}, \bibinfo{person}{T. {Gisselbrecht}}, {and}
  \bibinfo{person}{J. {Dureau}}.} \bibinfo{year}{2019}\natexlab{}.
\newblock \showarticletitle{Federated Learning for Keyword Spotting}. In
  \bibinfo{booktitle}{\emph{ICASSP 2019 - 2019 IEEE International Conference on
  Acoustics, Speech and Signal Processing (ICASSP)}}.
  \bibinfo{pages}{6341--6345}.
\newblock
\showISSN{1520-6149}


\bibitem[\protect\citeauthoryear{Li and Wang}{Li and Wang}{2019}]%
        {li2019fedmd}
\bibfield{author}{\bibinfo{person}{Daliang Li} {and} \bibinfo{person}{Junpu
  Wang}.} \bibinfo{year}{2019}\natexlab{}.
\newblock \showarticletitle{FedMD: Heterogenous Federated Learning via Model
  Distillation}.
\newblock \bibinfo{journal}{\emph{arXiv preprint arXiv:1910.03581}}
  (\bibinfo{year}{2019}).
\newblock


\bibitem[\protect\citeauthoryear{Li and Han}{Li and Han}{2019}]%
        {li2019endtoend}
\bibfield{author}{\bibinfo{person}{Hongyu Li} {and} \bibinfo{person}{Tianqi
  Han}.} \bibinfo{year}{2019}\natexlab{}.
\newblock \bibinfo{title}{An End-to-End Encrypted Neural Network for Gradient
  Updates Transmission in Federated Learning}.
\newblock
\newblock
\showeprint[arxiv]{1908.08340}~[cs.LG]


\bibitem[\protect\citeauthoryear{Li, Khodak, Caldas, and Talwalkar}{Li
  et~al\mbox{.}}{2019b}]%
        {li2019differentially}
\bibfield{author}{\bibinfo{person}{Jeffrey Li}, \bibinfo{person}{Mikhail
  Khodak}, \bibinfo{person}{Sebastian Caldas}, {and} \bibinfo{person}{Ameet
  Talwalkar}.} \bibinfo{year}{2019}\natexlab{b}.
\newblock \bibinfo{title}{Differentially Private Meta-Learning}.
\newblock
\newblock
\showeprint[arxiv]{1909.05830}~[cs.LG]


\bibitem[\protect\citeauthoryear{Li, Xiong, Guo, Wang, and Xu}{Li
  et~al\mbox{.}}{2019e}]%
        {Li2019}
\bibfield{author}{\bibinfo{person}{L. Li}, \bibinfo{person}{H. Xiong},
  \bibinfo{person}{Z. Guo}, \bibinfo{person}{J. Wang}, {and}
  \bibinfo{person}{C. Xu}.} \bibinfo{year}{2019}\natexlab{e}.
\newblock \showarticletitle{SmartPC: Hierarchical Pace Control in Real-Time
  Federated Learning System}, In \bibinfo{booktitle}{2019 IEEE Real-Time
  Systems Symposium (RTSS)}.
\newblock \bibinfo{journal}{\emph{2019 IEEE Real-Time Systems Symposium
  (RTSS)}}, \bibinfo{pages}{406--418}.
\newblock
\showISSN{2576-3172}


\bibitem[\protect\citeauthoryear{Li, Wen, Wu, Hu, Wang, and He}{Li
  et~al\mbox{.}}{2019d}]%
        {li2019survey}
\bibfield{author}{\bibinfo{person}{Qinbin Li}, \bibinfo{person}{Zeyi Wen},
  \bibinfo{person}{Zhaomin Wu}, \bibinfo{person}{Sixu Hu},
  \bibinfo{person}{Naibo Wang}, {and} \bibinfo{person}{Bingsheng He}.}
  \bibinfo{year}{2019}\natexlab{d}.
\newblock \bibinfo{title}{A Survey on Federated Learning Systems: Vision, Hype
  and Reality for Data Privacy and Protection}.
\newblock
\newblock
\showeprint[arxiv]{1907.09693}~[cs.LG]


\bibitem[\protect\citeauthoryear{Li, Cheng, Liu, Wang, and Chen}{Li
  et~al\mbox{.}}{2019a}]%
        {li2019abnormal}
\bibfield{author}{\bibinfo{person}{Suyi Li}, \bibinfo{person}{Yong Cheng},
  \bibinfo{person}{Yang Liu}, \bibinfo{person}{Wei Wang}, {and}
  \bibinfo{person}{Tianjian Chen}.} \bibinfo{year}{2019}\natexlab{a}.
\newblock \showarticletitle{Abnormal client behavior detection in federated
  learning}.
\newblock \bibinfo{journal}{\emph{arXiv preprint arXiv:1910.09933}}
  (\bibinfo{year}{2019}).
\newblock


\bibitem[\protect\citeauthoryear{Li, Sahu, Talwalkar, and Smith}{Li
  et~al\mbox{.}}{2020}]%
        {Li_2020}
\bibfield{author}{\bibinfo{person}{Tian Li}, \bibinfo{person}{Anit~Kumar Sahu},
  \bibinfo{person}{Ameet Talwalkar}, {and} \bibinfo{person}{Virginia Smith}.}
  \bibinfo{year}{2020}\natexlab{}.
\newblock \showarticletitle{Federated Learning: Challenges, Methods, and Future
  Directions}.
\newblock \bibinfo{journal}{\emph{IEEE Signal Processing Magazine}}
  \bibinfo{volume}{37}, \bibinfo{number}{3} (\bibinfo{date}{May}
  \bibinfo{year}{2020}), \bibinfo{pages}{50–60}.
\newblock
\showISSN{1558-0792}
\urldef\tempurl%
\url{https://doi.org/10.1109/msp.2020.2975749}
\showDOI{\tempurl}


\bibitem[\protect\citeauthoryear{Li, Sahu, Zaheer, Sanjabi, Talwalkar, and
  Smith}{Li et~al\mbox{.}}{2018}]%
        {li2018federated}
\bibfield{author}{\bibinfo{person}{Tian Li}, \bibinfo{person}{Anit~Kumar Sahu},
  \bibinfo{person}{Manzil Zaheer}, \bibinfo{person}{Maziar Sanjabi},
  \bibinfo{person}{Ameet Talwalkar}, {and} \bibinfo{person}{Virginia Smith}.}
  \bibinfo{year}{2018}\natexlab{}.
\newblock \bibinfo{title}{Federated Optimization in Heterogeneous Networks}.
\newblock
\newblock
\showeprint[arxiv]{1812.06127}~[cs.LG]


\bibitem[\protect\citeauthoryear{Li, Sanjabi, and Smith}{Li
  et~al\mbox{.}}{2019c}]%
        {li2019fair}
\bibfield{author}{\bibinfo{person}{Tian Li}, \bibinfo{person}{Maziar Sanjabi},
  {and} \bibinfo{person}{Virginia Smith}.} \bibinfo{year}{2019}\natexlab{c}.
\newblock \showarticletitle{Fair Resource Allocation in Federated Learning}.
\newblock \bibinfo{journal}{\emph{arXiv preprint arXiv:1905.10497}}
  (\bibinfo{year}{2019}).
\newblock


\bibitem[\protect\citeauthoryear{Lim, Luong, Hoang, Jiao, Liang, Yang, Niyato,
  and Miao}{Lim et~al\mbox{.}}{2019}]%
        {lim2019federated}
\bibfield{author}{\bibinfo{person}{Wei Yang~Bryan Lim},
  \bibinfo{person}{Nguyen~Cong Luong}, \bibinfo{person}{Dinh~Thai Hoang},
  \bibinfo{person}{Yutao Jiao}, \bibinfo{person}{Ying-Chang Liang},
  \bibinfo{person}{Qiang Yang}, \bibinfo{person}{Dusit Niyato}, {and}
  \bibinfo{person}{Chunyan Miao}.} \bibinfo{year}{2019}\natexlab{}.
\newblock \bibinfo{title}{Federated Learning in Mobile Edge Networks: A
  Comprehensive Survey}.
\newblock
\newblock
\showeprint[arxiv]{1909.11875}~[cs.NI]


\bibitem[\protect\citeauthoryear{{Liu}, {Wang}, and {Liu}}{{Liu}
  et~al\mbox{.}}{2019}]%
        {8772088}
\bibfield{author}{\bibinfo{person}{B. {Liu}}, \bibinfo{person}{L. {Wang}},
  {and} \bibinfo{person}{M. {Liu}}.} \bibinfo{year}{2019}\natexlab{}.
\newblock \showarticletitle{Lifelong Federated Reinforcement Learning: A
  Learning Architecture for Navigation in Cloud Robotic Systems}.
\newblock \bibinfo{journal}{\emph{IEEE Robotics and Automation Letters}}
  \bibinfo{volume}{4}, \bibinfo{number}{4} (\bibinfo{date}{Oct}
  \bibinfo{year}{2019}), \bibinfo{pages}{4555--4562}.
\newblock
\showISSN{2377-3774}


\bibitem[\protect\citeauthoryear{Liu, Chakraborty, and Verma}{Liu
  et~al\mbox{.}}{2019a}]%
        {Liu2019}
\bibfield{author}{\bibinfo{person}{Changchang Liu}, \bibinfo{person}{Supriyo
  Chakraborty}, {and} \bibinfo{person}{Dinesh Verma}.}
  \bibinfo{year}{2019}\natexlab{a}.
\newblock \showarticletitle{Secure Model Fusion for Distributed Learning Using
  Partial Homomorphic Encryption}.
\newblock In \bibinfo{booktitle}{\emph{Policy-Based Autonomic Data
  Governance}}, \bibfield{editor}{\bibinfo{person}{Seraphin Calo},
  \bibinfo{person}{Elisa Bertino}, {and} \bibinfo{person}{Dinesh Verma}}
  (Eds.). \bibinfo{publisher}{Springer International Publishing},
  \bibinfo{address}{Cham}, \bibinfo{pages}{154--179}.
\newblock
\urldef\tempurl%
\url{https://doi.org/10.1007/978-3-030-17277-0_9}
\showURL{%
\tempurl}


\bibitem[\protect\citeauthoryear{Liu, Zhang, Song, and Letaief}{Liu
  et~al\mbox{.}}{2019}]%
        {DBLP:journals/corr/abs-1905-06641}
\bibfield{author}{\bibinfo{person}{Lumin Liu}, \bibinfo{person}{Jun Zhang},
  \bibinfo{person}{S.~H. Song}, {and} \bibinfo{person}{Khaled~Ben Letaief}.}
  \bibinfo{year}{2019}\natexlab{}.
\newblock \showarticletitle{client-edge-cloud hierarchical federated learning}.
\newblock \bibinfo{journal}{\emph{CoRR}}  \bibinfo{volume}{abs/1905.06641}
  (\bibinfo{year}{2019}).
\newblock
\showeprint[arxiv]{1905.06641}
\urldef\tempurl%
\url{http://arxiv.org/abs/1905.06641}
\showURL{%
\tempurl}


\bibitem[\protect\citeauthoryear{Liu, Chen, and Yang}{Liu
  et~al\mbox{.}}{2018}]%
        {liu2018secure}
\bibfield{author}{\bibinfo{person}{Yang Liu}, \bibinfo{person}{Tianjian Chen},
  {and} \bibinfo{person}{Qiang Yang}.} \bibinfo{year}{2018}\natexlab{}.
\newblock \bibinfo{title}{Secure Federated Transfer Learning}.
\newblock
\newblock
\showeprint[arxiv]{1812.03337}~[cs.LG]


\bibitem[\protect\citeauthoryear{Liu, Kang, Zhang, Li, Cheng, Chen, Hong, and
  Yang}{Liu et~al\mbox{.}}{2019b}]%
        {liu2019communication}
\bibfield{author}{\bibinfo{person}{Yang Liu}, \bibinfo{person}{Yan Kang},
  \bibinfo{person}{Xinwei Zhang}, \bibinfo{person}{Liping Li},
  \bibinfo{person}{Yong Cheng}, \bibinfo{person}{Tianjian Chen},
  \bibinfo{person}{Mingyi Hong}, {and} \bibinfo{person}{Qiang Yang}.}
  \bibinfo{year}{2019}\natexlab{b}.
\newblock \showarticletitle{A Communication Efficient Vertical Federated
  Learning Framework}.
\newblock \bibinfo{journal}{\emph{arXiv preprint arXiv:1912.11187}}
  (\bibinfo{year}{2019}).
\newblock


\bibitem[\protect\citeauthoryear{Liu, Ma, Liu, Ma, Nepal, and Deng}{Liu
  et~al\mbox{.}}{2019c}]%
        {liu2019boosting}
\bibfield{author}{\bibinfo{person}{Yang Liu}, \bibinfo{person}{Zhuo Ma},
  \bibinfo{person}{Ximeng Liu}, \bibinfo{person}{Siqi Ma},
  \bibinfo{person}{Surya Nepal}, {and} \bibinfo{person}{Robert Deng}.}
  \bibinfo{year}{2019}\natexlab{c}.
\newblock \bibinfo{title}{Boosting Privately: Privacy-Preserving Federated
  Extreme Boosting for Mobile Crowdsensing}.
\newblock
\newblock
\showeprint[arxiv]{1907.10218}~[cs.CR]


\bibitem[\protect\citeauthoryear{{Lu}, {Huang}, {Dai}, {Maharjan}, and
  {Zhang}}{{Lu} et~al\mbox{.}}{2019a}]%
        {8843900}
\bibfield{author}{\bibinfo{person}{Y. {Lu}}, \bibinfo{person}{X. {Huang}},
  \bibinfo{person}{Y. {Dai}}, \bibinfo{person}{S. {Maharjan}}, {and}
  \bibinfo{person}{Y. {Zhang}}.} \bibinfo{year}{2019}\natexlab{a}.
\newblock \showarticletitle{Blockchain and Federated Learning for
  Privacy-preserved Data Sharing in Industrial IoT}.
\newblock \bibinfo{journal}{\emph{IEEE Transactions on Industrial Informatics}}
  (\bibinfo{year}{2019}), \bibinfo{pages}{1--1}.
\newblock
\showISSN{1941-0050}


\bibitem[\protect\citeauthoryear{{Lu}, {Huang}, {Dai}, {Maharjan}, and
  {Zhang}}{{Lu} et~al\mbox{.}}{2019b}]%
        {8843942}
\bibfield{author}{\bibinfo{person}{Y. {Lu}}, \bibinfo{person}{X. {Huang}},
  \bibinfo{person}{Y. {Dai}}, \bibinfo{person}{S. {Maharjan}}, {and}
  \bibinfo{person}{Y. {Zhang}}.} \bibinfo{year}{2019}\natexlab{b}.
\newblock \showarticletitle{Differentially Private Asynchronous Federated
  Learning for Mobile Edge Computing in Urban Informatics}.
\newblock \bibinfo{journal}{\emph{IEEE Transactions on Industrial Informatics}}
  (\bibinfo{year}{2019}), \bibinfo{pages}{1--1}.
\newblock
\showISSN{1941-0050}


\bibitem[\protect\citeauthoryear{{Lugan}, {Desbordes}, {Brion}, {Ramos Tormo},
  {Legay}, and {Macq}}{{Lugan} et~al\mbox{.}}{2019}]%
        {8932389}
\bibfield{author}{\bibinfo{person}{S. {Lugan}}, \bibinfo{person}{P.
  {Desbordes}}, \bibinfo{person}{E. {Brion}}, \bibinfo{person}{L.~X. {Ramos
  Tormo}}, \bibinfo{person}{A. {Legay}}, {and} \bibinfo{person}{B. {Macq}}.}
  \bibinfo{year}{2019}\natexlab{}.
\newblock \showarticletitle{Secure Architectures Implementing Trusted
  Coalitions for Blockchained Distributed Learning (TCLearn)}.
\newblock \bibinfo{journal}{\emph{IEEE Access}}  \bibinfo{volume}{7}
  (\bibinfo{year}{2019}), \bibinfo{pages}{181789--181799}.
\newblock
\showISSN{2169-3536}


\bibitem[\protect\citeauthoryear{Lyu, Yu, and Yang}{Lyu et~al\mbox{.}}{2020}]%
        {lyu2020threats}
\bibfield{author}{\bibinfo{person}{Lingjuan Lyu}, \bibinfo{person}{Han Yu},
  {and} \bibinfo{person}{Qiang Yang}.} \bibinfo{year}{2020}\natexlab{}.
\newblock \bibinfo{title}{Threats to Federated Learning: A Survey}.
\newblock
\newblock
\showeprint[arxiv]{2003.02133}~[cs.CR]


\bibitem[\protect\citeauthoryear{Ma, Zhang, Lou, Ho, Xiong, and Jiang}{Ma
  et~al\mbox{.}}{2019}]%
        {Ma2019}
\bibfield{author}{\bibinfo{person}{Jing Ma}, \bibinfo{person}{Qiuchen Zhang},
  \bibinfo{person}{Jian Lou}, \bibinfo{person}{Joyce~C. Ho},
  \bibinfo{person}{Li Xiong}, {and} \bibinfo{person}{Xiaoqian Jiang}.}
  \bibinfo{year}{2019}\natexlab{}.
\newblock \showarticletitle{Privacy-Preserving Tensor Factorization for
  Collaborative Health Data Analysis}. In \bibinfo{booktitle}{\emph{Proceedings
  of the 28th ACM International Conference on Information and Knowledge
  Management}}. \bibinfo{publisher}{Association for Computing Machinery},
  \bibinfo{address}{Beijing, China}, \bibinfo{pages}{1291–1300}.
\newblock


\bibitem[\protect\citeauthoryear{{Majeed} and {Hong}}{{Majeed} and
  {Hong}}{2019}]%
        {8892848}
\bibfield{author}{\bibinfo{person}{U. {Majeed}} {and} \bibinfo{person}{C.~S.
  {Hong}}.} \bibinfo{year}{2019}\natexlab{}.
\newblock \showarticletitle{FLchain: Federated Learning via MEC-enabled
  Blockchain Network}. In \bibinfo{booktitle}{\emph{2019 20th Asia-Pacific
  Network Operations and Management Symposium (APNOMS)}}.
  \bibinfo{pages}{1--4}.
\newblock


\bibitem[\protect\citeauthoryear{Mandal and Gong}{Mandal and Gong}{2019}]%
        {Mandal2019}
\bibfield{author}{\bibinfo{person}{Kalikinkar Mandal} {and}
  \bibinfo{person}{Guang Gong}.} \bibinfo{year}{2019}\natexlab{}.
\newblock \showarticletitle{PrivFL: Practical Privacy-preserving Federated
  Regressions on High-dimensional Data over Mobile Networks}. In
  \bibinfo{booktitle}{\emph{Proceedings of the 2019 ACM SIGSAC Conference on
  Cloud Computing Security Workshop}}. \bibinfo{publisher}{Association for
  Computing Machinery}, \bibinfo{address}{London, United Kingdom},
  \bibinfo{pages}{57–68}.
\newblock


\bibitem[\protect\citeauthoryear{{Martinez}, {Francis}, and {Hafid}}{{Martinez}
  et~al\mbox{.}}{2019}]%
        {8945913}
\bibfield{author}{\bibinfo{person}{I. {Martinez}}, \bibinfo{person}{S.
  {Francis}}, {and} \bibinfo{person}{A.~S. {Hafid}}.}
  \bibinfo{year}{2019}\natexlab{}.
\newblock \showarticletitle{Record and Reward Federated Learning Contributions
  with Blockchain}. In \bibinfo{booktitle}{\emph{2019 International Conference
  on Cyber-Enabled Distributed Computing and Knowledge Discovery (CyberC)}}.
  \bibinfo{pages}{50--57}.
\newblock


\bibitem[\protect\citeauthoryear{McMahan, Moore, Ramage, Hampson, and
  y~Arcas}{McMahan et~al\mbox{.}}{2016}]%
        {mcmahan2016communicationefficient}
\bibfield{author}{\bibinfo{person}{H.~Brendan McMahan}, \bibinfo{person}{Eider
  Moore}, \bibinfo{person}{Daniel Ramage}, \bibinfo{person}{Seth Hampson},
  {and} \bibinfo{person}{Blaise~Agüera y Arcas}.}
  \bibinfo{year}{2016}\natexlab{}.
\newblock \bibinfo{title}{Communication-Efficient Learning of Deep Networks
  from Decentralized Data}.
\newblock
\newblock
\showeprint[arxiv]{1602.05629}~[cs.LG]


\bibitem[\protect\citeauthoryear{{Mills}, {Hu}, and {Min}}{{Mills}
  et~al\mbox{.}}{2019}]%
        {8917724}
\bibfield{author}{\bibinfo{person}{J. {Mills}}, \bibinfo{person}{J. {Hu}},
  {and} \bibinfo{person}{G. {Min}}.} \bibinfo{year}{2019}\natexlab{}.
\newblock \showarticletitle{Communication-Efficient Federated Learning for
  Wireless Edge Intelligence in IoT}.
\newblock \bibinfo{journal}{\emph{IEEE Internet of Things Journal}}
  (\bibinfo{year}{2019}), \bibinfo{pages}{1--1}.
\newblock
\showISSN{2372-2541}


\bibitem[\protect\citeauthoryear{Mo and Haddadi}{Mo and Haddadi}{[n.d.]}]%
        {Mo}
\bibfield{author}{\bibinfo{person}{Fan Mo} {and} \bibinfo{person}{Hamed
  Haddadi}.} \bibinfo{year}{[n.d.]}\natexlab{}.
\newblock \showarticletitle{Efficient and Private Federated Learning using
  TEE}.
\newblock  (\bibinfo{year}{[n.\,d.]}).
\newblock


\bibitem[\protect\citeauthoryear{Mohri, Sivek, and Suresh}{Mohri
  et~al\mbox{.}}{2019}]%
        {mohri2019agnostic}
\bibfield{author}{\bibinfo{person}{Mehryar Mohri}, \bibinfo{person}{Gary
  Sivek}, {and} \bibinfo{person}{Ananda~Theertha Suresh}.}
  \bibinfo{year}{2019}\natexlab{}.
\newblock \showarticletitle{Agnostic federated learning}.
\newblock \bibinfo{journal}{\emph{arXiv preprint arXiv:1902.00146}}
  (\bibinfo{year}{2019}).
\newblock


\bibitem[\protect\citeauthoryear{{Mowla}, {Tran}, {Doh}, and {Chae}}{{Mowla}
  et~al\mbox{.}}{2020}]%
        {8945183}
\bibfield{author}{\bibinfo{person}{N.~I. {Mowla}}, \bibinfo{person}{N.~H.
  {Tran}}, \bibinfo{person}{I. {Doh}}, {and} \bibinfo{person}{K. {Chae}}.}
  \bibinfo{year}{2020}\natexlab{}.
\newblock \showarticletitle{Federated Learning-Based Cognitive Detection of
  Jamming Attack in Flying Ad-Hoc Network}.
\newblock \bibinfo{journal}{\emph{IEEE Access}}  \bibinfo{volume}{8}
  (\bibinfo{year}{2020}), \bibinfo{pages}{4338--4350}.
\newblock
\showISSN{2169-3536}


\bibitem[\protect\citeauthoryear{{Nadiger}, {Kumar}, and {Abdelhak}}{{Nadiger}
  et~al\mbox{.}}{2019}]%
        {8791693}
\bibfield{author}{\bibinfo{person}{C. {Nadiger}}, \bibinfo{person}{A. {Kumar}},
  {and} \bibinfo{person}{S. {Abdelhak}}.} \bibinfo{year}{2019}\natexlab{}.
\newblock \showarticletitle{Federated Reinforcement Learning for Fast
  Personalization}. In \bibinfo{booktitle}{\emph{2019 IEEE Second International
  Conference on Artificial Intelligence and Knowledge Engineering (AIKE)}}.
  \bibinfo{pages}{123--127}.
\newblock


\bibitem[\protect\citeauthoryear{{Nguyen}, {Marchal}, {Miettinen},
  {Fereidooni}, {Asokan}, and {Sadeghi}}{{Nguyen} et~al\mbox{.}}{2019}]%
        {8884802}
\bibfield{author}{\bibinfo{person}{T.~D. {Nguyen}}, \bibinfo{person}{S.
  {Marchal}}, \bibinfo{person}{M. {Miettinen}}, \bibinfo{person}{H.
  {Fereidooni}}, \bibinfo{person}{N. {Asokan}}, {and} \bibinfo{person}{A.
  {Sadeghi}}.} \bibinfo{year}{2019}\natexlab{}.
\newblock \showarticletitle{DÏoT: A Federated Self-learning Anomaly Detection
  System for IoT}. In \bibinfo{booktitle}{\emph{2019 IEEE 39th International
  Conference on Distributed Computing Systems (ICDCS)}}.
  \bibinfo{pages}{756--767}.
\newblock
\showISSN{1063-6927}


\bibitem[\protect\citeauthoryear{Niknam, Dhillon, and Reed}{Niknam
  et~al\mbox{.}}{2019}]%
        {niknam2019federated}
\bibfield{author}{\bibinfo{person}{Solmaz Niknam}, \bibinfo{person}{Harpreet~S.
  Dhillon}, {and} \bibinfo{person}{Jeffery~H. Reed}.}
  \bibinfo{year}{2019}\natexlab{}.
\newblock \bibinfo{title}{Federated Learning for Wireless Communications:
  Motivation, Opportunities and Challenges}.
\newblock
\newblock
\showeprint[arxiv]{1908.06847}~[eess.SP]


\bibitem[\protect\citeauthoryear{{Nishio} and {Yonetani}}{{Nishio} and
  {Yonetani}}{2019}]%
        {8761315}
\bibfield{author}{\bibinfo{person}{T. {Nishio}} {and} \bibinfo{person}{R.
  {Yonetani}}.} \bibinfo{year}{2019}\natexlab{}.
\newblock \showarticletitle{Client Selection for Federated Learning with
  Heterogeneous Resources in Mobile Edge}. In \bibinfo{booktitle}{\emph{ICC
  2019 - 2019 IEEE International Conference on Communications (ICC)}}.
  \bibinfo{pages}{1--7}.
\newblock
\showISSN{1550-3607}


\bibitem[\protect\citeauthoryear{Niu, Wu, Tang, Hua, Jia, Lv, Wu, and Chen}{Niu
  et~al\mbox{.}}{2019}]%
        {niu2019secure}
\bibfield{author}{\bibinfo{person}{Chaoyue Niu}, \bibinfo{person}{Fan Wu},
  \bibinfo{person}{Shaojie Tang}, \bibinfo{person}{Lifeng Hua},
  \bibinfo{person}{Rongfei Jia}, \bibinfo{person}{Chengfei Lv},
  \bibinfo{person}{Zhihua Wu}, {and} \bibinfo{person}{Guihai Chen}.}
  \bibinfo{year}{2019}\natexlab{}.
\newblock \showarticletitle{Secure federated submodel learning}.
\newblock \bibinfo{journal}{\emph{arXiv preprint arXiv:1911.02254}}
  (\bibinfo{year}{2019}).
\newblock


\bibitem[\protect\citeauthoryear{Nock, Hardy, Henecka, Ivey-Law, Patrini,
  Smith, and Thorne}{Nock et~al\mbox{.}}{2018}]%
        {nock2018entity}
\bibfield{author}{\bibinfo{person}{Richard Nock}, \bibinfo{person}{Stephen
  Hardy}, \bibinfo{person}{Wilko Henecka}, \bibinfo{person}{Hamish Ivey-Law},
  \bibinfo{person}{Giorgio Patrini}, \bibinfo{person}{Guillaume Smith}, {and}
  \bibinfo{person}{Brian Thorne}.} \bibinfo{year}{2018}\natexlab{}.
\newblock \showarticletitle{Entity Resolution and Federated Learning get a
  Federated Resolution}.
\newblock \bibinfo{journal}{\emph{arXiv preprint arXiv:1803.04035}}
  (\bibinfo{year}{2018}).
\newblock


\bibitem[\protect\citeauthoryear{Orekondy, Oh, Zhang, Schiele, and
  Fritz}{Orekondy et~al\mbox{.}}{2018}]%
        {orekondy2018gradientleaks}
\bibfield{author}{\bibinfo{person}{Tribhuvanesh Orekondy},
  \bibinfo{person}{Seong~Joon Oh}, \bibinfo{person}{Yang Zhang},
  \bibinfo{person}{Bernt Schiele}, {and} \bibinfo{person}{Mario Fritz}.}
  \bibinfo{year}{2018}\natexlab{}.
\newblock \bibinfo{title}{Gradient-Leaks: Understanding and Controlling
  Deanonymization in Federated Learning}.
\newblock
\newblock
\showeprint[arxiv]{1805.05838}~[cs.CR]


\bibitem[\protect\citeauthoryear{Pandey, Tran, Bennis, Tun, Han, and
  Hong}{Pandey et~al\mbox{.}}{2019}]%
        {Pandey2019}
\bibfield{author}{\bibinfo{person}{S.~R. Pandey}, \bibinfo{person}{N.~H. Tran},
  \bibinfo{person}{M. Bennis}, \bibinfo{person}{Y.~K. Tun}, \bibinfo{person}{Z.
  Han}, {and} \bibinfo{person}{C.~S. Hong}.} \bibinfo{year}{2019}\natexlab{}.
\newblock \showarticletitle{Incentivize to Build: A Crowdsourcing Framework for
  Federated Learning}, In \bibinfo{booktitle}{2019 IEEE Global Communications
  Conference (GLOBECOM)}.
\newblock \bibinfo{journal}{\emph{2019 IEEE Global Communications Conference
  (GLOBECOM)}}, \bibinfo{pages}{1--6}.
\newblock
\showISSN{2576-6813}


\bibitem[\protect\citeauthoryear{Pandey, Tran, Bennis, Tun, Manzoor, and
  Hong}{Pandey et~al\mbox{.}}{2020}]%
        {Pandey2020}
\bibfield{author}{\bibinfo{person}{S.~R. Pandey}, \bibinfo{person}{N.~H. Tran},
  \bibinfo{person}{M. Bennis}, \bibinfo{person}{Y.~K. Tun}, \bibinfo{person}{A.
  Manzoor}, {and} \bibinfo{person}{C.~S. Hong}.}
  \bibinfo{year}{2020}\natexlab{}.
\newblock \showarticletitle{A Crowdsourcing Framework for On-Device Federated
  Learning}.
\newblock \bibinfo{journal}{\emph{IEEE Transactions on Wireless
  Communications}} \bibinfo{volume}{19}, \bibinfo{number}{5}
  (\bibinfo{year}{2020}), \bibinfo{pages}{3241--3256}.
\newblock
\showISSN{1558-2248}


\bibitem[\protect\citeauthoryear{Peng, Huang, Zhu, and Saenko}{Peng
  et~al\mbox{.}}{2019}]%
        {peng2019federated}
\bibfield{author}{\bibinfo{person}{Xingchao Peng}, \bibinfo{person}{Zijun
  Huang}, \bibinfo{person}{Yizhe Zhu}, {and} \bibinfo{person}{Kate Saenko}.}
  \bibinfo{year}{2019}\natexlab{}.
\newblock \bibinfo{title}{Federated Adversarial Domain Adaptation}.
\newblock
\newblock
\showeprint[arxiv]{1911.02054}~[cs.CV]


\bibitem[\protect\citeauthoryear{Peterson, Kanani, and Marathe}{Peterson
  et~al\mbox{.}}{2019}]%
        {peterson2019private}
\bibfield{author}{\bibinfo{person}{Daniel Peterson}, \bibinfo{person}{Pallika
  Kanani}, {and} \bibinfo{person}{Virendra~J Marathe}.}
  \bibinfo{year}{2019}\natexlab{}.
\newblock \showarticletitle{Private Federated Learning with Domain Adaptation}.
\newblock \bibinfo{journal}{\emph{arXiv preprint arXiv:1912.06733}}
  (\bibinfo{year}{2019}).
\newblock


\bibitem[\protect\citeauthoryear{Pillutla, Kakade, and Harchaoui}{Pillutla
  et~al\mbox{.}}{2019}]%
        {pillutla2019robust}
\bibfield{author}{\bibinfo{person}{Krishna Pillutla}, \bibinfo{person}{Sham~M
  Kakade}, {and} \bibinfo{person}{Zaid Harchaoui}.}
  \bibinfo{year}{2019}\natexlab{}.
\newblock \showarticletitle{Robust aggregation for federated learning}.
\newblock \bibinfo{journal}{\emph{arXiv preprint arXiv:1912.13445}}
  (\bibinfo{year}{2019}).
\newblock


\bibitem[\protect\citeauthoryear{Preuveneers, Rimmer, Tsingenopoulos, Spooren,
  Joosen, and Ilie-Zudor}{Preuveneers et~al\mbox{.}}{2018}]%
        {preuveneers2018chained}
\bibfield{author}{\bibinfo{person}{Davy Preuveneers}, \bibinfo{person}{Vera
  Rimmer}, \bibinfo{person}{Ilias Tsingenopoulos}, \bibinfo{person}{Jan
  Spooren}, \bibinfo{person}{Wouter Joosen}, {and} \bibinfo{person}{Elisabeth
  Ilie-Zudor}.} \bibinfo{year}{2018}\natexlab{}.
\newblock \showarticletitle{Chained anomaly detection models for federated
  learning: an intrusion detection case study}.
\newblock \bibinfo{journal}{\emph{Applied Sciences}} \bibinfo{volume}{8},
  \bibinfo{number}{12} (\bibinfo{year}{2018}), \bibinfo{pages}{2663}.
\newblock


\bibitem[\protect\citeauthoryear{{Qian}, {Gochhayat}, and {Hansen}}{{Qian}
  et~al\mbox{.}}{2019}]%
        {8854053}
\bibfield{author}{\bibinfo{person}{J. {Qian}}, \bibinfo{person}{S.~P.
  {Gochhayat}}, {and} \bibinfo{person}{L.~K. {Hansen}}.}
  \bibinfo{year}{2019}\natexlab{}.
\newblock \showarticletitle{Distributed Active Learning Strategies on Edge
  Computing}. In \bibinfo{booktitle}{\emph{2019 6th IEEE International
  Conference on Cyber Security and Cloud Computing (CSCloud)/ 2019 5th IEEE
  International Conference on Edge Computing and Scalable Cloud (EdgeCom)}}.
  \bibinfo{pages}{221--226}.
\newblock


\bibitem[\protect\citeauthoryear{Qian, Sengupta, and Hansen}{Qian
  et~al\mbox{.}}{2019b}]%
        {qian2019active}
\bibfield{author}{\bibinfo{person}{Jia Qian}, \bibinfo{person}{Sayantan
  Sengupta}, {and} \bibinfo{person}{Lars~Kai Hansen}.}
  \bibinfo{year}{2019}\natexlab{b}.
\newblock \bibinfo{title}{Active Learning Solution on Distributed Edge
  Computing}.
\newblock
\newblock
\showeprint[arxiv]{1906.10718}~[cs.DC]


\bibitem[\protect\citeauthoryear{Qian, Hu, Chen, Guan, Hassan, and
  Alelaiwi}{Qian et~al\mbox{.}}{2019a}]%
        {QIAN2019562}
\bibfield{author}{\bibinfo{person}{Yongfeng Qian}, \bibinfo{person}{Long Hu},
  \bibinfo{person}{Jing Chen}, \bibinfo{person}{Xin Guan},
  \bibinfo{person}{Mohammad~Mehedi Hassan}, {and} \bibinfo{person}{Abdulhameed
  Alelaiwi}.} \bibinfo{year}{2019}\natexlab{a}.
\newblock \showarticletitle{Privacy-aware service placement for mobile edge
  computing via federated learning}.
\newblock \bibinfo{journal}{\emph{Information Sciences}}  \bibinfo{volume}{505}
  (\bibinfo{year}{2019}), \bibinfo{pages}{562 -- 570}.
\newblock
\showISSN{0020-0255}
\urldef\tempurl%
\url{http://www.sciencedirect.com/science/article/pii/S0020025519306814}
\showURL{%
\tempurl}


\bibitem[\protect\citeauthoryear{Ramaswamy, Mathews, Rao, and
  Beaufays}{Ramaswamy et~al\mbox{.}}{2019}]%
        {ramaswamy2019federated}
\bibfield{author}{\bibinfo{person}{Swaroop Ramaswamy}, \bibinfo{person}{Rajiv
  Mathews}, \bibinfo{person}{Kanishka Rao}, {and}
  \bibinfo{person}{Fran{\c{c}}oise Beaufays}.} \bibinfo{year}{2019}\natexlab{}.
\newblock \showarticletitle{Federated learning for emoji prediction in a mobile
  keyboard}.
\newblock \bibinfo{journal}{\emph{arXiv preprint arXiv:1906.04329}}
  (\bibinfo{year}{2019}).
\newblock


\bibitem[\protect\citeauthoryear{Reisizadeh, Mokhtari, Hassani, Jadbabaie, and
  Pedarsani}{Reisizadeh et~al\mbox{.}}{2019}]%
        {reisizadeh2019fedpaq}
\bibfield{author}{\bibinfo{person}{Amirhossein Reisizadeh},
  \bibinfo{person}{Aryan Mokhtari}, \bibinfo{person}{Hamed Hassani},
  \bibinfo{person}{Ali Jadbabaie}, {and} \bibinfo{person}{Ramtin Pedarsani}.}
  \bibinfo{year}{2019}\natexlab{}.
\newblock \showarticletitle{Fedpaq: A communication-efficient federated
  learning method with periodic averaging and quantization}.
\newblock \bibinfo{journal}{\emph{arXiv preprint arXiv:1909.13014}}
  (\bibinfo{year}{2019}).
\newblock


\bibitem[\protect\citeauthoryear{{Ren}, {Wang}, {Hou}, {Zheng}, and
  {Tang}}{{Ren} et~al\mbox{.}}{2019}]%
        {8728285}
\bibfield{author}{\bibinfo{person}{J. {Ren}}, \bibinfo{person}{H. {Wang}},
  \bibinfo{person}{T. {Hou}}, \bibinfo{person}{S. {Zheng}}, {and}
  \bibinfo{person}{C. {Tang}}.} \bibinfo{year}{2019}\natexlab{}.
\newblock \showarticletitle{Federated Learning-Based Computation Offloading
  Optimization in Edge Computing-Supported Internet of Things}.
\newblock \bibinfo{journal}{\emph{IEEE Access}}  \bibinfo{volume}{7}
  (\bibinfo{year}{2019}), \bibinfo{pages}{69194--69201}.
\newblock
\showISSN{2169-3536}


\bibitem[\protect\citeauthoryear{Ren, Yu, and Ding}{Ren et~al\mbox{.}}{2019}]%
        {ren2019accelerating}
\bibfield{author}{\bibinfo{person}{Jinke Ren}, \bibinfo{person}{Guanding Yu},
  {and} \bibinfo{person}{Guangyao Ding}.} \bibinfo{year}{2019}\natexlab{}.
\newblock \bibinfo{title}{Accelerating DNN Training in Wireless Federated Edge
  Learning System}.
\newblock
\newblock
\showeprint[arxiv]{1905.09712}~[cs.LG]


\bibitem[\protect\citeauthoryear{Roy, Siddiqui, P{\"o}lsterl, Navab, and
  Wachinger}{Roy et~al\mbox{.}}{2019}]%
        {roy2019braintorrent}
\bibfield{author}{\bibinfo{person}{Abhijit~Guha Roy}, \bibinfo{person}{Shayan
  Siddiqui}, \bibinfo{person}{Sebastian P{\"o}lsterl}, \bibinfo{person}{Nassir
  Navab}, {and} \bibinfo{person}{Christian Wachinger}.}
  \bibinfo{year}{2019}\natexlab{}.
\newblock \showarticletitle{BrainTorrent: A Peer-to-Peer Environment for
  Decentralized Federated Learning}.
\newblock \bibinfo{journal}{\emph{arXiv preprint arXiv:1905.06731}}
  (\bibinfo{year}{2019}).
\newblock


\bibitem[\protect\citeauthoryear{Ryffel, Trask, Dahl, Wagner, Mancuso,
  Rueckert, and Passerat-Palmbach}{Ryffel et~al\mbox{.}}{2018}]%
        {ryffel2018generic}
\bibfield{author}{\bibinfo{person}{Theo Ryffel}, \bibinfo{person}{Andrew
  Trask}, \bibinfo{person}{Morten Dahl}, \bibinfo{person}{Bobby Wagner},
  \bibinfo{person}{Jason Mancuso}, \bibinfo{person}{Daniel Rueckert}, {and}
  \bibinfo{person}{Jonathan Passerat-Palmbach}.}
  \bibinfo{year}{2018}\natexlab{}.
\newblock \bibinfo{title}{A generic framework for privacy preserving deep
  learning}.
\newblock
\newblock
\showeprint[arxiv]{1811.04017}~[cs.LG]


\bibitem[\protect\citeauthoryear{{Sarikaya} and {Ercetin}}{{Sarikaya} and
  {Ercetin}}{2019}]%
        {8867906}
\bibfield{author}{\bibinfo{person}{Y. {Sarikaya}} {and} \bibinfo{person}{O.
  {Ercetin}}.} \bibinfo{year}{2019}\natexlab{}.
\newblock \showarticletitle{Motivating Workers in Federated Learning: A
  Stackelberg Game Perspective}.
\newblock \bibinfo{journal}{\emph{IEEE Networking Letters}}
  (\bibinfo{year}{2019}), \bibinfo{pages}{1--1}.
\newblock
\showISSN{2576-3156}


\bibitem[\protect\citeauthoryear{Sattler, M{\"u}ller, and Samek}{Sattler
  et~al\mbox{.}}{2019}]%
        {sattler2019clustered}
\bibfield{author}{\bibinfo{person}{Felix Sattler},
  \bibinfo{person}{Klaus-Robert M{\"u}ller}, {and} \bibinfo{person}{Wojciech
  Samek}.} \bibinfo{year}{2019}\natexlab{}.
\newblock \showarticletitle{Clustered Federated Learning: Model-Agnostic
  Distributed Multi-Task Optimization under Privacy Constraints}.
\newblock \bibinfo{journal}{\emph{arXiv preprint arXiv:1910.01991}}
  (\bibinfo{year}{2019}).
\newblock


\bibitem[\protect\citeauthoryear{{Sattler}, {Wiedemann}, {Müller}, and
  {Samek}}{{Sattler} et~al\mbox{.}}{2019}]%
        {8889996}
\bibfield{author}{\bibinfo{person}{F. {Sattler}}, \bibinfo{person}{S.
  {Wiedemann}}, \bibinfo{person}{K. {Müller}}, {and} \bibinfo{person}{W.
  {Samek}}.} \bibinfo{year}{2019}\natexlab{}.
\newblock \showarticletitle{Robust and Communication-Efficient Federated
  Learning From Non-i.i.d. Data}.
\newblock \bibinfo{journal}{\emph{IEEE Transactions on Neural Networks and
  Learning Systems}} (\bibinfo{year}{2019}), \bibinfo{pages}{1--14}.
\newblock
\showISSN{2162-2388}


\bibitem[\protect\citeauthoryear{{Savazzi}, {Nicoli}, and {Rampa}}{{Savazzi}
  et~al\mbox{.}}{2020}]%
        {8950073}
\bibfield{author}{\bibinfo{person}{S. {Savazzi}}, \bibinfo{person}{M.
  {Nicoli}}, {and} \bibinfo{person}{V. {Rampa}}.}
  \bibinfo{year}{2020}\natexlab{}.
\newblock \showarticletitle{Federated Learning with Cooperating Devices: A
  Consensus Approach for Massive IoT Networks}.
\newblock \bibinfo{journal}{\emph{IEEE Internet of Things Journal}}
  (\bibinfo{year}{2020}), \bibinfo{pages}{1--1}.
\newblock
\showISSN{2372-2541}


\bibitem[\protect\citeauthoryear{Shayan, Fung, Yoon, and Beschastnikh}{Shayan
  et~al\mbox{.}}{2018}]%
        {shayan2018biscotti}
\bibfield{author}{\bibinfo{person}{Muhammad Shayan}, \bibinfo{person}{Clement
  Fung}, \bibinfo{person}{Chris J.~M. Yoon}, {and} \bibinfo{person}{Ivan
  Beschastnikh}.} \bibinfo{year}{2018}\natexlab{}.
\newblock \bibinfo{title}{Biscotti: A Ledger for Private and Secure
  Peer-to-Peer Machine Learning}.
\newblock
\newblock
\showeprint[arxiv]{1811.09904}~[cs.LG]


\bibitem[\protect\citeauthoryear{Sheller, Reina, Edwards, Martin, and
  Bakas}{Sheller et~al\mbox{.}}{2019}]%
        {Sheller2019}
\bibfield{author}{\bibinfo{person}{Micah~J. Sheller},
  \bibinfo{person}{G.~Anthony Reina}, \bibinfo{person}{Brandon Edwards},
  \bibinfo{person}{Jason Martin}, {and} \bibinfo{person}{Spyridon Bakas}.}
  \bibinfo{year}{2019}\natexlab{}.
\newblock \showarticletitle{Multi-institutional Deep Learning Modeling Without
  Sharing Patient Data: A Feasibility Study on Brain Tumor Segmentation}.
  \bibinfo{publisher}{Springer International Publishing},
  \bibinfo{address}{Cham}, \bibinfo{pages}{92--104}.
\newblock


\bibitem[\protect\citeauthoryear{Shen, Han, Wang, and Wang}{Shen
  et~al\mbox{.}}{2019}]%
        {Shen2019}
\bibfield{author}{\bibinfo{person}{Shihao Shen}, \bibinfo{person}{Yiwen Han},
  \bibinfo{person}{Xiaofei Wang}, {and} \bibinfo{person}{Yan Wang}.}
  \bibinfo{year}{2019}\natexlab{}.
\newblock \showarticletitle{Computation Offloading with Multiple Agents in
  Edge-Computing–Supported IoT}.
\newblock \bibinfo{journal}{\emph{ACM Trans. Sen. Netw.}} \bibinfo{volume}{16},
  \bibinfo{number}{1} (\bibinfo{year}{2019}), \bibinfo{pages}{Article 8}.
\newblock


\bibitem[\protect\citeauthoryear{Shi, Zhou, and Niu}{Shi et~al\mbox{.}}{2019}]%
        {shi2019device}
\bibfield{author}{\bibinfo{person}{Wenqi Shi}, \bibinfo{person}{Sheng Zhou},
  {and} \bibinfo{person}{Zhisheng Niu}.} \bibinfo{year}{2019}\natexlab{}.
\newblock \showarticletitle{Device Scheduling with Fast Convergence for
  Wireless Federated Learning}.
\newblock \bibinfo{journal}{\emph{arXiv preprint arXiv:1911.00856}}
  (\bibinfo{year}{2019}).
\newblock


\bibitem[\protect\citeauthoryear{{Silva}, {Gutman}, {Romero}, {Thompson},
  {Altmann}, and {Lorenzi}}{{Silva} et~al\mbox{.}}{2019}]%
        {8759317}
\bibfield{author}{\bibinfo{person}{S. {Silva}}, \bibinfo{person}{B.~A.
  {Gutman}}, \bibinfo{person}{E. {Romero}}, \bibinfo{person}{P.~M. {Thompson}},
  \bibinfo{person}{A. {Altmann}}, {and} \bibinfo{person}{M. {Lorenzi}}.}
  \bibinfo{year}{2019}\natexlab{}.
\newblock \showarticletitle{Federated Learning in Distributed Medical
  Databases: Meta-Analysis of Large-Scale Subcortical Brain Data}. In
  \bibinfo{booktitle}{\emph{2019 IEEE 16th International Symposium on
  Biomedical Imaging (ISBI 2019)}}. \bibinfo{pages}{270--274}.
\newblock
\showISSN{1945-7928}


\bibitem[\protect\citeauthoryear{Smith, Chiang, Sanjabi, and Talwalkar}{Smith
  et~al\mbox{.}}{2017}]%
        {Smith2017}
\bibfield{author}{\bibinfo{person}{Virginia Smith}, \bibinfo{person}{Chao-Kai
  Chiang}, \bibinfo{person}{Maziar Sanjabi}, {and} \bibinfo{person}{Ameet
  Talwalkar}.} \bibinfo{year}{2017}\natexlab{}.
\newblock \showarticletitle{Federated multi-task learning}. In
  \bibinfo{booktitle}{\emph{Proceedings of the 31st International Conference on
  Neural Information Processing Systems}}. \bibinfo{publisher}{Curran
  Associates Inc.}, \bibinfo{address}{Long Beach, California, USA},
  \bibinfo{pages}{4427–4437}.
\newblock


\bibitem[\protect\citeauthoryear{Song, Li, Huang, Wang, and Zeng}{Song
  et~al\mbox{.}}{2019}]%
        {Song2019b}
\bibfield{author}{\bibinfo{person}{Chunhe Song}, \bibinfo{person}{Tong Li},
  \bibinfo{person}{Xu Huang}, \bibinfo{person}{Zhongfeng Wang}, {and}
  \bibinfo{person}{Peng Zeng}.} \bibinfo{year}{2019}\natexlab{}.
\newblock \showarticletitle{Towards Edge Computing Based Distributed Data
  Analytics Framework in Smart Grids}. \bibinfo{publisher}{Springer
  International Publishing}, \bibinfo{address}{Cham},
  \bibinfo{pages}{283--292}.
\newblock


\bibitem[\protect\citeauthoryear{{Sozinov}, {Vlassov}, and
  {Girdzijauskas}}{{Sozinov} et~al\mbox{.}}{2018}]%
        {8672262}
\bibfield{author}{\bibinfo{person}{K. {Sozinov}}, \bibinfo{person}{V.
  {Vlassov}}, {and} \bibinfo{person}{S. {Girdzijauskas}}.}
  \bibinfo{year}{2018}\natexlab{}.
\newblock \showarticletitle{Human Activity Recognition Using Federated
  Learning}. In \bibinfo{booktitle}{\emph{2018 IEEE Intl Conf on Parallel
  Distributed Processing with Applications, Ubiquitous Computing
  Communications, Big Data Cloud Computing, Social Computing Networking,
  Sustainable Computing Communications
  (ISPA/IUCC/BDCloud/SocialCom/SustainCom)}}. \bibinfo{pages}{1103--1111}.
\newblock


\bibitem[\protect\citeauthoryear{Sun, Bommert, Pfisterer, R{\"a}henf{\"u}rher,
  Lang, and Bischl}{Sun et~al\mbox{.}}{2020}]%
        {10.1007/978-3-030-29516-5_48}
\bibfield{author}{\bibinfo{person}{Xudong Sun}, \bibinfo{person}{Andrea
  Bommert}, \bibinfo{person}{Florian Pfisterer}, \bibinfo{person}{J{\"o}rg
  R{\"a}henf{\"u}rher}, \bibinfo{person}{Michel Lang}, {and}
  \bibinfo{person}{Bernd Bischl}.} \bibinfo{year}{2020}\natexlab{}.
\newblock \showarticletitle{High Dimensional Restrictive Federated Model
  Selection with Multi-objective Bayesian Optimization over Shifted
  Distributions}. In \bibinfo{booktitle}{\emph{Intelligent Systems and
  Applications}}, \bibfield{editor}{\bibinfo{person}{Yaxin Bi},
  \bibinfo{person}{Rahul Bhatia}, {and} \bibinfo{person}{Supriya Kapoor}}
  (Eds.). \bibinfo{publisher}{Springer International Publishing},
  \bibinfo{address}{Cham}, \bibinfo{pages}{629--647}.
\newblock
\showISBNx{978-3-030-29516-5}


\bibitem[\protect\citeauthoryear{Sun, Zhou, and G{\"u}nd{\"u}z}{Sun
  et~al\mbox{.}}{2019}]%
        {sun2019energy}
\bibfield{author}{\bibinfo{person}{Yuxuan Sun}, \bibinfo{person}{Sheng Zhou},
  {and} \bibinfo{person}{Deniz G{\"u}nd{\"u}z}.}
  \bibinfo{year}{2019}\natexlab{}.
\newblock \showarticletitle{Energy-Aware Analog Aggregation for Federated
  Learning with Redundant Data}.
\newblock \bibinfo{journal}{\emph{arXiv preprint arXiv:1911.00188}}
  (\bibinfo{year}{2019}).
\newblock


\bibitem[\protect\citeauthoryear{Thomas, Abraham, and Liu}{Thomas
  et~al\mbox{.}}{2018}]%
        {thomas2018federated}
\bibfield{author}{\bibinfo{person}{Manoj~A Thomas},
  \bibinfo{person}{Diya~Suzanne Abraham}, {and} \bibinfo{person}{Dapeng Liu}.}
  \bibinfo{year}{2018}\natexlab{}.
\newblock \showarticletitle{Federated Machine Learning for Translational
  Research}.
\newblock \bibinfo{journal}{\emph{AMCIS2018}} (\bibinfo{year}{2018}).
\newblock


\bibitem[\protect\citeauthoryear{Triastcyn and Faltings}{Triastcyn and
  Faltings}{2019a}]%
        {Triastcyn2019}
\bibfield{author}{\bibinfo{person}{Aleksei Triastcyn} {and}
  \bibinfo{person}{Boi Faltings}.} \bibinfo{year}{2019}\natexlab{a}.
\newblock \bibinfo{title}{Federated Generative Privacy}.
\newblock
\newblock
\showeprint[arxiv]{1910.08385}~[stat.ML]


\bibitem[\protect\citeauthoryear{Triastcyn and Faltings}{Triastcyn and
  Faltings}{2019b}]%
        {triastcyn2019federated}
\bibfield{author}{\bibinfo{person}{Aleksei Triastcyn} {and}
  \bibinfo{person}{Boi Faltings}.} \bibinfo{year}{2019}\natexlab{b}.
\newblock \showarticletitle{Federated Learning with Bayesian Differential
  Privacy}.
\newblock \bibinfo{journal}{\emph{arXiv preprint arXiv:1911.10071}}
  (\bibinfo{year}{2019}).
\newblock


\bibitem[\protect\citeauthoryear{{Troglia}, {Melcher}, {Zheng}, {Anthony},
  {Yang}, and {Yang}}{{Troglia} et~al\mbox{.}}{2019}]%
        {8935736}
\bibfield{author}{\bibinfo{person}{M. {Troglia}}, \bibinfo{person}{J.
  {Melcher}}, \bibinfo{person}{Y. {Zheng}}, \bibinfo{person}{D. {Anthony}},
  \bibinfo{person}{A. {Yang}}, {and} \bibinfo{person}{T. {Yang}}.}
  \bibinfo{year}{2019}\natexlab{}.
\newblock \showarticletitle{FaIR: Federated Incumbent Detection in CBRS Band}.
  In \bibinfo{booktitle}{\emph{2019 IEEE International Symposium on Dynamic
  Spectrum Access Networks (DySPAN)}}. \bibinfo{pages}{1--6}.
\newblock
\showISSN{2334-3125}


\bibitem[\protect\citeauthoryear{Truex, Baracaldo, Anwar, Steinke, Ludwig,
  Zhang, and Zhou}{Truex et~al\mbox{.}}{2019}]%
        {Truex2019}
\bibfield{author}{\bibinfo{person}{Stacey Truex}, \bibinfo{person}{Nathalie
  Baracaldo}, \bibinfo{person}{Ali Anwar}, \bibinfo{person}{Thomas Steinke},
  \bibinfo{person}{Heiko Ludwig}, \bibinfo{person}{Rui Zhang}, {and}
  \bibinfo{person}{Yi Zhou}.} \bibinfo{year}{2019}\natexlab{}.
\newblock \showarticletitle{A Hybrid Approach to Privacy-Preserving Federated
  Learning}. In \bibinfo{booktitle}{\emph{Proceedings of the 12th ACM Workshop
  on Artificial Intelligence and Security}}. \bibinfo{publisher}{Association
  for Computing Machinery}, \bibinfo{address}{London, United Kingdom},
  \bibinfo{pages}{1–11}.
\newblock


\bibitem[\protect\citeauthoryear{Ulm, Gustavsson, and Jirstrand}{Ulm
  et~al\mbox{.}}{2019}]%
        {Ulm2019}
\bibfield{author}{\bibinfo{person}{Gregor Ulm}, \bibinfo{person}{Emil
  Gustavsson}, {and} \bibinfo{person}{Mats Jirstrand}.}
  \bibinfo{year}{2019}\natexlab{}.
\newblock \showarticletitle{Functional Federated Learning in Erlang (ffl-erl)}.
  \bibinfo{publisher}{Springer International Publishing},
  \bibinfo{address}{Cham}, \bibinfo{pages}{162--178}.
\newblock


\bibitem[\protect\citeauthoryear{{Verma}, {White}, and {de Mel}}{{Verma}
  et~al\mbox{.}}{2019}]%
        {8818446}
\bibfield{author}{\bibinfo{person}{D. {Verma}}, \bibinfo{person}{G. {White}},
  {and} \bibinfo{person}{G. {de Mel}}.} \bibinfo{year}{2019}\natexlab{}.
\newblock \showarticletitle{Federated AI for the Enterprise: A Web Services
  Based Implementation}. In \bibinfo{booktitle}{\emph{2019 IEEE International
  Conference on Web Services (ICWS)}}. \bibinfo{pages}{20--27}.
\newblock


\bibitem[\protect\citeauthoryear{Verma, White, Julier, Pasteris, Chakraborty,
  and Cirincione}{Verma et~al\mbox{.}}{2019}]%
        {verma2019approaches}
\bibfield{author}{\bibinfo{person}{Dinesh~C Verma}, \bibinfo{person}{Graham
  White}, \bibinfo{person}{Simon Julier}, \bibinfo{person}{Stepehen Pasteris},
  \bibinfo{person}{Supriyo Chakraborty}, {and} \bibinfo{person}{Greg
  Cirincione}.} \bibinfo{year}{2019}\natexlab{}.
\newblock \showarticletitle{Approaches to address the data skew problem in
  federated learning}. In \bibinfo{booktitle}{\emph{Artificial Intelligence and
  Machine Learning for Multi-Domain Operations Applications}},
  Vol.~\bibinfo{volume}{11006}. International Society for Optics and Photonics,
  \bibinfo{pages}{110061I}.
\newblock


\bibitem[\protect\citeauthoryear{Vu, Ngo, Tran, Ngo, Dao, and Middleton}{Vu
  et~al\mbox{.}}{2019}]%
        {vu2019cell}
\bibfield{author}{\bibinfo{person}{Tung~T Vu}, \bibinfo{person}{Duy~T Ngo},
  \bibinfo{person}{Nguyen~H Tran}, \bibinfo{person}{Hien~Quoc Ngo},
  \bibinfo{person}{Minh~N Dao}, {and} \bibinfo{person}{Richard~H Middleton}.}
  \bibinfo{year}{2019}\natexlab{}.
\newblock \showarticletitle{Cell-Free Massive MIMO for Wireless Federated
  Learning}.
\newblock \bibinfo{journal}{\emph{arXiv preprint arXiv:1909.12567}}
  (\bibinfo{year}{2019}).
\newblock


\bibitem[\protect\citeauthoryear{{Wan}, {Xia}, {Lo}, and {Murphy}}{{Wan}
  et~al\mbox{.}}{2019}]%
        {8812912}
\bibfield{author}{\bibinfo{person}{Z. {Wan}}, \bibinfo{person}{X. {Xia}},
  \bibinfo{person}{D. {Lo}}, {and} \bibinfo{person}{G.~C. {Murphy}}.}
  \bibinfo{year}{2019}\natexlab{}.
\newblock \showarticletitle{How does Machine Learning Change Software
  Development Practices?}
\newblock \bibinfo{journal}{\emph{IEEE Transactions on Software Engineering}}
  (\bibinfo{year}{2019}), \bibinfo{pages}{1--1}.
\newblock


\bibitem[\protect\citeauthoryear{Wang}{Wang}{2019}]%
        {wang2019interpret}
\bibfield{author}{\bibinfo{person}{Guan Wang}.}
  \bibinfo{year}{2019}\natexlab{}.
\newblock \showarticletitle{Interpret Federated Learning with Shapley Values}.
\newblock \bibinfo{journal}{\emph{arXiv preprint arXiv:1905.04519}}
  (\bibinfo{year}{2019}).
\newblock


\bibitem[\protect\citeauthoryear{Wang, Dang, and Zhou}{Wang
  et~al\mbox{.}}{2019a}]%
        {wang2019measure}
\bibfield{author}{\bibinfo{person}{Guan Wang},
  \bibinfo{person}{Charlie~Xiaoqian Dang}, {and} \bibinfo{person}{Ziye Zhou}.}
  \bibinfo{year}{2019}\natexlab{a}.
\newblock \showarticletitle{Measure Contribution of Participants in Federated
  Learning}.
\newblock \bibinfo{journal}{\emph{arXiv preprint arXiv:1909.08525}}
  (\bibinfo{year}{2019}).
\newblock


\bibitem[\protect\citeauthoryear{Wang, Mathews, Kiddon, Eichner, Beaufays, and
  Ramage}{Wang et~al\mbox{.}}{2019b}]%
        {wang2019federated}
\bibfield{author}{\bibinfo{person}{Kangkang Wang}, \bibinfo{person}{Rajiv
  Mathews}, \bibinfo{person}{Chloé Kiddon}, \bibinfo{person}{Hubert Eichner},
  \bibinfo{person}{Françoise Beaufays}, {and} \bibinfo{person}{Daniel
  Ramage}.} \bibinfo{year}{2019}\natexlab{b}.
\newblock \bibinfo{title}{Federated Evaluation of On-device Personalization}.
\newblock
\newblock
\showeprint[arxiv]{1910.10252}~[cs.LG]


\bibitem[\protect\citeauthoryear{{WANG}, {WANG}, and {LI}}{{WANG}
  et~al\mbox{.}}{2019}]%
        {8885054}
\bibfield{author}{\bibinfo{person}{L. {WANG}}, \bibinfo{person}{W. {WANG}},
  {and} \bibinfo{person}{B. {LI}}.} \bibinfo{year}{2019}\natexlab{}.
\newblock \showarticletitle{CMFL: Mitigating Communication Overhead for
  Federated Learning}. In \bibinfo{booktitle}{\emph{2019 IEEE 39th
  International Conference on Distributed Computing Systems (ICDCS)}}.
  \bibinfo{pages}{954--964}.
\newblock
\showISSN{1063-6927}


\bibitem[\protect\citeauthoryear{{Wang}, {Tuor}, {Salonidis}, {Leung},
  {Makaya}, {He}, and {Chan}}{{Wang} et~al\mbox{.}}{2019}]%
        {8664630}
\bibfield{author}{\bibinfo{person}{S. {Wang}}, \bibinfo{person}{T. {Tuor}},
  \bibinfo{person}{T. {Salonidis}}, \bibinfo{person}{K.~K. {Leung}},
  \bibinfo{person}{C. {Makaya}}, \bibinfo{person}{T. {He}}, {and}
  \bibinfo{person}{K. {Chan}}.} \bibinfo{year}{2019}\natexlab{}.
\newblock \showarticletitle{Adaptive Federated Learning in Resource Constrained
  Edge Computing Systems}.
\newblock \bibinfo{journal}{\emph{IEEE Journal on Selected Areas in
  Communications}} \bibinfo{volume}{37}, \bibinfo{number}{6}
  (\bibinfo{date}{June} \bibinfo{year}{2019}), \bibinfo{pages}{1205--1221}.
\newblock
\showISSN{1558-0008}


\bibitem[\protect\citeauthoryear{Wei, Li, Ding, Ma, Yang, Farhad, Jin, Quek,
  and Poor}{Wei et~al\mbox{.}}{2019a}]%
        {wei2019federated}
\bibfield{author}{\bibinfo{person}{Kang Wei}, \bibinfo{person}{Jun Li},
  \bibinfo{person}{Ming Ding}, \bibinfo{person}{Chuan Ma},
  \bibinfo{person}{Howard~H. Yang}, \bibinfo{person}{Farokhi Farhad},
  \bibinfo{person}{Shi Jin}, \bibinfo{person}{Tony Q.~S. Quek}, {and}
  \bibinfo{person}{H.~Vincent Poor}.} \bibinfo{year}{2019}\natexlab{a}.
\newblock \bibinfo{title}{Federated Learning with Differential Privacy:
  Algorithms and Performance Analysis}.
\newblock
\newblock
\showeprint[arxiv]{1911.00222}~[cs.LG]


\bibitem[\protect\citeauthoryear{Wei, Li, Liu, Yu, Chen, and Yang}{Wei
  et~al\mbox{.}}{2019b}]%
        {wei2019multi}
\bibfield{author}{\bibinfo{person}{Xiguang Wei}, \bibinfo{person}{Quan Li},
  \bibinfo{person}{Yang Liu}, \bibinfo{person}{Han Yu},
  \bibinfo{person}{Tianjian Chen}, {and} \bibinfo{person}{Qiang Yang}.}
  \bibinfo{year}{2019}\natexlab{b}.
\newblock \showarticletitle{Multi-agent visualization for explaining federated
  learning}. In \bibinfo{booktitle}{\emph{Proceedings of the 28th International
  Joint Conference on Artificial Intelligence}}. AAAI Press,
  \bibinfo{pages}{6572--6574}.
\newblock


\bibitem[\protect\citeauthoryear{{Weng}, {Weng}, {Zhang}, {Li}, {Zhang}, and
  {Luo}}{{Weng} et~al\mbox{.}}{2019}]%
        {8894364}
\bibfield{author}{\bibinfo{person}{J. {Weng}}, \bibinfo{person}{J. {Weng}},
  \bibinfo{person}{J. {Zhang}}, \bibinfo{person}{M. {Li}}, \bibinfo{person}{Y.
  {Zhang}}, {and} \bibinfo{person}{W. {Luo}}.} \bibinfo{year}{2019}\natexlab{}.
\newblock \showarticletitle{DeepChain: Auditable and Privacy-Preserving Deep
  Learning with Blockchain-based Incentive}.
\newblock \bibinfo{journal}{\emph{IEEE Transactions on Dependable and Secure
  Computing}} (\bibinfo{year}{2019}), \bibinfo{pages}{1--1}.
\newblock
\showISSN{2160-9209}


\bibitem[\protect\citeauthoryear{Wieringa, Maiden, Mead, and Rolland}{Wieringa
  et~al\mbox{.}}{2005}]%
        {10.1007/s00766-005-0021-6}
\bibfield{author}{\bibinfo{person}{Roel Wieringa}, \bibinfo{person}{Neil
  Maiden}, \bibinfo{person}{Nancy Mead}, {and} \bibinfo{person}{Colette
  Rolland}.} \bibinfo{year}{2005}\natexlab{}.
\newblock \showarticletitle{Requirements Engineering Paper Classification and
  Evaluation Criteria: A Proposal and a Discussion}.
\newblock \bibinfo{journal}{\emph{Requir. Eng.}} \bibinfo{volume}{11},
  \bibinfo{number}{1} (\bibinfo{date}{Dec.} \bibinfo{year}{2005}),
  \bibinfo{pages}{102–107}.
\newblock
\showISSN{0947-3602}
\urldef\tempurl%
\url{https://doi.org/10.1007/s00766-005-0021-6}
\showURL{%
\tempurl}


\bibitem[\protect\citeauthoryear{Wu, He, Lin, Jarvis, et~al\mbox{.}}{Wu
  et~al\mbox{.}}{2019}]%
        {wu2019safa}
\bibfield{author}{\bibinfo{person}{Wentai Wu}, \bibinfo{person}{Ligang He},
  \bibinfo{person}{Weiwei Lin}, \bibinfo{person}{Stephen Jarvis},
  {et~al\mbox{.}}} \bibinfo{year}{2019}\natexlab{}.
\newblock \showarticletitle{SAFA: a Semi-Asynchronous Protocol for Fast
  Federated Learning with Low Overhead}.
\newblock \bibinfo{journal}{\emph{arXiv preprint arXiv:1910.01355}}
  (\bibinfo{year}{2019}).
\newblock


\bibitem[\protect\citeauthoryear{Xie, Koyejo, and Gupta}{Xie
  et~al\mbox{.}}{2019}]%
        {xie2019asynchronous}
\bibfield{author}{\bibinfo{person}{Cong Xie}, \bibinfo{person}{Sanmi Koyejo},
  {and} \bibinfo{person}{Indranil Gupta}.} \bibinfo{year}{2019}\natexlab{}.
\newblock \bibinfo{title}{Asynchronous Federated Optimization}.
\newblock
\newblock
\showeprint[arxiv]{1903.03934}~[cs.DC]


\bibitem[\protect\citeauthoryear{{Xu}, {Li}, {Liu}, {Yang}, and {Lin}}{{Xu}
  et~al\mbox{.}}{2020}]%
        {8765347}
\bibfield{author}{\bibinfo{person}{G. {Xu}}, \bibinfo{person}{H. {Li}},
  \bibinfo{person}{S. {Liu}}, \bibinfo{person}{K. {Yang}}, {and}
  \bibinfo{person}{X. {Lin}}.} \bibinfo{year}{2020}\natexlab{}.
\newblock \showarticletitle{VerifyNet: Secure and Verifiable Federated
  Learning}.
\newblock \bibinfo{journal}{\emph{IEEE Transactions on Information Forensics
  and Security}}  \bibinfo{volume}{15} (\bibinfo{year}{2020}),
  \bibinfo{pages}{911--926}.
\newblock
\showISSN{1556-6021}


\bibitem[\protect\citeauthoryear{Xu and Wang}{Xu and Wang}{2019}]%
        {xu2019federated}
\bibfield{author}{\bibinfo{person}{Jie Xu} {and} \bibinfo{person}{Fei Wang}.}
  \bibinfo{year}{2019}\natexlab{}.
\newblock \bibinfo{title}{Federated Learning for Healthcare Informatics}.
\newblock
\newblock
\showeprint[arxiv]{1911.06270}~[cs.LG]


\bibitem[\protect\citeauthoryear{Xu, Qian, Mei, Huang, and Liu}{Xu
  et~al\mbox{.}}{2018}]%
        {Xu2018}
\bibfield{author}{\bibinfo{person}{Mengwei Xu}, \bibinfo{person}{Feng Qian},
  \bibinfo{person}{Qiaozhu Mei}, \bibinfo{person}{Kang Huang}, {and}
  \bibinfo{person}{Xuanzhe Liu}.} \bibinfo{year}{2018}\natexlab{}.
\newblock \showarticletitle{DeepType: On-Device Deep Learning for Input
  Personalization Service with Minimal Privacy Concern}.
\newblock \bibinfo{journal}{\emph{Proc. ACM Interact. Mob. Wearable Ubiquitous
  Technol.}} \bibinfo{volume}{2}, \bibinfo{number}{4} (\bibinfo{year}{2018}),
  \bibinfo{pages}{Article 197}.
\newblock


\bibitem[\protect\citeauthoryear{Xu, Baracaldo, Zhou, Anwar, and Ludwig}{Xu
  et~al\mbox{.}}{2019a}]%
        {Xu2019a}
\bibfield{author}{\bibinfo{person}{Runhua Xu}, \bibinfo{person}{Nathalie
  Baracaldo}, \bibinfo{person}{Yi Zhou}, \bibinfo{person}{Ali Anwar}, {and}
  \bibinfo{person}{Heiko Ludwig}.} \bibinfo{year}{2019}\natexlab{a}.
\newblock \showarticletitle{HybridAlpha: An Efficient Approach for
  Privacy-Preserving Federated Learning}. In
  \bibinfo{booktitle}{\emph{Proceedings of the 12th ACM Workshop on Artificial
  Intelligence and Security}}. \bibinfo{publisher}{Association for Computing
  Machinery}, \bibinfo{address}{London, United Kingdom},
  \bibinfo{pages}{13–23}.
\newblock


\bibitem[\protect\citeauthoryear{Xu, Li, and Zou}{Xu et~al\mbox{.}}{2019b}]%
        {Xu2019}
\bibfield{author}{\bibinfo{person}{Zichen Xu}, \bibinfo{person}{Li Li}, {and}
  \bibinfo{person}{Wenting Zou}.} \bibinfo{year}{2019}\natexlab{b}.
\newblock \showarticletitle{Exploring federated learning on battery-powered
  devices}. In \bibinfo{booktitle}{\emph{Proceedings of the ACM Turing
  Celebration Conference - China}}. \bibinfo{publisher}{Association for
  Computing Machinery}, \bibinfo{address}{Chengdu, China},
  \bibinfo{pages}{Article 6}.
\newblock


\bibitem[\protect\citeauthoryear{Yang, Arafa, Quek, and Poor}{Yang
  et~al\mbox{.}}{2019a}]%
        {yang2019age}
\bibfield{author}{\bibinfo{person}{Howard~H Yang}, \bibinfo{person}{Ahmed
  Arafa}, \bibinfo{person}{Tony~QS Quek}, {and} \bibinfo{person}{H~Vincent
  Poor}.} \bibinfo{year}{2019}\natexlab{a}.
\newblock \showarticletitle{Age-Based Scheduling Policy for Federated Learning
  in Mobile Edge Networks}.
\newblock \bibinfo{journal}{\emph{arXiv preprint arXiv:1910.14648}}
  (\bibinfo{year}{2019}).
\newblock


\bibitem[\protect\citeauthoryear{Yang, Fan, Chen, Shi, and Yang}{Yang
  et~al\mbox{.}}{2019c}]%
        {yang2019quasi}
\bibfield{author}{\bibinfo{person}{Kai Yang}, \bibinfo{person}{Tao Fan},
  \bibinfo{person}{Tianjian Chen}, \bibinfo{person}{Yuanming Shi}, {and}
  \bibinfo{person}{Qiang Yang}.} \bibinfo{year}{2019}\natexlab{c}.
\newblock \showarticletitle{A Quasi-Newton Method Based Vertical Federated
  Learning Framework for Logistic Regression}.
\newblock \bibinfo{journal}{\emph{arXiv preprint arXiv:1912.00513}}
  (\bibinfo{year}{2019}).
\newblock


\bibitem[\protect\citeauthoryear{{Yang}, {Jiang}, {Shi}, and {Ding}}{{Yang}
  et~al\mbox{.}}{2020}]%
        {8952884}
\bibfield{author}{\bibinfo{person}{K. {Yang}}, \bibinfo{person}{T. {Jiang}},
  \bibinfo{person}{Y. {Shi}}, {and} \bibinfo{person}{Z. {Ding}}.}
  \bibinfo{year}{2020}\natexlab{}.
\newblock \showarticletitle{Federated Learning via Over-the-Air Computation}.
\newblock \bibinfo{journal}{\emph{IEEE Transactions on Wireless
  Communications}} (\bibinfo{year}{2020}), \bibinfo{pages}{1--1}.
\newblock
\showISSN{1558-2248}


\bibitem[\protect\citeauthoryear{Yang, Liu, Chen, and Tong}{Yang
  et~al\mbox{.}}{2019d}]%
        {10.1145/3298981}
\bibfield{author}{\bibinfo{person}{Qiang Yang}, \bibinfo{person}{Yang Liu},
  \bibinfo{person}{Tianjian Chen}, {and} \bibinfo{person}{Yongxin Tong}.}
  \bibinfo{year}{2019}\natexlab{d}.
\newblock \showarticletitle{Federated Machine Learning: Concept and
  Applications}.
\newblock \bibinfo{journal}{\emph{ACM Trans. Intell. Syst. Technol.}}
  \bibinfo{volume}{10}, \bibinfo{number}{2}, Article \bibinfo{articleno}{12}
  (\bibinfo{date}{Jan.} \bibinfo{year}{2019}), \bibinfo{numpages}{19}~pages.
\newblock
\showISSN{2157-6904}
\urldef\tempurl%
\url{https://doi.org/10.1145/3298981}
\showURL{%
\tempurl}


\bibitem[\protect\citeauthoryear{Yang, Ren, Zhou, and Liu}{Yang
  et~al\mbox{.}}{2019e}]%
        {yang2019parallel}
\bibfield{author}{\bibinfo{person}{Shengwen Yang}, \bibinfo{person}{Bing Ren},
  \bibinfo{person}{Xuhui Zhou}, {and} \bibinfo{person}{Liping Liu}.}
  \bibinfo{year}{2019}\natexlab{e}.
\newblock \showarticletitle{Parallel Distributed Logistic Regression for
  Vertical Federated Learning without Third-Party Coordinator}.
\newblock \bibinfo{journal}{\emph{arXiv preprint arXiv:1911.09824}}
  (\bibinfo{year}{2019}).
\newblock


\bibitem[\protect\citeauthoryear{Yang, Andrew, Eichner, Sun, Li, Kong, Ramage,
  and Beaufays}{Yang et~al\mbox{.}}{2018}]%
        {yang2018applied}
\bibfield{author}{\bibinfo{person}{Timothy Yang}, \bibinfo{person}{Galen
  Andrew}, \bibinfo{person}{Hubert Eichner}, \bibinfo{person}{Haicheng Sun},
  \bibinfo{person}{Wei Li}, \bibinfo{person}{Nicholas Kong},
  \bibinfo{person}{Daniel Ramage}, {and} \bibinfo{person}{Fran{\c{c}}oise
  Beaufays}.} \bibinfo{year}{2018}\natexlab{}.
\newblock \showarticletitle{Applied federated learning: Improving google
  keyboard query suggestions}.
\newblock \bibinfo{journal}{\emph{arXiv preprint arXiv:1812.02903}}
  (\bibinfo{year}{2018}).
\newblock


\bibitem[\protect\citeauthoryear{Yang, Zhang, Ye, Li, and Xu}{Yang
  et~al\mbox{.}}{2019f}]%
        {Yang2019}
\bibfield{author}{\bibinfo{person}{Wensi Yang}, \bibinfo{person}{Yuhang Zhang},
  \bibinfo{person}{Kejiang Ye}, \bibinfo{person}{Li Li}, {and}
  \bibinfo{person}{Cheng-Zhong Xu}.} \bibinfo{year}{2019}\natexlab{f}.
\newblock \showarticletitle{FFD: A Federated Learning Based Method for Credit
  Card Fraud Detection}. \bibinfo{publisher}{Springer International
  Publishing}, \bibinfo{address}{Cham}, \bibinfo{pages}{18--32}.
\newblock


\bibitem[\protect\citeauthoryear{Yang, Chen, Saad, Hong, and Shikh-Bahaei}{Yang
  et~al\mbox{.}}{2019b}]%
        {yang2019energy}
\bibfield{author}{\bibinfo{person}{Zhaohui Yang}, \bibinfo{person}{Mingzhe
  Chen}, \bibinfo{person}{Walid Saad}, \bibinfo{person}{Choong~Seon Hong},
  {and} \bibinfo{person}{Mohammad Shikh-Bahaei}.}
  \bibinfo{year}{2019}\natexlab{b}.
\newblock \showarticletitle{Energy Efficient Federated Learning Over Wireless
  Communication Networks}.
\newblock \bibinfo{journal}{\emph{arXiv preprint arXiv:1911.02417}}
  (\bibinfo{year}{2019}).
\newblock


\bibitem[\protect\citeauthoryear{{Yao}, {Huang}, and {Sun}}{{Yao}
  et~al\mbox{.}}{2018}]%
        {8698609}
\bibfield{author}{\bibinfo{person}{X. {Yao}}, \bibinfo{person}{C. {Huang}},
  {and} \bibinfo{person}{L. {Sun}}.} \bibinfo{year}{2018}\natexlab{}.
\newblock \showarticletitle{Two-Stream Federated Learning: Reduce the
  Communication Costs}. In \bibinfo{booktitle}{\emph{2018 IEEE Visual
  Communications and Image Processing (VCIP)}}. \bibinfo{pages}{1--4}.
\newblock
\showISSN{1018-8770}


\bibitem[\protect\citeauthoryear{{Yao}, {Huang}, {Wu}, {Zhang}, and
  {Sun}}{{Yao} et~al\mbox{.}}{2019}]%
        {8803001}
\bibfield{author}{\bibinfo{person}{X. {Yao}}, \bibinfo{person}{T. {Huang}},
  \bibinfo{person}{C. {Wu}}, \bibinfo{person}{R. {Zhang}}, {and}
  \bibinfo{person}{L. {Sun}}.} \bibinfo{year}{2019}\natexlab{}.
\newblock \showarticletitle{Towards Faster and Better Federated Learning: A
  Feature Fusion Approach}. In \bibinfo{booktitle}{\emph{2019 IEEE
  International Conference on Image Processing (ICIP)}}.
  \bibinfo{pages}{175--179}.
\newblock
\showISSN{1522-4880}


\bibitem[\protect\citeauthoryear{Ye, Yu, Pan, and Han}{Ye
  et~al\mbox{.}}{2020}]%
        {Ye2020}
\bibfield{author}{\bibinfo{person}{D. Ye}, \bibinfo{person}{R. Yu},
  \bibinfo{person}{M. Pan}, {and} \bibinfo{person}{Z. Han}.}
  \bibinfo{year}{2020}\natexlab{}.
\newblock \showarticletitle{Federated Learning in Vehicular Edge Computing: A
  Selective Model Aggregation Approach}.
\newblock \bibinfo{journal}{\emph{IEEE Access}} (\bibinfo{year}{2020}),
  \bibinfo{pages}{1--1}.
\newblock
\showISSN{2169-3536}


\bibitem[\protect\citeauthoryear{Yin, Yin, Wu, and Jiang}{Yin
  et~al\mbox{.}}{2020}]%
        {Yin2020}
\bibfield{author}{\bibinfo{person}{B. Yin}, \bibinfo{person}{H. Yin},
  \bibinfo{person}{Y. Wu}, {and} \bibinfo{person}{Z. Jiang}.}
  \bibinfo{year}{2020}\natexlab{}.
\newblock \showarticletitle{FDC: A Secure Federated Deep Learning Mechanism for
  Data Collaborations in the Internet of Things}.
\newblock \bibinfo{journal}{\emph{IEEE Internet of Things Journal}}
  (\bibinfo{year}{2020}), \bibinfo{pages}{1--1}.
\newblock
\showISSN{2327-4662}


\bibitem[\protect\citeauthoryear{{Yu}, {Hu}, {Min}, {Lu}, {Zhao}, {Wang}, and
  {Georgalas}}{{Yu} et~al\mbox{.}}{2018}]%
        {8647616}
\bibfield{author}{\bibinfo{person}{Z. {Yu}}, \bibinfo{person}{J. {Hu}},
  \bibinfo{person}{G. {Min}}, \bibinfo{person}{H. {Lu}}, \bibinfo{person}{Z.
  {Zhao}}, \bibinfo{person}{H. {Wang}}, {and} \bibinfo{person}{N.
  {Georgalas}}.} \bibinfo{year}{2018}\natexlab{}.
\newblock \showarticletitle{Federated Learning Based Proactive Content Caching
  in Edge Computing}. In \bibinfo{booktitle}{\emph{2018 IEEE Global
  Communications Conference (GLOBECOM)}}. \bibinfo{pages}{1--6}.
\newblock
\showISSN{1930-529X}


\bibitem[\protect\citeauthoryear{Yufeng~Zhan}{Yufeng~Zhan}{2020}]%
        {Zhan2020ExperienceDrivenCR}
\bibfield{author}{\bibinfo{person}{Song~Guo Yufeng~Zhan, Peng~Li}.}
  \bibinfo{year}{2020}\natexlab{}.
\newblock \showarticletitle{Experience-Driven Computational Resource Allocation
  of Federated Learning by Deep Reinforcement Learning}.
\newblock \bibinfo{journal}{\emph{IEEE IPDPS}} (\bibinfo{year}{2020}).
\newblock


\bibitem[\protect\citeauthoryear{Yurochkin, Agarwal, Ghosh, Greenewald, Hoang,
  and Khazaeni}{Yurochkin et~al\mbox{.}}{2019}]%
        {yurochkin2019bayesian}
\bibfield{author}{\bibinfo{person}{Mikhail Yurochkin}, \bibinfo{person}{Mayank
  Agarwal}, \bibinfo{person}{Soumya Ghosh}, \bibinfo{person}{Kristjan
  Greenewald}, \bibinfo{person}{Trong~Nghia Hoang}, {and}
  \bibinfo{person}{Yasaman Khazaeni}.} \bibinfo{year}{2019}\natexlab{}.
\newblock \showarticletitle{Bayesian nonparametric federated learning of neural
  networks}.
\newblock \bibinfo{journal}{\emph{arXiv preprint arXiv:1905.12022}}
  (\bibinfo{year}{2019}).
\newblock


\bibitem[\protect\citeauthoryear{Zeng, Du, Leung, and Huang}{Zeng
  et~al\mbox{.}}{2019}]%
        {DBLP:journals/corr/abs-1907-06040}
\bibfield{author}{\bibinfo{person}{Qunsong Zeng}, \bibinfo{person}{Yuqing Du},
  \bibinfo{person}{Kin~K. Leung}, {and} \bibinfo{person}{Kaibin Huang}.}
  \bibinfo{year}{2019}\natexlab{}.
\newblock \showarticletitle{Energy-Efficient Radio Resource Allocation for
  Federated Edge Learning}.
\newblock \bibinfo{journal}{\emph{CoRR}}  \bibinfo{volume}{abs/1907.06040}
  (\bibinfo{year}{2019}).
\newblock
\showeprint[arxiv]{1907.06040}
\urldef\tempurl%
\url{http://arxiv.org/abs/1907.06040}
\showURL{%
\tempurl}


\bibitem[\protect\citeauthoryear{Zhan, Li, Qu, Zeng, and Guo}{Zhan
  et~al\mbox{.}}{2020}]%
        {Zhan2020}
\bibfield{author}{\bibinfo{person}{Y. Zhan}, \bibinfo{person}{P. Li},
  \bibinfo{person}{Z. Qu}, \bibinfo{person}{D. Zeng}, {and} \bibinfo{person}{S.
  Guo}.} \bibinfo{year}{2020}\natexlab{}.
\newblock \showarticletitle{A Learning-based Incentive Mechanism for Federated
  Learning}.
\newblock \bibinfo{journal}{\emph{IEEE Internet of Things Journal}}
  (\bibinfo{year}{2020}), \bibinfo{pages}{1--1}.
\newblock
\showISSN{2372-2541}


\bibitem[\protect\citeauthoryear{Zhang, Wang, Zhao, and Chen}{Zhang
  et~al\mbox{.}}{2019}]%
        {Zhang2019a}
\bibfield{author}{\bibinfo{person}{Jiale Zhang}, \bibinfo{person}{Junyu Wang},
  \bibinfo{person}{Yanchao Zhao}, {and} \bibinfo{person}{Bing Chen}.}
  \bibinfo{year}{2019}\natexlab{}.
\newblock \showarticletitle{An Efficient Federated Learning Scheme with
  Differential Privacy in Mobile Edge Computing}. \bibinfo{publisher}{Springer
  International Publishing}, \bibinfo{address}{Cham},
  \bibinfo{pages}{538--550}.
\newblock


\bibitem[\protect\citeauthoryear{{Zhang}, {Chen}, {Liu}, and {Xiang}}{{Zhang}
  et~al\mbox{.}}{2019a}]%
        {8836609}
\bibfield{author}{\bibinfo{person}{X. {Zhang}}, \bibinfo{person}{X. {Chen}},
  \bibinfo{person}{J. {Liu}}, {and} \bibinfo{person}{Y. {Xiang}}.}
  \bibinfo{year}{2019}\natexlab{a}.
\newblock \showarticletitle{DeepPAR and DeepDPA: Privacy-Preserving and
  Asynchronous Deep Learning for Industrial IoT}.
\newblock \bibinfo{journal}{\emph{IEEE Transactions on Industrial Informatics}}
  (\bibinfo{year}{2019}), \bibinfo{pages}{1--1}.
\newblock
\showISSN{1941-0050}


\bibitem[\protect\citeauthoryear{{Zhang}, {Peng}, {Yan}, and {Sun}}{{Zhang}
  et~al\mbox{.}}{2019b}]%
        {8944302}
\bibfield{author}{\bibinfo{person}{X. {Zhang}}, \bibinfo{person}{M. {Peng}},
  \bibinfo{person}{S. {Yan}}, {and} \bibinfo{person}{Y. {Sun}}.}
  \bibinfo{year}{2019}\natexlab{b}.
\newblock \showarticletitle{Deep Reinforcement Learning Based Mode Selection
  and Resource Allocation for Cellular V2X Communications}.
\newblock \bibinfo{journal}{\emph{IEEE Internet of Things Journal}}
  (\bibinfo{year}{2019}), \bibinfo{pages}{1--1}.
\newblock
\showISSN{2372-2541}


\bibitem[\protect\citeauthoryear{Zhao, Chen, Wu, Teng, and Yu}{Zhao
  et~al\mbox{.}}{2019a}]%
        {Zhao2019}
\bibfield{author}{\bibinfo{person}{Ying Zhao}, \bibinfo{person}{Junjun Chen},
  \bibinfo{person}{Di Wu}, \bibinfo{person}{Jian Teng}, {and}
  \bibinfo{person}{Shui Yu}.} \bibinfo{year}{2019}\natexlab{a}.
\newblock \showarticletitle{Multi-Task Network Anomaly Detection using
  Federated Learning}. In \bibinfo{booktitle}{\emph{Proceedings of the Tenth
  International Symposium on Information and Communication Technology}}.
  \bibinfo{publisher}{Association for Computing Machinery},
  \bibinfo{address}{Hanoi, Ha Long Bay, Viet Nam}, \bibinfo{pages}{273–279}.
\newblock


\bibitem[\protect\citeauthoryear{Zhao, Chen, Zhang, Wu, Teng, and Yu}{Zhao
  et~al\mbox{.}}{2020}]%
        {Zhao2020}
\bibfield{author}{\bibinfo{person}{Ying Zhao}, \bibinfo{person}{Junjun Chen},
  \bibinfo{person}{Jiale Zhang}, \bibinfo{person}{Di Wu}, \bibinfo{person}{Jian
  Teng}, {and} \bibinfo{person}{Shui Yu}.} \bibinfo{year}{2020}\natexlab{}.
\newblock \showarticletitle{PDGAN: A Novel Poisoning Defense Method in
  Federated Learning Using Generative Adversarial Network}. In
  \bibinfo{booktitle}{\emph{Algorithms and Architectures for Parallel
  Processing}}, \bibfield{editor}{\bibinfo{person}{Sheng Wen},
  \bibinfo{person}{Albert Zomaya}, {and} \bibinfo{person}{Laurence~T. Yang}}
  (Eds.). \bibinfo{publisher}{Springer International Publishing},
  \bibinfo{address}{Cham}, \bibinfo{pages}{595--609}.
\newblock
\showISSN{978-3-030-38991-8}


\bibitem[\protect\citeauthoryear{Zhao, Li, Lai, Suda, Civin, and Chandra}{Zhao
  et~al\mbox{.}}{2018}]%
        {zhao2018federated}
\bibfield{author}{\bibinfo{person}{Yue Zhao}, \bibinfo{person}{Meng Li},
  \bibinfo{person}{Liangzhen Lai}, \bibinfo{person}{Naveen Suda},
  \bibinfo{person}{Damon Civin}, {and} \bibinfo{person}{Vikas Chandra}.}
  \bibinfo{year}{2018}\natexlab{}.
\newblock \showarticletitle{Federated learning with non-iid data}.
\newblock \bibinfo{journal}{\emph{arXiv preprint arXiv:1806.00582}}
  (\bibinfo{year}{2018}).
\newblock


\bibitem[\protect\citeauthoryear{Zhao, Zhao, Jiang, Tan, and Niyato}{Zhao
  et~al\mbox{.}}{2019b}]%
        {zhao2019mobile}
\bibfield{author}{\bibinfo{person}{Yang Zhao}, \bibinfo{person}{Jun Zhao},
  \bibinfo{person}{Linshan Jiang}, \bibinfo{person}{Rui Tan}, {and}
  \bibinfo{person}{Dusit Niyato}.} \bibinfo{year}{2019}\natexlab{b}.
\newblock \showarticletitle{Mobile Edge Computing, Blockchain and
  Reputation-based Crowdsourcing IoT Federated Learning: A Secure,
  Decentralized and Privacy-preserving System}.
\newblock \bibinfo{journal}{\emph{arXiv preprint arXiv:1906.10893}}
  (\bibinfo{year}{2019}).
\newblock


\bibitem[\protect\citeauthoryear{{Zhou}, {Wang}, {Guo}, {Gong}, and
  {Zheng}}{{Zhou} et~al\mbox{.}}{2019}]%
        {8807242}
\bibfield{author}{\bibinfo{person}{P. {Zhou}}, \bibinfo{person}{K. {Wang}},
  \bibinfo{person}{L. {Guo}}, \bibinfo{person}{S. {Gong}}, {and}
  \bibinfo{person}{B. {Zheng}}.} \bibinfo{year}{2019}\natexlab{}.
\newblock \showarticletitle{A Privacy-Preserving Distributed Contextual
  Federated Online Learning Framework with Big Data Support in Social
  Recommender Systems}.
\newblock \bibinfo{journal}{\emph{IEEE Transactions on Knowledge and Data
  Engineering}} (\bibinfo{year}{2019}), \bibinfo{pages}{1--1}.
\newblock
\showISSN{2326-3865}


\bibitem[\protect\citeauthoryear{{Zhou}, {Li}, {Chen}, and {Ding}}{{Zhou}
  et~al\mbox{.}}{2018}]%
        {8560084}
\bibfield{author}{\bibinfo{person}{W. {Zhou}}, \bibinfo{person}{Y. {Li}},
  \bibinfo{person}{S. {Chen}}, {and} \bibinfo{person}{B. {Ding}}.}
  \bibinfo{year}{2018}\natexlab{}.
\newblock \showarticletitle{Real-Time Data Processing Architecture for
  Multi-Robots Based on Differential Federated Learning}. In
  \bibinfo{booktitle}{\emph{2018 IEEE SmartWorld, Ubiquitous Intelligence
  Computing, Advanced Trusted Computing, Scalable Computing Communications,
  Cloud Big Data Computing, Internet of People and Smart City Innovation
  (SmartWorld/SCALCOM/UIC/ATC/CBDCom/IOP/SCI)}}. \bibinfo{pages}{462--471}.
\newblock


\bibitem[\protect\citeauthoryear{{Zhu}, {Wang}, and {Huang}}{{Zhu}
  et~al\mbox{.}}{2020}]%
        {8870236}
\bibfield{author}{\bibinfo{person}{G. {Zhu}}, \bibinfo{person}{Y. {Wang}},
  {and} \bibinfo{person}{K. {Huang}}.} \bibinfo{year}{2020}\natexlab{}.
\newblock \showarticletitle{Broadband Analog Aggregation for Low-Latency
  Federated Edge Learning}.
\newblock \bibinfo{journal}{\emph{IEEE Transactions on Wireless
  Communications}} \bibinfo{volume}{19}, \bibinfo{number}{1}
  (\bibinfo{date}{Jan} \bibinfo{year}{2020}), \bibinfo{pages}{491--506}.
\newblock
\showISSN{1558-2248}


\bibitem[\protect\citeauthoryear{{Zhu} and {Jin}}{{Zhu} and {Jin}}{2019}]%
        {8744465}
\bibfield{author}{\bibinfo{person}{H. {Zhu}} {and} \bibinfo{person}{Y. {Jin}}.}
  \bibinfo{year}{2019}\natexlab{}.
\newblock \showarticletitle{Multi-Objective Evolutionary Federated Learning}.
\newblock \bibinfo{journal}{\emph{IEEE Transactions on Neural Networks and
  Learning Systems}} (\bibinfo{year}{2019}), \bibinfo{pages}{1--13}.
\newblock
\showISSN{2162-2388}


\bibitem[\protect\citeauthoryear{Zhu, Li, and Yu}{Zhu et~al\mbox{.}}{2019}]%
        {Zhu2019}
\bibfield{author}{\bibinfo{person}{Xudong Zhu}, \bibinfo{person}{Hui Li}, {and}
  \bibinfo{person}{Yang Yu}.} \bibinfo{year}{2019}\natexlab{}.
\newblock \showarticletitle{Blockchain-Based Privacy Preserving Deep Learning}.
  \bibinfo{publisher}{Springer International Publishing},
  \bibinfo{address}{Cham}, \bibinfo{pages}{370--383}.
\newblock


\bibitem[\protect\citeauthoryear{{Zou}, {Feng}, {Niyato}, {Jiao}, {Gong}, and
  {Cheng}}{{Zou} et~al\mbox{.}}{2019}]%
        {8875353}
\bibfield{author}{\bibinfo{person}{Y. {Zou}}, \bibinfo{person}{S. {Feng}},
  \bibinfo{person}{D. {Niyato}}, \bibinfo{person}{Y. {Jiao}},
  \bibinfo{person}{S. {Gong}}, {and} \bibinfo{person}{W. {Cheng}}.}
  \bibinfo{year}{2019}\natexlab{}.
\newblock \showarticletitle{Mobile Device Training Strategies in Federated
  Learning: An Evolutionary Game Approach}. In \bibinfo{booktitle}{\emph{2019
  International Conference on Internet of Things (iThings) and IEEE Green
  Computing and Communications (GreenCom) and IEEE Cyber, Physical and Social
  Computing (CPSCom) and IEEE Smart Data (SmartData)}}.
  \bibinfo{pages}{874--879}.
\newblock


\end{thebibliography}

\appendix

\section{Appendix}
\subsection{Data extraction sheet of selected primary studies} \label{appendix_1}
\url{https://drive.google.com/file/d/10yYG8W1FW0qVQOru_kMyS86owuKnZPnz/view?usp=sharing}

\end{document}